\documentclass[epsfig,12pt]{article}
\usepackage{makeidx}
\usepackage{amsmath}
\usepackage{amsfonts}
\usepackage{amssymb}
\usepackage{graphicx}

\input epsf.sty
\newtheorem{theorem}{Theorem}
\newtheorem{acknowledgement}[theorem]{Acknowledgement}

\newtheorem{axiom}[theorem]{Axiom}

\newtheorem{conjecture}[theorem]{Conjecture}
\newtheorem{corollary}[theorem]{Corollary}

\newtheorem{definition}[theorem]{Definition}
\newtheorem{example}[theorem]{Example}
\newtheorem{exercise}[theorem]{Exercise}
\newtheorem{lemma}[theorem]{Lemma}

\newtheorem{proposition}[theorem]{Proposition}
\newtheorem{remark}[theorem]{Remark}

\newenvironment{proof}[1][Proof]{\noindent\textbf{#1.} }{\ \rule{0.5em}{0.5em}}
\makeatletter \@addtoreset{equation}{section}
\renewcommand{\theequation}{\thesection.\arabic{equation}}

\newcount\@hour\newcount\@minute\chardef\@x10\chardef\@xv60
\def\tcitime{
\def\@time{%
  \@minute\time\@hour\@minute\divide\@hour\@xv
  \ifnum\@hour<\@x 0\fi\the\@hour:%
  \multiply\@hour\@xv\advance\@minute-\@hour
  \ifnum\@minute<\@x 0\fi\the\@minute
  }}%

\def\x@hyperref#1#2#3{%
   \catcode`\~ = 12
   \catcode`\$ = 12
   \catcode`\_ = 12
   \catcode`\# = 12
   \catcode`\& = 12
   \y@hyperref{#1}{#2}{#3}%
}

\def\y@hyperref#1#2#3#4{%
   #2\ref{#4}#3
   \catcode`\~ = 13
   \catcode`\$ = 3
   \catcode`\_ = 8
   \catcode`\# = 6
   \catcode`\& = 4
}

\@ifundefined{hyperref}{\let\hyperref\x@hyperref}{}
\@ifundefined{msihyperref}{\let\msihyperref\x@hyperref}{}

\@ifundefined{qExtProgCall}{\def\qExtProgCall#1#2#3#4#5#6{\relax}}{}
\def\QCTOpt[#1]#2{%
  \def\QCTOptB{#1}
  \def\QCTOptA{#2}
}
\def\QCTNOpt#1{%
  \def\QCTOptA{#1}
  \let\QCTOptB\empty
}
\def\Qct{%
  \@ifnextchar[{%
    \QCTOpt}{\QCTNOpt}
}
\def\QCBOpt[#1]#2{%
  \def\QCBOptB{#1}%
  \def\QCBOptA{#2}%
}
\def\QCBNOpt#1{%
  \def\QCBOptA{#1}%
  \let\QCBOptB\empty
}
\def\Qcb{%
  \@ifnextchar[{%
    \QCBOpt}{\QCBNOpt}%
}
\def\PrepCapArgs{%
  \ifx\QCBOptA\empty
    \ifx\QCTOptA\empty
      {}%
    \else
      \ifx\QCTOptB\empty
        {\QCTOptA}%
      \else
        [\QCTOptB]{\QCTOptA}%
      \fi
    \fi
  \else
    \ifx\QCBOptA\empty
      {}%
    \else
      \ifx\QCBOptB\empty
        {\QCBOptA}%
      \else
        [\QCBOptB]{\QCBOptA}%
      \fi
    \fi
  \fi
}
\newcount\GRAPHICSTYPE
\GRAPHICSTYPE=\z@
\def\GRAPHICSPS#1{%
 \ifcase\GRAPHICSTYPE
   \special{ps: #1}%
 \or
   \special{language "PS", include "#1"}%
 \fi
}%
\def\graffile#1#2#3#4{%
    \bgroup
       \@inlabelfalse
       \leavevmode
       \@ifundefined{bbl@deactivate}{\def~{\string~}}{\activesoff}%
        \raise -#4 \BOXTHEFRAME{%
           \hbox to #2{\raise #3\hbox to #2{\null #1\hfil}}}%
    \egroup
}%
\def\draftbox#1#2#3#4{%
 \leavevmode\raise -#4 \hbox{%
  \frame{\rlap{\protect\tiny #1}\hbox to #2%
   {\vrule height#3 width\z@ depth\z@\hfil}%
  }%
 }%
}%
\newcount\@msidraft
\@msidraft=\z@
\let\nographics=\@msidraft
\newif\ifwasdraft
\wasdraftfalse

\def\GRAPHIC#1#2#3#4#5{%
   \ifnum\@msidraft=\@ne\draftbox{#2}{#3}{#4}{#5}%
   \else\graffile{#1}{#3}{#4}{#5}%
   \fi
}
\def\addtoLaTeXparams#1{%
    \edef\LaTeXparams{\LaTeXparams #1}}%
\newif\ifBoxFrame \BoxFramefalse
\newif\ifOverFrame \OverFramefalse
\newif\ifUnderFrame \UnderFramefalse

\def\BOXTHEFRAME#1{%
   \hbox{%
      \ifBoxFrame
         \frame{#1}%
      \else
         {#1}%
      \fi
   }%
}

\def\doFRAMEparams#1{\BoxFramefalse\OverFramefalse\UnderFramefalse\readFRAMEparams#1\end}%
\def\readFRAMEparams#1{%
 \ifx#1\end%
  \let\next=\relax
  \else
  \ifx#1i\dispkind=\z@\fi
  \ifx#1d\dispkind=\@ne\fi
  \ifx#1f\dispkind=\tw@\fi
  \ifx#1t\addtoLaTeXparams{t}\fi
  \ifx#1b\addtoLaTeXparams{b}\fi
  \ifx#1p\addtoLaTeXparams{p}\fi
  \ifx#1h\addtoLaTeXparams{h}\fi
  \ifx#1X\BoxFrametrue\fi
  \ifx#1O\OverFrametrue\fi
  \ifx#1U\UnderFrametrue\fi
  \ifx#1w
    \ifnum\@msidraft=1\wasdrafttrue\else\wasdraftfalse\fi
    \@msidraft=\@ne
  \fi
  \let\next=\readFRAMEparams
  \fi
 \next
 }%
\def\IFRAME#1#2#3#4#5#6{%
      \bgroup
      \let\QCTOptA\empty
      \let\QCTOptB\empty
      \let\QCBOptA\empty
      \let\QCBOptB\empty
      #6%
      \parindent=0pt
      \leftskip=0pt
      \rightskip=0pt
      \setbox0=\hbox{\QCBOptA}%
      \@tempdima=#1\relax
      \ifOverFrame
          \typeout{This is not implemented yet}%
          \show\HELP
      \else
         \ifdim\wd0>\@tempdima
            \advance\@tempdima by \@tempdima
            \ifdim\wd0 >\@tempdima
               \setbox1 =\vbox{%
                  \unskip\hbox to \@tempdima{\hfill\GRAPHIC{#5}{#4}{#1}{#2}{#3}\hfill}%
                  \unskip\hbox to \@tempdima{\parbox[b]{\@tempdima}{\QCBOptA}}%
               }%
               \wd1=\@tempdima
            \else
               \textwidth=\wd0
               \setbox1 =\vbox{%
                 \noindent\hbox to \wd0{\hfill\GRAPHIC{#5}{#4}{#1}{#2}{#3}\hfill}\\%
                 \noindent\hbox{\QCBOptA}%
               }%
               \wd1=\wd0
            \fi
         \else
            \ifdim\wd0>0pt
              \hsize=\@tempdima
              \setbox1=\vbox{%
                \unskip\GRAPHIC{#5}{#4}{#1}{#2}{0pt}%
                \break
                \unskip\hbox to \@tempdima{\hfill \QCBOptA\hfill}%
              }%
              \wd1=\@tempdima
           \else
              \hsize=\@tempdima
              \setbox1=\vbox{%
                \unskip\GRAPHIC{#5}{#4}{#1}{#2}{0pt}%
              }%
              \wd1=\@tempdima
           \fi
         \fi
         \@tempdimb=\ht1
         \advance\@tempdimb by -#2
         \advance\@tempdimb by #3
         \leavevmode
         \raise -\@tempdimb \hbox{\box1}%
      \fi
      \egroup%
}%
\def\DFRAME#1#2#3#4#5{%
  \hfil\break
  \bgroup
     \leftskip\@flushglue
     \rightskip\@flushglue
     \parindent\z@
     \parfillskip\z@skip
     \let\QCTOptA\empty
     \let\QCTOptB\empty
     \let\QCBOptA\empty
     \let\QCBOptB\empty
     \vbox\bgroup
        \ifOverFrame
           #5\QCTOptA\par
        \fi
        \GRAPHIC{#4}{#3}{#1}{#2}{\z@}%
        \ifUnderFrame
           \break#5\QCBOptA
        \fi
     \egroup
   \egroup
   \break
}%
\def\FFRAME#1#2#3#4#5#6#7{%
  \@ifundefined{floatstyle}
    {
     \begin{figure}[#1]%
    }
    {
     \ifx#1h
      \begin{figure}[H]%
     \else
      \begin{figure}[#1]%
     \fi
    }
  \let\QCTOptA\empty
  \let\QCTOptB\empty
  \let\QCBOptA\empty
  \let\QCBOptB\empty
  \ifOverFrame
    #4
    \ifx\QCTOptA\empty
    \else
      \ifx\QCTOptB\empty
        \caption{\QCTOptA}%
      \else
        \caption[\QCTOptB]{\QCTOptA}%
      \fi
    \fi
    \ifUnderFrame\else
      \label{#5}%
    \fi
  \else
    \UnderFrametrue%
  \fi
  \begin{center}\GRAPHIC{#7}{#6}{#2}{#3}{\z@}\end{center}%
  \ifUnderFrame
    #4
    \ifx\QCBOptA\empty
      \caption{}%
    \else
      \ifx\QCBOptB\empty
        \caption{\QCBOptA}%
      \else
        \caption[\QCBOptB]{\QCBOptA}%
      \fi
    \fi
    \label{#5}%
  \fi
  \end{figure}%
 }%
\newcount\dispkind%

\def\makeactives{
  \catcode`\"=\active
  \catcode`\;=\active
  \catcode`\:=\active
  \catcode`\'=\active
  \catcode`\~=\active
} \bgroup
   \makeactives
   \gdef\activesoff{%
      \def"{\string"}
      \def;{\string;}
      \def:{\string:}
      \def'{\string'}
      \def~{\string~}
    }
\egroup

\def\FRAME#1#2#3#4#5#6#7#8{%
 \bgroup
 \ifnum\@msidraft=\@ne
   \wasdrafttrue
 \else
   \wasdraftfalse%
 \fi
 \def\LaTeXparams{}%
 \dispkind=\z@
 \def\LaTeXparams{}%
 \doFRAMEparams{#1}%
 \ifnum\dispkind=\z@\IFRAME{#2}{#3}{#4}{#7}{#8}{#5}\else
  \ifnum\dispkind=\@ne\DFRAME{#2}{#3}{#7}{#8}{#5}\else
   \ifnum\dispkind=\tw@
    \edef\@tempa{\noexpand\FFRAME{\LaTeXparams}}%
    \@tempa{#2}{#3}{#5}{#6}{#7}{#8}%
    \fi
   \fi
  \fi
  \ifwasdraft\@msidraft=1\else\@msidraft=0\fi{}%
  \egroup
 }%
\def\TEXUX#1{"texux"}

\long\def\QQQ#1#2{%
     \long\expandafter\def\csname#1\endcsname{#2}}%
\@ifundefined{QTP}{\def\QTP#1{}}{}
\@ifundefined{QEXCLUDE}{\def\QEXCLUDE#1{}}{}
\@ifundefined{Qlb}{}{}
\@ifundefined{Qlt}{}{}
\long\def\QQA#1#2{}%
\def\QTR#1#2{{\csname#1\endcsname #2}}
\def\EXPAND#1[#2]#3{}%
\def\NOEXPAND#1[#2]#3{}%
\def\LaTeXparent#1{}%
\def\ChildStyles#1{}%
\def\ChildDefaults#1{}%
\def\QTagDef#1#2#3{}%

\@ifundefined{correctchoice}{}{}
\@ifundefined{HTML}{\def\HTML#1{\relax}}{}
\@ifundefined{TCIIcon}{\def\TCIIcon#1#2#3#4{\relax}}{}
\if@compatibility
\else
  \providecommand{\UNICODE}[2][]{\protect\rule{.1in}{.1in}}
  \providecommand{\U}[1]{\protect\rule{.1in}{.1in}}
  
\fi

\@ifundefined{lambdabar}{
      
   }{}

\@ifundefined{StyleEditBeginDoc}{}{}
\def\QQfnmark#1{\footnotemark}

\@ifundefined{TCIMAKEINDEX}{}{\makeindex}%
\@ifundefined{abstract}{%
 \def\abstract{%
  \if@twocolumn
   \section*{Abstract (Not appropriate in this style!)}%
   \else \small
   \begin{center}{\bf Abstract\vspace{-.5em}\vspace{\z@}}\end{center}%
   \quotation
   \fi
  }%
 }{%
 }%
\@ifundefined{endabstract}{\def\endabstract
  {\if@twocolumn\else\endquotation\fi}}{}%
\@ifundefined{maketitle}{\def\maketitle#1{}}{}%
\@ifundefined{affiliation}{\def\affiliation#1{}}{}%
\@ifundefined{proof}{}{}%
\@ifundefined{endproof}{}{}%
\@ifundefined{newfield}{\def\newfield#1#2{}}{}%
\@ifundefined{chapter}{\def\chapter#1{\par(Chapter head:)#1\par }%
 \newcount\c@chapter}{}%
\@ifundefined{part}{\def\part#1{\par(Part head:)#1\par }}{}%
\@ifundefined{section}{\def\section#1{\par(Section head:)#1\par }}{}%
\@ifundefined{subsection}{\def\subsection#1%
 {\par(Subsection head:)#1\par }}{}%
\@ifundefined{subsubsection}{\def\subsubsection#1%
 {\par(Subsubsection head:)#1\par }}{}%
\@ifundefined{paragraph}{\def\paragraph#1%
 {\par(Subsubsubsection head:)#1\par }}{}%
\@ifundefined{subparagraph}{\def\subparagraph#1%
 {\par(Subsubsubsubsection head:)#1\par }}{}%
\@ifundefined{therefore}{}{}%
\@ifundefined{backepsilon}{}{}%
\@ifundefined{yen}{}{}%
\@ifundefined{registered}{%
   \def\registered{\relax\ifmmode{}\r@gistered
                    \else$\m@th\r@gistered$\fi}%
 \def\r@gistered{^{\ooalign
  {\hfil\raise.07ex\hbox{$\scriptstyle\rm\text{R}$}\hfil\crcr
  \mathhexbox20D}}}}{}%
\@ifundefined{Eth}{}{}%
\@ifundefined{eth}{}{}%
\@ifundefined{Thorn}{}{}%
\@ifundefined{thorn}{}{}%
\@ifundefined{degree}{}{}%
\newdimen\theight
\@ifundefined{Column}{\def\Column{%
 \vadjust{\setbox\z@=\hbox{\scriptsize\quad\quad tcol}%
  \theight=\ht\z@\advance\theight by \dp\z@\advance\theight by \lineskip
  \kern -\theight \vbox to \theight{%
   \rightline{\rlap{\box\z@}}%
   \vss
   }%
  }%
 }}{}%
\@ifundefined{qed}{\def\qed{%
 \ifhmode\unskip\nobreak\fi\ifmmode\ifinner\else\hskip5\p@\fi\fi
 \hbox{\hskip5\p@\vrule width4\p@ height6\p@ depth1.5\p@\hskip\p@}%
 }}{}%
\@ifundefined{cents}{}{}%
\@ifundefined{tciLaplace}{}{}%
\@ifundefined{tciFourier}{}{}%
\@ifundefined{textcurrency}{}{}%
\@ifundefined{texteuro}{}{}%
\@ifundefined{textfranc}{}{}%
\@ifundefined{textlira}{}{}%
\@ifundefined{textpeseta}{}{}%
\@ifundefined{miss}{\def\miss{\hbox{\vrule height2\p@ width 2\p@ depth\z@}}}{}%
\@ifundefined{vvert}{}{}
\@ifundefined{tcol}{\def\tcol#1{{\baselineskip=6\p@ \vcenter{#1}} \Column}}{}%
\@ifundefined{dB}{}{}
\@ifundefined{mB}{}{}
\@ifundefined{nB}{}{}
\@ifundefined{note}{}{}%
\def\newfmtname{LaTeX2e}
\ifx\fmtname\newfmtname
  \DeclareOldFontCommand{\rm}{\normalfont\rmfamily}{\mathrm}
  \DeclareOldFontCommand{\sf}{\normalfont\sffamily}{\mathsf}
  \DeclareOldFontCommand{\tt}{\normalfont\ttfamily}{\mathtt}
  \DeclareOldFontCommand{\bf}{\normalfont\bfseries}{\mathbf}
  \DeclareOldFontCommand{\it}{\normalfont\itshape}{\mathit}
  \DeclareOldFontCommand{\sl}{\normalfont\slshape}{\@nomath\sl}
  \DeclareOldFontCommand{\sc}{\normalfont\scshape}{\@nomath\sc}
\fi

\def\alpha{{\Greekmath 010B}}%
\def\beta{{\Greekmath 010C}}%
\def\gamma{{\Greekmath 010D}}%
\def\delta{{\Greekmath 010E}}%
\def\epsilon{{\Greekmath 010F}}%
\def\zeta{{\Greekmath 0110}}%
\def\eta{{\Greekmath 0111}}%
\def\theta{{\Greekmath 0112}}%
\def\iota{{\Greekmath 0113}}%
\def\kappa{{\Greekmath 0114}}%
\def\lambda{{\Greekmath 0115}}%
\def\mu{{\Greekmath 0116}}%
\def\nu{{\Greekmath 0117}}%
\def\xi{{\Greekmath 0118}}%
\def\pi{{\Greekmath 0119}}%
\def\rho{{\Greekmath 011A}}%
\def\sigma{{\Greekmath 011B}}%
\def\tau{{\Greekmath 011C}}%
\def\upsilon{{\Greekmath 011D}}%
\def\phi{{\Greekmath 011E}}%
\def\chi{{\Greekmath 011F}}%
\def\psi{{\Greekmath 0120}}%
\def\omega{{\Greekmath 0121}}%
\def\varepsilon{{\Greekmath 0122}}%
\def\vartheta{{\Greekmath 0123}}%
\def\varpi{{\Greekmath 0124}}%
\def\varrho{{\Greekmath 0125}}%
\def\varsigma{{\Greekmath 0126}}%
\def\varphi{{\Greekmath 0127}}%

\def\nabla{{\Greekmath 0272}}
\def\FindBoldGroup{%
   {\setbox0=\hbox{$\mathbf{x\global\edef\theboldgroup{\the\mathgroup}}$}}%
}

\def\Greekmath#1#2#3#4{%
    \if@compatibility
        \ifnum\mathgroup=\symbold
           \mathchoice{\mbox{\boldmath$\displaystyle\mathchar"#1#2#3#4$}}%
                      {\mbox{\boldmath$\textstyle\mathchar"#1#2#3#4$}}%
                      {\mbox{\boldmath$\scriptstyle\mathchar"#1#2#3#4$}}%
                      {\mbox{\boldmath$\scriptscriptstyle\mathchar"#1#2#3#4$}}%
        \else
           \mathchar"#1#2#3#4%
        \fi
    \else
        \FindBoldGroup
        \ifnum\mathgroup=\theboldgroup 
           \mathchoice{\mbox{\boldmath$\displaystyle\mathchar"#1#2#3#4$}}%
                      {\mbox{\boldmath$\textstyle\mathchar"#1#2#3#4$}}%
                      {\mbox{\boldmath$\scriptstyle\mathchar"#1#2#3#4$}}%
                      {\mbox{\boldmath$\scriptscriptstyle\mathchar"#1#2#3#4$}}%
        \else
           \mathchar"#1#2#3#4%
        \fi
      \fi}

\newif\ifGreekBold  \GreekBoldfalse
\let\SAVEPBF=\pbf
\def\pbf{\GreekBoldtrue\SAVEPBF}%

\@ifundefined{theorem}{\newtheorem{theorem}{Theorem}}{}
\@ifundefined{lemma}{}{}
\@ifundefined{corollary}{}{}
\@ifundefined{conjecture}{}{}
\@ifundefined{proposition}{}{}
\@ifundefined{axiom}{}{}
\@ifundefined{remark}{}{}
\@ifundefined{example}{}{}
\@ifundefined{exercise}{}{}
\@ifundefined{definition}{}{}

\@ifundefined{mathletters}{%
  \newcounter{equationnumber}
  \def\mathletters{%
     \addtocounter{equation}{1}
     \edef\@currentlabel{\theequation}%
     \setcounter{equationnumber}{\c@equation}
     \setcounter{equation}{0}%
     \edef\theequation{\@currentlabel\noexpand\alph{equation}}%
  }
  
}{}

\@ifundefined{BibTeX}{%
    \def\BibTeX{{\rm B\kern-.05em{\sc i\kern-.025em b}\kern-.08em
                 T\kern-.1667em\lower.7ex\hbox{E}\kern-.125emX}}}{}%
\@ifundefined{AmS}%
    {\def\AmS{{\protect\usefont{OMS}{cmsy}{m}{n}%
                A\kern-.1667em\lower.5ex\hbox{M}\kern-.125emS}}}{}%
\@ifundefined{AmSTeX}{}{}%
\def\@@eqncr{\let\@tempa\relax
    \ifcase\@eqcnt \def\@tempa{& & &}\or \def\@tempa{& &}%
      \else \def\@tempa{&}\fi
     \@tempa
     \if@eqnsw
        \iftag@
           \@taggnum
        \else
           \@eqnnum\stepcounter{equation}%
        \fi
     \fi
     \global\tag@false
     \global\@eqnswtrue
     \global\@eqcnt\z@\cr}

\def\TCItag{\@ifnextchar*{\@TCItagstar}{\@TCItag}}
\def\@TCItag#1{%
    \global\tag@true
    \global\def\@taggnum{(#1)}}
\def\@TCItagstar*#1{%
    \global\tag@true
    \global\def\@taggnum{#1}}
\def\dint{\displaystyle \int}%
\def\dsum{\mathop{\displaystyle \sum }}%
\def\dprod{\mathop{\displaystyle \prod }}%
\def\dbigoplus{\mathop{\displaystyle \bigoplus }}%
\def\dbigcup{\mathop{\displaystyle \bigcup }}%

\begin{document}

\title{\vspace{-2cm}\rightline{\mbox{\small
{Lab/UFR-HEP0902/GNPHE/0902}}} \textbf{\ Quiver Gauge Models in F-Theory on}%
\\
\textbf{Local Tetrahedron }}
\author{Lalla Btissam Drissi\thanks{{\small Permanent address: Institute for
Nanomaterials and Nanotechnology, Rabat, Morocco}}\thanks{%
drissilb@gmail.com}$^{\text{ }1,2}$, Leila Medari\thanks{%
mml.leila@yahoo.fr}$^{3}$, El Hassan Saidi{\small \thanks{%
h-saidi@fsr.ac.ma}}$^{\text{ }1,2}${\small \ } \\
{\small 1. Lab/UFR- Physique des Hautes Energies, Facult\'{e} des Sciences,
Rabat, Morocco,}\\
{\small 2. GNPHE, focal point, Facult\'{e} des  Sciences Rabat, Morocco,}\\
{\small 3. LPHEA, Physics Department, Faculty of Science Semlalia,
Marrakesh, Morocco}\\
}
\maketitle

\begin{abstract}
We study a class of 4D $\mathcal{N}=1$ supersymmetric GUT- type models in
the framework of the Beasley-Heckman-Vafa theory. We first review general
results on MSSM and supersymmetric GUT; and we describe useful tools on 4D
quiver gauge theories in F- theory set up. Then we study the effective
supersymmetric gauge theory in the 7-brane wrapping 4-cycles in F-theory on\
local elliptic CY4s based on a complex tetrahedral surface $\mathcal{T}$ and
its blown ups $\mathcal{T}_{n}$. The complex 2d geometries $\mathcal{T}$ and
$\mathcal{T}_{n}$ are \emph{non planar} projective surfaces that extend the
projective plane $\mathbb{P}^{2}$ and the del Pezzos. Using the power of
toric geometry encoding the toric data of the base of the local CY4, we
build a class of \emph{4D} $\mathcal{N}=1$ non minimal GUT- type models
based on $\mathcal{T}$ and $\mathcal{T}_{n}$. An explicit construction is
given for the SU$\left( 5\right) $ GUT-type model.\newline
\end{abstract}


\section{Introduction}

\begin{figure}[tbph]
\begin{center}
\hspace{0cm} \includegraphics[width=6cm]{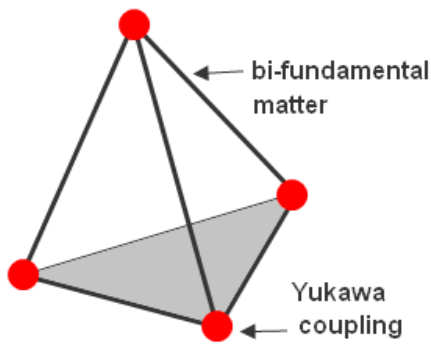}
\end{center}
\par
\vspace{-0.5 cm}
\caption{{\protect\small Toric graph of the tetrahedral surface }$\mathcal{T}%
={\protect\small \cup }_{a=1}^{4}{\protect\small S}_{a}$ with $%
{\protect\small S}_{a}{\protect\small \cap S}_{b}{\protect\small =\Sigma }%
_{ab}${\protect\small . The toric fibration of }$\mathcal{T}${\protect\small %
\ degenerate once on the six edges }${\protect\small \Sigma }_{ab}$%
{\protect\small \ and twice at the vertices }${\protect\small P}_{abc}%
{\protect\small =S}_{a}{\protect\small \cap S}_{b}{\protect\small \cap S}%
_{c} ${\protect\small .}}
\label{by}
\end{figure}

In the last few years an increasing interest has been given to linking
superstring theory to the low energy elementary particle physics
phenomenology\textrm{\ \cite{A1}-\cite{A4}}. Several attempts have been
particulary focusing on type II superstrings and M- theory to engineer
extensions of the Minimal Supersymmetric Standard Model (\emph{MSSM}) of
elementary particles at TeV- scale \textrm{\cite{C1}-\cite{D4}.} This
interest in physics beyond standard model is also motivated by the Large
Hadron Collider (LHC) event whose ATLAS and CMS detectors are expected to
capture new signals beyond the theory of electroweak interactions\textrm{\
\cite{B1}-\cite{B4}}. Recall that the energy magnitude used in the LHC and
the power of the grid computing constitute the beginning of a new era for
testing several ideas and proposals such as supersymmetry and extra
dimensions \textrm{\cite{E1}-\cite{E4}}. The access to the TeV energy band
will allow to check early phenomenological prototypes beyond the $%
SU_{C}\left( {\small 3}\right) \times SU_{L}\left( {\small 2}\right) \times
U_{Y}\left( {\small 1}\right) $ Standard Model such as the $SU_{C}\left(
4\right) \times SU_{L}\left( 2\right) \times SU_{R}\left( 2\right) $
Pati-Salam model treating quarks and leptons on equal footing \textrm{\cite%
{F1,F2}} and the $SU^{3}\left( 3\right) $ tri- unification \textrm{\cite%
{F3,F4,F5}}. The TeV band allows as well to shed more light on grand unified
theory (GUT) proposals; especially those based on gauge symmetry groups like
$SU\left( 5\right) $, flipped $SU\left( 5\right) $, $SO\left( 10\right) $
and $E_{6}$ GUT models \textrm{\cite{G1,G2,G3,G4}}.\newline
Recently a model has been proposed to linking quantum physics at TeV
energies to twelve dimensional F- theory compactified on a local Calabi-Yau
four- folds in the limit of decoupled supergravity \textrm{\cite{H1}}. In
this proposal, to which we shall refer to as the \textrm{BHV} theory, and
which has been further developed in a series of seminal papers \textrm{\cite%
{H2, H3, H4, H5}}, the visible $\mathcal{N}=1$ supersymmetric local GUT
models in the \emph{4D} space time is given by an effective non abelian
gauge theory living on a seven brane wrapping 4- cycles in F- theory on
local elliptically K3 fibered Calabi-Yau four- folds $X_{4}$,%
\begin{equation}
\begin{tabular}{lll}
$Y$ & $\rightarrow $ & $X_{4}$ \\
&  & $\downarrow \pi _{s}$ \\
&  & $S$%
\end{tabular}
\label{F}
\end{equation}%
where, to fix the ideas, the base surface $S$ is thought of as the del Pezzo
complex base surface $dP_{8}$. Together with the nature of the singularity
in the fiber Y (type A, type D or type E) which engineer the gauge
invariance that we see in the 4D space time, the del Pezzo surface $dP_{8}$
and its $dP_{n}$ sisters with $n\geq 5$ are used to engineer chiral matter
and Yukawa couplings of the 4D space time standard model and beyond. Notice
that besides $\mathcal{N}=1$ supersymmetry in 4D space time, the $dP_{n}$
base surfaces play as well a central role in the \textrm{BHV theory} due to
their special features; in particular the two following:\newline
(\textbf{1}) the $dP_{n}$'s are in some sense artificial surfaces engineered
by performing blow ups of the complex projective plane $\mathbb{P}^{2}$ at n
isolated points ($n\leq 8$) \textrm{\cite{DP, DQ, DR}}. In addition to the
hyperline class H of projective plane $\mathbb{P}^{2}$, the blow ups are
generated by n exceptional curves $E_{i}$, which altogether with H, generate
the $\left( 1+n\right) $ dimensional homology group $H_{2}\left( dP_{n},%
\mathbb{Z}\right) $ of real 2- cycles in the complex $dP_{n}$ surfaces.%
\newline
(\textbf{2}) the $dP_{n}$'s are also remarkably linked to the "exceptional"
\textbf{E}$_{n}$ Lie algebras \textrm{\cite{H1,DP, DR} }which are known to
exist in the non perturbative regime of type IIB superstrings realized as
F-theory. The real 2- cycle homology group $H_{2}\left( dP_{n},\mathbb{Z}%
\right) $ decomposes as the direct sum,
\begin{equation}
H_{2}\left( dP_{n},\mathbb{Z}\right) =\Omega _{n}\oplus \mathcal{L}_{n},
\label{L}
\end{equation}%
with $\Omega _{n}$\ being the anticanonical class of $dP_{n}$ and the
orthogonal class $\mathcal{L}_{n}$ is a n dimensional sublattice that is
isomorphic to the root space of the exceptional Lie algebras \textbf{E}$_{n}$%
. \newline
These two properties make the $dP_{n}$s very special complex surfaces which
allow an explicit geometric engineering of: \newline
(\textbf{a}) chiral matter localizing on complex curves $\Sigma _{i}$ at the
intersections of seven branes wrapping $dP_{n}$,\newline
(\textbf{b}) the MSSM and GUT tri-fields Yukawa couplings localizing at
isolated points in the del Pezzo surface $dP_{n}$ with $n\geq 5$ where
matter curves intersect and where the bulk gauge invariance gets enhanced
\textrm{\cite{H1}}.\newline
The aim of this paper is to contribute to the efforts for the study of
embedding the \emph{MSSM} and $\mathcal{N}=1$ supersymmetric GUT models in
F-theory compactification on local Calabi-Yau four- folds (CY4). More
precisely, we focus on \emph{4D} $\mathcal{N}=1$ supersymmetric GUT- type
models along the line of the \textrm{BHV theory}; but by considering a seven
brane wrapping 4-cycles in F-theory local CY4- folds based on a \emph{%
tetrahedral} surface $\mathcal{T}$ of the figure (\ref{by}) and its toric
\emph{blown ups }$\mathcal{T}_{n}$. Using these backgrounds, we first
engineer unrealistic $\mathcal{N}=1$ supersymmetric GUT- type models based
on the tetrahedron $\mathcal{T}$. Then we consider extensions based on a
particular class of blow ups\ of the tetrahedron namely the sub-family $%
\mathcal{T}_{n}^{{\small toric}}$ of toric blow ups of $\mathcal{T}$. These
extensions, which involve exotic matter, constitute a step towards
engineering non minimal quasi-realistic \emph{4D}$\ \mathcal{N}=1$
supersymmetric GUT on $\mathcal{T}$ \ and $\mathcal{T}_{n}$. To fix the
ideas, we shall mainly focus on the engineering of supersymmetric $SU\left(
5\right) $ GUT- type models based on $\mathcal{T}$ and $\mathcal{T}_{n}$;
but the method works as well for the other GUT gauge groups.

\ \ \newline
Before proceeding it is interesting to say few words about the motivations
behind our interest into the tetrahedral surface $\mathcal{T}$ and its blown
ups $\mathcal{T}_{n}$ as a base surface of the local CY4-folds. Besides the
fact of being a particular\emph{\ non planar} complex surface, our interest
into the tetrahedron and the cousin geometries has been motivated by the two
following features:\newline
\textbf{(i)} the complex tetrahedral surface $\mathcal{T}$ , viewed as a
toric surface, has a natural toric fibration given by a 2- torus $\mathbb{T}%
^{2}$\ fibered over a real two dimension base $B_{2}$,
\begin{equation}
\begin{tabular}{lll}
$\mathbb{T}^{2}$ & $\rightarrow $ & $\mathcal{T}$ \\
&  & $\downarrow \pi _{_{B}}$ \\
&  & $B_{2}$%
\end{tabular}
\label{fb}
\end{equation}%
The complex surface $\mathcal{T}$ is nicely represented by a toric graph $%
\Delta _{\mathcal{T}}$ which is precisely the usual real tetrahedron given
by the figure (\ref{by}). The polytope $\Delta _{\mathcal{T}}$ encodes the
toric data on the shrinking cycles of the toric fibration (\ref{fb}).
\newline
(\textbf{ii}) the toric geometry of the tetrahedral surface $\mathcal{T}$
has a set of remarkable properties that have an interpretation in F- theory
GUT models building. Below, we describe three of these features:\newline
($\mathbf{\alpha }$) the 2-torus fibration (\ref{fb}) has an inherent $%
U\left( 1\right) \times U\left( 1\right) $ gauge symmetry which may be
interpreted in F-theory compactifications in terms of abelian gauge
symmetries. Each $U\left( 1\right) $ factor describes gauge translation
along compact 1-cycles in $\mathbb{T}^{2}$. \newline
($\mathbf{\beta }$) the $\mathbb{T}^{2}$ fiber has two shrinking properties:
first down to 1- cycles on the six edges of the tetrahedron and second down
to 0-cycles at its four vertices. \newline
These properties capture in a remarkable way the enhancement of gauge
symmetry used in the engineering of the F-theory GUT-models \`{a} la BHV.
\newline
Notice moreover that the non planar tetrahedral surface $\mathcal{T}$
involves:

\begin{itemize}
\item four intersecting planar faces $S_{a}$, $a=1,2,3,4$, with different 2-
torus fibers $\mathbb{T}_{a}^{2}$,

\item six intersecting edges $\Sigma _{ab}=S_{a}\cap S_{b}$ having different
1- cycle fibers $\mathbb{S}_{ab}^{1}$,

\item four vertices $P_{abc}$ given by the curves intersection $\Sigma
_{ab}\cap \Sigma _{bc}\cap \Sigma _{ca}$. At these special points, the 2-
torus fiber eq\ref{fb}) shrinks to zero.
\end{itemize}

\ \ \newline
From the view of the F-theory- GUT models building, the faces $S_{a}$ of the
tetrahedron correspond roughly to 4- cycles wrapped by seven branes. These
faces intersect mutually along six edges $\Sigma _{ab}$ on which the fibers $%
\mathbb{T}_{a}^{2}$ and $\mathbb{T}_{b}^{2}$ shrink down to $\mathbb{S}%
_{ab}^{1}$. Along these curves seven branes intersect and give rise to
bi-fundamental matter. Moreover, three faces $S_{a}$, $S_{b}$ and $S_{c}$
intersect at a point $P_{abc}$ corresponding to a vertex of the figure (\ref%
{by}). From this picture, it follows that the vertices of the tetrahedron
are good candidates to host the tri-fields Yukawa couplings such as those of
the 4D supersymmetric SU$\left( 5\right) $- GUT model namely,%
\begin{equation}
\begin{tabular}{llllll}
$H_{u}Q_{10}Q_{10}$ & $\rightarrow $ & $5_{H}\otimes 10_{M}\otimes 10_{M}$ &
$\rightarrow $ & $P_{1}$ & , \\
$\digamma H_{u}H_{d}$ & $\rightarrow $ & $1_{E}\otimes 5_{H}\otimes \bar{5}%
_{H}$ & $\rightarrow $ & $P_{2}$ & , \\
$N_{R}H_{u}Q_{\bar{5}}$ & $\rightarrow $ & $1_{M}\otimes 5_{H}\otimes \bar{5}%
_{M}$ & $\rightarrow $ & $P_{3}$ & , \\
$H_{d}Q_{\bar{5}}Q_{10}$ & $\rightarrow $ & $\bar{5}_{H}\otimes \bar{5}%
_{M}\otimes 10_{M}$ & $\rightarrow $ & $P_{4}$ & . \\
&  &  &  &  &
\end{tabular}
\label{ykc}
\end{equation}%
In these relations, $5_{H}$ refers to Higgs fields and $\bar{5}_{M}$, $%
10_{M} $ to matter. The vertex $P_{1}$ stands for $P_{\left( 234\right) }$
and similarly for the others.\newline
($\mathbf{\gamma }$) The third feature behind the study of this local
Calabi-Yau four- folds geometry is that the tetrahedral surface $\mathcal{T}$
shares also some basic properties with the del Pezzo surfaces $dP_{n}$ used
in the \textrm{BHV\ theory}. The point is that each one of the four faces $%
S_{a}$ of the tetrahedron is in one to one with the four projective plane $%
\mathbb{P}_{a}^{2}$ in the complex three dimension projective space $\mathbb{%
P}^{3}$,%
\begin{equation}
\begin{tabular}{llllll}
$S_{a}$ & $\leftrightarrow $ & $\mathbb{P}_{a}^{2}$ & $,$ & $a=1,...,4$ & $.$%
\end{tabular}%
\end{equation}%
On each of these $\mathbb{P}_{a}^{2}$s, one may a priori perform blow ups
leading to a $\mathcal{T}_{n}$ family of blown tetrahedrons. The number of
blow ups of the tetrahedron are obviously richer than the ones encountered
in the del Pezzo surfaces since the tetrahedron involves four kinds of
projective planes; for more details see \textrm{\cite{SA,AS}}. From this
view blown ups of the tetrahedral surface may be thought of as given by
intersections of del Pezzo surfaces and thereby F-theory GUT models based on
blown ups tetrahedron could incorporate the BHV ones based on del Pezzo
surfaces.\newline
The presentation of this paper is as follows: In section 2, we review
briefly the main lines of MSSM and supersymmetric GUT models in \emph{4D }%
space time. Comments using quiver gauge theory ideas and intersecting brane
realizations are also given. In section 3, we review general results on
F-theory and we study the engineering of the \emph{non abelian} gauge
symmetries in the frame work of F- theory on local CY4- folds. An heuristic
classification of pure and hybrid colliding singularities in CY4s is also
made. In section 4, we first review $\mathcal{N}=1$ supergravity theory
coupled to super Yang-Mills. Then, we focus on the gauge theory in the seven
branes wrapping 4- cycles and study the engineering of the effective $%
\mathcal{N}=1$ supersymmetric gauge theory in \emph{4D} obtained by using
topological twisted ideas. In section 5, we study the engineering of
F-theory GUT- model along the line of the \textrm{BHV}\ approach. We take
this opportunity to give a brane realization of SU$\left( 5\right) $ GUT
model by using five stacks of intersecting seven branes. In section 6, we
study F- theory on local CY4- folds based on the complex tetrahedral surface
$\mathcal{T}$ and its $\mathcal{T}_{n}$ blown ups and we develop a first
class of F- theory GUT- type models based on the $\mathcal{T}$. In section
7, we build a second class of F- theory GUT- type models based on $\mathcal{T%
}_{n}$ blown ups and fractional bundle ideas. In section 8, we give our
conclusion and in section 9, we give an appendix on the engineering of bi-
fundamental matter in F- theory GUT- models building.

\section{General on MSSM and GUT}

In this section we review briefly some useful tools on the \emph{MSSM} and
the \emph{4D} $\mathcal{N}=1$ Supersymmetric Grand Unified Theories (\emph{%
SGUT}) as well as general links to superstrings. These tools are helpful to
fix the ideas on: (\textbf{1}) how fundamental matter and gauge particles
get unified into group representations method and (\textbf{2}) how the
geometric tri-fields Yukawa couplings (\ref{ykc}) are handled in the \emph{4D%
} $\mathcal{N}=1$ superfield theory set up. These materials are also needed
for later use when we consider the embedding \emph{SGUT}- type models into
the effective non abelian twisted gauge theory \textrm{\cite{H1}} on the
seven brane wrapping 4-cycles in the twelve dimensional F-theory on local
Calabi-Yau four- folds.

\subsection{MSSM}

We start by recalling some general aspects on Standard Model of electroweak
interactions. The basic elements in this model are as follows: \newline
\textbf{(a)} the elementary particles namely: quarks, leptons, gauge bosons
and Higgs particles,\newline
\textbf{(b)} the $SU_{C}\left( {\small 3}\right) $ $\times $ $SU_{L}\left(
{\small 2}\right) $ $\times $ $U_{Y}\left( {\small 1}\right) $ gauge
symmetry to be denoted as $G_{str}$,\newline
(\textbf{c}) the $G_{str}$ representations unifying the particles into gauge
group multiplets. \newline
In the Cartan basis, the Lie algebra of the Standard Model group $G_{str}$
is generated by the following matrices,%
\begin{equation}
\begin{tabular}{lll|ll|ll}
&  & ${\small SU}_{C}\left( {\small 3}\right) $ &  & ${\small SU}_{L}\left(
{\small 2}\right) $ &  & ${\small U}_{Y}\left( {\small 1}\right) $ \\
\hline\hline
$\text{{\small Cartan} {\small operators}}$ & $:$ & $H_{{\small su}\left(
{\small 3}\right) }^{1},\text{ \ }H_{{\small su}\left( {\small 3}\right)
}^{2}$ &  & $\text{ }H_{{\small su}\left( {\small 2}\right) }^{0}$ &  & $Y_{%
{\small u}\left( {\small 1}\right) }$ \\
&  &  &  &  &  & - \\
$\text{{\small step operators}}$ & $:$ & $E_{{\small su}\left( {\small 3}%
\right) }^{\pm \alpha }$ &  & $E_{{\small su}\left( {\small 2}\right) }^{\pm
}$ &  &  \\ \hline
\end{tabular}%
\end{equation}%
where $H_{{\small su}\left( {\small 3}\right) }^{1},$ $H_{{\small su}\left(
{\small 3}\right) }^{2},$ $H_{{\small su}\left( {\small 2}\right) }^{0}$, $%
Y_{{\small u}\left( {\small 1}\right) }$ are commuting Cartan generators and
$E_{{\small su}\left( {\small 2}\right) }^{\pm }$, $E_{{\small su}\left(
{\small 3}\right) }^{\pm \alpha }$ are step operators with $\alpha $ being a
generic positive root of the $SU_{C}\left( {\small 3}\right) $ root system $%
\Delta _{su\left( 3\right) }$. \newline
The fundamental particles of the Standard Model are of two kinds:\newline
\textbf{(i)} Elementary fermions forming three hierarchical families \emph{F}%
$\left( e\right) $, \emph{F}$\left( \mu \right) $ and \emph{F}$\left( \tau
\right) $; each one containing quarks and leptons in different
representations of the $G_{str}$ group. For the family \emph{F}$\left(
e\right) $ of the electron $e^{-}\equiv e$, the \emph{sixteen} left- handed
fermions are packaged into smaller representations $R_{{\small su}_{C}\left(
{\small 3}\right) }\times R_{{\small su}_{L}\left( {\small 2}\right) }\times
R_{{\small u}_{Y}\left( {\small 1}\right) }$ of the $G_{str}$ gauge symmetry
as given below,%
\begin{equation}
\begin{tabular}{ll|llll}
{\small Quarks} & ${\small \qquad }$ & \qquad & \qquad & {\small Leptons} &
\\ \hline\hline
&  &  &  &  &  \\
$q=\left(
\begin{array}{c}
u \\
d%
\end{array}%
\right) $ &  &  &  & $l=\left(
\begin{array}{c}
\nu _{e} \\
e%
\end{array}%
\right) $ &  \\
$u^{c}$ \ \ $,\ \ \ d^{c}$ &  &  &  & $\nu ^{c}$ $\ \ ,$ $\ \ e^{c}$ &  \\
&  &  &  &  &  \\ \hline
\end{tabular}
\label{ql}
\end{equation}%
Using the conventional notation $\left( n,m\right) _{y}$ with $m=\dim
{\small R}_{{\small su}_{C}\left( {\small 3}\right) }$, $n=\dim {\small R}_{%
{\small su}_{L}\left( {\small 2}\right) }$ and $y$ being the eigenvalue of
the hypercharge charge ${\small R}_{{\small Y}}$, the group theoretical
description of the \emph{F}$\left( e\right) $ family is as follows:

\begin{equation}
\begin{tabular}{llllll}
$\ \ q$ & $\ \ u^{c}$ & $\ \ d^{c}$ & $\ \ l$ & $\ \ \nu ^{c}$ & $\ \ e^{c}$
\\ \hline\hline
&  &  &  &  &  \\
$\left( {\small 3,2}\right) _{\frac{{\small 1}}{{\small 3}}}$ & $\left(
{\small \bar{3}},1\right) _{\frac{-{\small 4}}{{\small 3}}}$ & $\left(
{\small \bar{3}},{\small 1}\right) _{\frac{{\small 2}}{{\small 3}}}$ & $%
\left( {\small 1},{\small 2}\right) _{{\small -1}}$ & $\left( {\small 1,1}%
\right) _{{\small 0}}$ & $\left( {\small 1,1}\right) _{{\small 2}}$ \\
&  &  &  &  &  \\ \hline
\end{tabular}
\label{ta}
\end{equation}

\ \ \ \newline
The usual $U_{em}\left( 1\right) $ electric charge operator is given by $%
Q_{U_{em}\left( 1\right) }=H_{{\small su}_{L}\left( {\small 2}\right) }^{0}+%
\frac{Y}{2}$. \newline
Later on (\textrm{see section 5} ), we find as well that these matter fields
and their group theoretical configurations get a nice geometric
interpretation in the framework of F-theory on Calabi-Yau four- folds and
intersecting seven branes wrapping 4- cycles and filling the non compact
space time directions. \newline
For completeness, notice that implementation of the two other generations of
flavors \emph{F}$_{i}$ with,
\begin{equation}
\begin{tabular}{llllll}
{\small family} &  & {\small Quarks} &  & \multicolumn{1}{|l}{\qquad \qquad}
& {\small Leptons} \\ \hline\hline
&  &  &  & \multicolumn{1}{|l}{} &  \\
\emph{F}$_{2}=$\emph{F}$\left( \mu \right) $ & : & $\left(
\begin{array}{c}
c \\
s%
\end{array}%
\right) $ $\ ,$ \ $c^{c}$ \ $,$ \ $s^{c}$ &  & \multicolumn{1}{|l}{} & $%
\left(
\begin{array}{c}
\nu _{\mu } \\
\mu%
\end{array}%
\right) $ $\ ,$ \ $\nu _{\mu }^{c}$ $\ ,$ \ $\mu ^{c}$ \\
&  &  &  & \multicolumn{1}{|l}{} &  \\
\emph{F}$_{3}=$\emph{F}$\left( \tau \right) $ & : & $\left(
\begin{array}{c}
t \\
b%
\end{array}%
\right) $ $\ ,$ \ $t^{c}$ \ $,$ \ $b^{c}$ &  & \multicolumn{1}{|l}{} & $%
\left(
\begin{array}{c}
\nu _{\tau } \\
\tau%
\end{array}%
\right) $ $\ ,$ \ $\nu _{\tau }^{c}$ $\ ,$ \ $\tau ^{c}$ \\
&  &  &  & \multicolumn{1}{|l}{} &  \\ \hline
&  &  &  &  &
\end{tabular}%
\end{equation}%
is achieved by help of inserting a flavor index $i$ running as $i=1,2,3$
with \emph{F}$_{1}=$\emph{F}$\left( e\right) $. As such, the full set of
\emph{3}$\times $\emph{16} elementary fermionic fields will be denoted as $%
q_{i}$, $u_{i}^{c}$, $d_{i}^{c}$, $l_{i}$, $\nu _{i}^{c}$ and $e_{i}^{c}$.
\newline
In the \emph{MSSM}, the \emph{F}$_{i}$ families get promoted to super-
families $\mathcal{F}_{i}$ where the above \emph{3}$\times $\emph{16}
elementary fermionic fields are now promoted to \emph{3}$\times $\emph{16}
chiral superfields
\begin{equation}
\begin{tabular}{llllll}
$Q_{i},$ & $U_{i}^{c},$ & $D_{i}^{c},$ & $L_{i},$ & $N_{i}^{c},$ & $%
E_{i}^{c},$%
\end{tabular}%
\end{equation}%
with same gauge quantum numbers as in the non supersymmetric case. \newline
Below we denote collectively these chiral superfields by $\Psi \left(
\mathrm{y},\theta \right) $ living on the chiral superspace $\left( \mathrm{y%
},\theta \right) $ with 4D space time coordinates $x^{\mu }$ shifted $%
\mathrm{y}^{\mu }=x^{\mu }-i\theta \sigma ^{\mu }\bar{\theta}$ and Grassmann
odd variable given by a $SO\left( 1,3\right) $ Weyl spinor. Since $\theta $
is nilpotent ($\theta ^{3}=0$), the $\Psi \left( \mathrm{y},\theta \right) $
admits then the following finite $\theta $- expansion

\begin{equation}
\Psi \left( \mathrm{y},\theta \right) =\tilde{\phi}\left( \mathrm{y}\right) +%
\sqrt{2}\theta ^{\alpha }\psi _{\alpha }\left( \mathrm{y}\right) +\theta
^{2}F\left( x\right) ,  \label{tex}
\end{equation}%
where the left handed fermion $\psi _{\alpha }$ is one of the fields in
\textrm{eq(\ref{ta})}, $\tilde{\phi}$ the corresponding \emph{sparticle} and
$F$ the usual auxiliary field which, amongst others, plays a central role in
the study of supersymmetry breaking and in the geometric interpretation of
supersymmetric quiver gauge theories embedded in type II superstrings.%
\newline
(\textbf{ii}) Bosons are of two types namely Higgs scalars and vector
particles. In the \emph{MSSM}, we need two space time Higgs scalars $%
h_{u}=\left( h^{+},h^{0}\right) $ and $h_{d}=\left( \bar{h}^{0},\bar{h}%
^{-}\right) $ together with their superpartners. These fields form chiral
multiplet denoted by $H_{u}$ and $H_{d}$ with $\theta $- expansion as in eq(%
\ref{tex}). Regarding the vector particles, we have in addition to the \emph{%
twelve} space time 4- vector potentials%
\begin{equation}
\begin{tabular}{llllll}
$\mathcal{A}_{\mu }^{{\small su}\left( 3\right) }$ & $,$ & $\mathcal{A}_{\mu
}^{{\small su}\left( 2\right) }\equiv \left( W_{\mu }^{\pm },Z_{\mu
}^{0}\right) $ & $,$ & $\mathcal{A}_{\mu }^{{\small Y}}\equiv B_{\mu }$ & $,$%
\end{tabular}%
\end{equation}%
the gaugino partners described by four dimensional space time Majorana
spinors.\newline
In 4D $\mathcal{N}=1$ superspace, the Higgs sector is described by two
doublets of chiral Higgs superfield,%
\begin{equation}
\begin{tabular}{l|ll}
Higgs &  & ${\small G}_{str}$ group \\ \hline\hline
&  &  \\
$H_{u}=\left(
\begin{array}{c}
H^{+} \\
H^{0}%
\end{array}%
\right) $ &  & $\left( 1,2\right) _{+1}$ \\
$H_{d}=\left(
\begin{array}{c}
\bar{H}^{0} \\
\bar{H}^{-}%
\end{array}%
\right) $ &  & $\left( 1,2\right) _{-1}$ \\
&  &  \\ \hline
\end{tabular}
\label{hi}
\end{equation}%
These chiral superfields are needed to break the $G_{str}$ gauge symmetry
down to $SU_{C}\left( {\small 3}\right) $ $\times $ $U_{em}\left( {\small 1}%
\right) $. The gauge fields involve in addition to the space time gauge
bosons
\begin{equation}
A_{\mu }^{{\small su}\left( {\small 3}\right) }\oplus A_{\mu }^{{\small su}%
\left( {\small 2}\right) }\oplus \frac{Y}{2}B_{\mu },
\end{equation}%
the gauginos
\begin{equation}
\tilde{\lambda}_{{\small su}\left( {\small 3}\right) }\oplus \tilde{\lambda}%
_{su\left( 2\right) }\oplus \frac{Y}{2}\tilde{\lambda}_{u_{Y}\left( 1\right)
}.
\end{equation}%
These 4D space time fields are combined altogether in $4D$ $\mathcal{N}=1$
real superspace $\left( x,\theta ,\bar{\theta}\right) $ to form real 4D
superfield
\begin{equation}
V=V_{_{{\small su}\left( {\small 3}\right) }}\oplus V_{_{{\small su}\left(
{\small 2}\right) }}\oplus \frac{Y}{2}V_{_{{\small Y}}}\text{ \ ,}
\end{equation}%
valued in the Lie algebra ${\small su}_{C}\left( {\small 3}\right) $ $%
{\small \oplus }$ ${\small su}_{L}\left( {\small 2}\right) {\small \oplus u}%
_{Y}\left( {\small 1}\right) $. The real superfields $V_{_{{\small su}\left(
{\small 3}\right) }}$, $V_{_{{\small su}\left( {\small 2}\right) }}$ and $%
V_{_{{\small Y}}}$ mediate the gauge interactions with superspace dynamics
described by the following lagrangian density%
\begin{eqnarray}
\mathcal{L}_{{\small MSSM}} &=&+\int d^{4}\theta \sum_{\text{superfields }%
\Psi }\Psi ^{+}\left( e^{-2\left[ g_{{\small su}\left( {\small 3}\right)
}V_{_{{\small su}\left( {\small 3}\right) }}+g_{{\small su}\left( {\small 2}%
\right) }V_{_{{\small su}\left( {\small 2}\right) }}+g_{{\small Y}}\frac{Y}{2%
}V_{_{{\small Y}}}\right] }\right) \Psi  \notag \\
&&+\int d^{2}\theta \left( \frac{1}{8g_{{\small su}\left( {\small 3}\right) }%
}Tr\mathcal{W}_{_{{\small su}\left( {\small 3}\right) }}^{2}+\frac{1}{8g_{%
{\small su}\left( {\small 2}\right) }}Tr\mathcal{W}_{_{{\small su}\left(
{\small 2}\right) }}^{2}+\frac{1}{8g_{Y}}\mathcal{W}_{Y}^{2}\right) +hc
\notag \\
&&+\int d^{2}\theta W+\int d^{2}\bar{\theta}\bar{W},
\end{eqnarray}%
where $W$ is the chiral superpotential. This is a gauge invariant
superfunction depending on the matter chiral superfields and describing mass
terms and Yukawa tri-couplings as shown below,%
\begin{eqnarray}
W &=&-\mathrm{\mu }H_{u}H_{d}-\sum_{i,j=1}^{3}\frac{\mathrm{m}^{ij}}{2}%
N_{i}^{c}N_{j}^{c}  \notag \\
&&+\sum_{i,j=1}^{3}\frac{\lambda _{e}^{ij}}{3}L_{i}E_{j}^{c}H_{d}+%
\sum_{i,j=1}^{3}\frac{\lambda _{\nu }^{ij}}{3}L_{i}N_{j}^{c}H_{u} \\
&&+\sum_{i,j=1}^{3}\frac{\lambda _{d}^{ij}}{3}Q_{i}D_{j}^{c}H_{d}+%
\sum_{i,j=1}^{3}\frac{\lambda _{u}^{ij}}{3}Q_{i}U_{j}^{c}H_{u},  \notag
\end{eqnarray}%
where the numbers $\mathrm{\mu }$ and $\mathrm{m}^{ij}$ scale as mass and
where the dimensionless complex numbers $\lambda _{e}^{ij}$, $\lambda _{\nu
}^{ij}$, $\lambda _{d}^{ij}$\ and $\lambda _{u}^{ij}$ are complex Yukawa
couplings.

\subsection{Beyond MSSM}

In the \emph{MSSM}, quarks and leptons, together with their superpartners,
belong to several irreducible representations of the $G_{str}$ gauge
symmetry involving three gauge coupling constants $g_{su_{C}\left( 3\right)
} $, $g_{su_{L}\left( 2\right) }$ and $g_{Y}$. A true unification model
requires however packaging all the fundamental particles in a unique
irreducible representation of a simple gauge symmetry group. This is the
basic idea behind grand unified theories (GUT) of strong and electroweak
interactions using the real $24$ dimensional unitary $SU\left( 5\right) $,
the $45$ dimensional orthogonal $SO\left( 10\right) $ and the $78$
dimensional exceptional $E_{6}$. As a first step towards this goal, we
distinguish below two main ways in getting the GUT gauge groups, either by
using \emph{physical imagination} \`{a} la Pati-Salam; or by using \emph{%
group theoretical methods} \`{a} la Georgi-Glashow by looking for the
smallest simple gauge group containing $G_{str}$ as a maximal gauge
subgroup. let us describe briefly these two ways.

\subsubsection{Pati-Salam model and $SO\left( 10\right) $ GUT}

In the Pati-Salam model, the gauge symmetry is given by ${\small SU}%
_{C}\left( {\small 4}\right) $ ${\small \times }$ ${\small SU}_{L}\left(
{\small 2}\right) $ ${\small \times }$ ${\small SU}_{R}\left( {\small 2}%
\right) $. There, the quarks and leptons supermultiplets of each one of the
three super-families $\mathcal{F}_{i}$ are packaged in \emph{two}
irreducible representations $Q$ and $Q^{c}$ of this group. The basic idea
behind this packaging is to think about the lepton number as the fourth
color so that the previous ${\small SU}_{C}\left( {\small 3}\right) $ color
gauge symmetry gets promoted to a ${\small SU}_{C}\left( {\small 4}\right) $
gauge invariance. In this way, the quarks and the leptons of the standard
model family (\ref{ql}) are now put into two ${\small SU}_{C}\left( {\small 4%
}\right) $ quartets as follows,%
\begin{equation}
\begin{tabular}{l|ll}
{\small quarks and leptons} &  & ${\small SU}_{C}\left( {\small 4}\right)
{\small \times SU}_{L}\left( {\small 2}\right) {\small \times SU}_{R}\left(
{\small 2}\right) $ \\ \hline\hline
$Q=\left( q,l\right) $ &  & \qquad \qquad $\left( 4,2,1\right) $ \\
$Q^{c}=\left( q^{c},l^{c}\right) $ &  & \qquad \qquad $\left( \bar{4},1,\bar{%
2}\right) $ \\ \hline
\end{tabular}%
\end{equation}%
with $q$ and $l$ as in eqs(\ref{ql}) and
\begin{equation}
\begin{tabular}{llll}
$q^{c}=\left( u^{c},d^{c}\right) $ & , & $l^{c}=\left( \nu ^{c},e^{c}\right)
$ & .%
\end{tabular}%
\end{equation}%
The baryon number $B$ \emph{minus} the lepton number $L$ and the electric
charge operator $Q_{em}$ act on the representation $\underline{4}$ of $%
SU\left( 4\right) $ as follows,%
\begin{equation}
\begin{tabular}{llllll}
$B-L$ & $=$ & $\frac{1}{3}$ $diag\left( 1,1,1,-3\right) $ & , & $Tr_{%
\underline{4}}\left( B-L\right) =0$ & , \\
&  &  &  &  &  \\
$Q_{em}$ & $=$ & $H_{{\small su}_{{\small L}}\left( {\small 2}\right)
}^{0}+H_{{\small su}_{{\small R}}\left( {\small 2}\right) }^{0}+\frac{\left(
B-L\right) }{2}$ & , & $Tr_{\underline{4}}\left( Q_{em}\right) =0$ & .%
\end{tabular}%
\end{equation}%
Regarding bosons, we have a quite similar picture. Besides the gauge
particles transforming in the adjoint of the gauge symmetry, the two Higgs
doublets H$_{u}$ and H$_{d}$ of eq(\ref{hi}) are also combined into one
irreducible quartet Higgs multiplet%
\begin{equation}
\mathcal{H}=\left( H_{u},H_{d}\right) ,
\end{equation}%
transforming under the $SU_{C}\left( 4\right) \times SU_{L}\left( 2\right)
\times SU_{R}\left( 2\right) $ Pati-Salam group like $\left( 1,2,\bar{2}%
\right) $ and so allowing the following unique gauge invariant trilinear
Yukawa coupling term
\begin{equation}
\mathcal{L}_{Yukawa}=\lambda _{{\small qdh}}\int d^{2}\theta \text{ }Q^{c}%
\mathcal{H}Q\text{ \ }+\text{ \ }hc,
\end{equation}%
where $\lambda _{{\small qdh}}$ is the Yukawa coupling constant. Notice that
Pati Salam group $SU_{C}\left( 4\right) \times SU_{L}\left( 2\right) \times
SU_{R}\left( 2\right) $ which is homomorphic to $SO\left( 6\right) \times
SO\left( 4\right) $ is not a grand unified gauge symmetry as it still
involves three gauge coupling constants $g_{{\small SU}_{C}\left( {\small 4}%
\right) }$, $g_{{\small SU}_{L}\left( {\small 2}\right) }$ and $g_{{\small SU%
}_{R}\left( {\small 2}\right) }$. However, this gauge symmetry can be
embedded into the simple $SO\left( 10\right) $ group. In this larger simple
group, the reducible $\underline{16}$ dimensional matter representation $%
\left( 4,2,1\right) {\small \oplus }\left( \bar{4},1,\bar{2}\right) $ of the
Pati-Salam group gets interpreted as the left handed $SO\left( {\small 10}%
\right) $ spinor representation%
\begin{equation}
\begin{tabular}{llllll}
${\small SO}\left( 10\right) $ & $\rightarrow $ & ${\small SO}\left(
6\right) {\small \times SO}\left( 4\right) $ & $\rightarrow $ & ${\small SU}%
_{C}\left( 3\right) {\small \times SU}\left( 2\right) {\small \times U}%
_{Y}\left( 1\right) $ &  \\
$16_{+}$ & $\rightarrow $ & $\left( 4,2,1\right) {\small \oplus }\left( \bar{%
4},1,\bar{2}\right) $ & $\rightarrow $ & $\left( 3,2\right) _{\frac{1}{3}}%
{\small \oplus }\left( \bar{3},1\right) _{-\frac{4}{3}}{\small \oplus }%
\left( \bar{3},1\right) _{\frac{2}{3}}{\small \oplus }$ &  \\
&  &  &  & $\left( 1,2\right) _{-1}{\small \oplus }\left( 1,1\right) _{0}%
{\small \oplus }\left( 1,1\right) _{2}$ & .%
\end{tabular}%
\end{equation}%
In this embedding, we have a lepton-quark unification as well as a gauge
coupling unification. This feature makes the $SO\left( 10\right) $ gauge
symmetry as one of the most attractant \emph{GUT} models for gauge
unification of strong and electroweak interactions.

\subsubsection{Georgi Glashow model}

\begin{figure}[tbph]
\begin{center}
\hspace{0cm} \includegraphics[width=12cm]{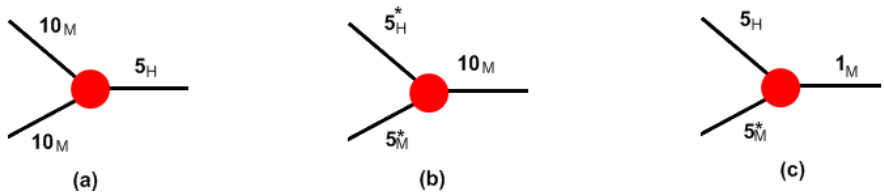}
\end{center}
\par
\vspace{-0.5 cm}
\caption{Yukawa couplings {\protect\small \ in supersymmetric SU}$\left(
5\right) ${\protect\small \ GUT model}: three {\protect\small kind of chiral
superfield tri- vertices namely }${\protect\small 5}\times {\protect\small 10%
}\times {\protect\small 10}${\protect\small , }${\protect\small \bar{5}}%
\times {\protect\small \bar{5}}\times {\protect\small 10}${\protect\small \
and }${\protect\small 5}\times {\protect\small \bar{5}}\times
{\protect\small 1}${\protect\small . Tri-coupling involving Higgs superfield
in the 24 adjoint, which is also allowed, is not reported.}}
\label{ou}
\end{figure}

In the Georgi-Glashow model based on group theory analysis, the GUT symmetry
is given by the simple rank four unitary group $SU\left( 5\right) $. There,
the leptons and quarks of each \emph{sixteen} dimensional family of the
standard model are packaged into three $SU\left( 5\right) $ irreducible
representations: the singlet, the anti- fundamental $\bar{5}$ and the
antisymmetric $10=\left[ 5\otimes 5\right] _{A}$ representations. This
property follows from the decomposition
\begin{equation}
\begin{tabular}{llllll}
${\small SO}\left( 10\right) $ &  & $\rightarrow $ &  & ${\small SU}\left(
5\right) $ &  \\
$16_{M}$ &  & $\rightarrow $ &  & $1_{M}\oplus \bar{5}_{M}\oplus 10_{M}$ & ,%
\end{tabular}%
\end{equation}%
where the sub-index M refers for matter. To get more insight in the field
content of these elementary particles unification, we recall that the
singlet stands for the anti- neutrino, $1_{M}\sim \nu ^{c}$ while the $\bar{5%
}_{M}$ and $10_{M}$ correspond to
\begin{equation}
\begin{tabular}{lllllllll}
$\bar{5}_{M}$ & $\sim $ & $\left(
\begin{array}{c}
d_{1}^{c} \\
d_{2}^{c} \\
d_{3}^{c} \\
e \\
\nu%
\end{array}%
\right) $ & , & $10_{M}$ & $\sim $ & $\left(
\begin{array}{ccccc}
0 & u_{3}^{c} & -u_{2}^{c} & u^{1} & d^{1} \\
-u_{3}^{c} & 0 & u_{1}^{c} & u^{2} & d^{2} \\
u_{2}^{c} & -u_{1}^{c} & 0 & u^{3} & d^{3} \\
-u^{1} & -u^{2} & -u^{3} & 0 & e^{c} \\
-d^{1} & -d^{2} & -d^{3} & -e^{c} & 0%
\end{array}%
\right) $ &  & .%
\end{tabular}%
\end{equation}%
Similar representations are valid for the two other families and their
supersymmetric extensions. Later on, we will see that, along with this group
theoretic representation, these matter fields have a nice geometric
representation in terms of Riemann surfaces $\Sigma $ (complex curves)
inside the internal space used in the F- theory compactification on real
eight dimensional Calabi-Yau four- folds; for illustration, think about this
feature as "corresponding" to the matter localized on the edges of the
figure (\ref{by}),%
\begin{equation}
\begin{tabular}{llll}
$\bar{5}_{M}\rightarrow \Sigma _{M}^{\left( \bar{5}\right) }$ & , & $%
10_{M}\rightarrow \Sigma _{M}^{\left( 10\right) }$ & .%
\end{tabular}%
\end{equation}%
Notice moreover that altogether with these chiral matter representations,
which are promoted to chiral superfield in $SU\left( 5\right) $\ \emph{SGUT}
model, we have moreover: \newline
(\textbf{1}) two Higgs multiplets H$_{u}$ and H$_{d}$ transforming
respectively in the $5_{H}$ and $\bar{5}_{H}$ representations,\newline
(\textbf{2}) twenty four \emph{4D} $\mathcal{N}=1$ gauge multiplets $V^{a}$
transforming the adjoint representation of the SU$\left( 5\right) $ gauge
symmetry.\newline
Furthermore, the $SU\left( 5\right) $ gauge invariant chiral superpotential $%
W$ between two matter superfields and one Higgs superfield has the following
structure,%
\begin{equation}
\begin{tabular}{llllll}
$W_{Yukawa}$ & $=$ & $+$ $\frac{\lambda _{1}}{3}\ \left( 5_{H}\otimes
10_{M}\otimes 10_{M}\right) $ & $+$ & $\frac{\lambda _{2}}{3}\ \left( \bar{5}%
_{H}\otimes \bar{5}_{M}\otimes 10_{M}\right) $ &  \\
&  &  &  &  &  \\
&  & $+$ $\frac{\lambda _{3}}{3}\ \left( 5_{H}\otimes \bar{5}_{M}\otimes
1_{M}\right) $ & $+$ & $\mathrm{\mu }\ \left( 5_{H}\otimes \bar{5}%
_{H}\right) $ & ,%
\end{tabular}%
\end{equation}%
where $\mathrm{\mu }$\ is a mass constant and the $\lambda _{i}$\ 's are
Yukawa coupling constants. This chiral superpotential involves \emph{three}
kinds of chiral superfield vertices as depicted in the figure (\ref{ou}).

\subsection{\emph{MSSM} as a quiver gauge theory}

In the \emph{MSSM }with $SU_{C}\left( 3\right) \times SU_{L}\left( 2\right)
\times U_{Y}\left( 1\right) $ gauge invariance, the matter fields are
generally charged under representations of the groups factors; that is under
$SU_{C}\left( 3\right) $, $SU_{L}\left( 2\right) $ and $U_{Y}\left( 1\right)
$. This property suggests that \emph{MSSM} might be thought of as quiver
gauge theory that can be embedded in superstrings compactifications. Recall
that \emph{4D} $\mathcal{N}=1$ supersymmetric quiver gauge theories have
been subject to an intensive interest during last decade \textrm{\cite{I1}-%
\cite{I4}}. These theories, which may be engineered in different, but dual,
ways appear as low energy effective field theory of \emph{10D} superstrings
on CY3- folds, \emph{11D} M- theory compactification on G2 manifolds and
\emph{12D} F -theory on CY4-folds preserving four supersymmetries \textrm{%
\cite{J1}-\cite{J4}}. \newline
In this subsection, we explore rapidly what kind of quiver diagram one gets
in the engineering of \emph{MSSM} as a supersymmetric quiver gauge theory.

\subsubsection{Engineering the MSQSM}

\begin{figure}[tbph]
\begin{center}
\hspace{0cm} \includegraphics[width=5cm]{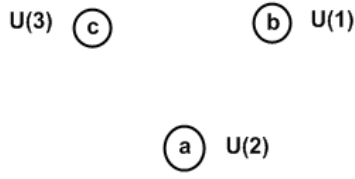}
\end{center}
\par
\vspace{-0.5 cm}
\caption{A quiver gauge diagram for pure gauge theory: Without matter, the
gauge group factors of the ${\protect\small U}\left( 3\right)
{\protect\small \times U}\left( 2\right) {\protect\small \times U}\left(
1\right) $ are represented by three nodes.}
\label{abc}
\end{figure}

One of the main actors in the Minimal Supersymmetric Quiver Standard Model (%
\emph{MSQSM}) is that supersymmetric chiral matter in the three $\mathcal{F}%
_{i}$ families of elementary particles transform in specific representations
of the $SU_{C}\left( 3\right) $ $\times $ $SU_{L}\left( 2\right) $ $\times $
$U_{Y}\left( 1\right) $ gauge symmetry. These representations are mainly
given by:\newline
\textbf{(1)} the hermitian adjoint representation of each factor of the
\emph{MSSM} gauge symmetry where transform the twelve \emph{MSSM} gauge
superfields $\mathcal{V}_{{\small MSSM}}^{a}$, that is:
\begin{equation}
\begin{tabular}{llll}
$\mathcal{V}_{{\small MSSM}}^{a}$ & $\mathcal{\sim }$ & $\left( 8,1\right)
_{0}\oplus \left( 1,3\right) _{0}\oplus \left( 1,1\right) _{0}$ & .%
\end{tabular}%
\end{equation}%
These hermitian representations have a nice interpretation in terms of
massless excitations of open superstrings ending on stacks of D-branes of
\emph{10D} closed type II superstrings. In this regards, it is interesting
to note that in the D- brane setting, a stack of $N$ coincident D- branes of
type II superstings involves $U\left( N\right) =U\left( 1\right) \times
SU\left( N\right) $ gauge invariance in \emph{4D} space time \textrm{\cite%
{A1,A2}}. As such the gauge symmetry in the \emph{MSQSM} is, instead of $%
G_{str}$, is rather given by,
\begin{equation}
U_{a}\left( 3\right) \times U_{b}\left( 2\right) \times U_{c}\left( 1\right)
,
\end{equation}%
involving two extra undesired $U\left( 1\right) $ gauge factors namely%
\begin{equation}
\begin{tabular}{llll}
$U_{a}\left( 1\right) $ & $=$ & $U_{a}\left( 3\right) /SU_{C}\left( 3\right)
$ & , \\
$U_{b}\left( 1\right) $ & $=$ & $U_{b}\left( 2\right) /SU_{L}\left( 2\right)
$ & ,%
\end{tabular}%
\end{equation}%
which may be interpreted as baryon and lepton numbers. The $U_{Y}\left(
1\right) $ hypercharge in the G$_{str}$ group should be then given by the
non anomalous combination of the three $U_{i}\left( 1\right) $s with a
massless gauge field. The two other combinations are anomalous; but
following \textrm{\cite{A1,A2}}, these anomalies may be canceled by a
generalized Green-Schwarz mechanism which at the same time gives large
masses to the corresponding gauge bosons. As such these abelian symmetries
remain as global symmetries in the effective Lagrangian of the theory.
\newline
If forgetting for a while about the right handed leptons that are charged
under the $U_{Y}\left( 1\right) $ hypercharge, the quiver graph that would
describe this supersymmetric quiver gauge theory without fundamental matter
would involve three separated nodes as depicted in the figure (\ref{abc}).
Each node\textrm{\footnote{%
In fact it is the requirement that the lepton \ doublets remain charged
under the $SU_{L}\left( 2\right) $ factor but transform as singlets under $%
SU_{C}\left( 3\right) $ which implies that any minimal embedding will
possess at least three quiver nodes.}} refers to a gauge group factor and
represents the world volume of the branes at some fix points under some
given orbifold action on the internal manifold.

\begin{figure}[tbph]
\begin{center}
\hspace{0cm} \includegraphics[width=6cm]{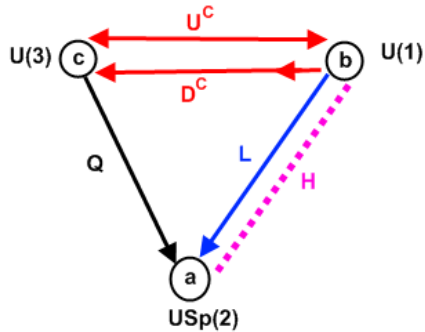}
\end{center}
\par
\vspace{-0.5 cm}
\caption{Quiver graph of MQSM: directed lines denote three generations of
left-handed chiral fermions. Lines with two arrows determine fermions
charged under $U\left( 3\right) \times U\left( 1\right) $. Dashed \ line
refers to the SM Higgs doublet. In the supersymmetric version MSQSM,
oriented line denotes a chiral superfield and dashed line a vector-like pair
of fields.}
\label{F1}
\end{figure}

\ \ \newline
\textbf{(2)} Matter of the \emph{MSQSM} is in several complex
representations of the gauge invariance as shown below,%
\begin{equation}
\begin{tabular}{lllllll}
\hline
quark multiplet & : & ${\small Q=}\left( {\small 3,2}\right) _{\frac{{\small %
1}}{{\small 3}}}$ & , & ${\small U}^{c}{\small =}\left( {\small \bar{3}}%
,1\right) _{\frac{-{\small 4}}{{\small 3}}}$ & , & ${\small D}^{c}{\small =}%
\left( {\small \bar{3}},{\small 1}\right) _{\frac{{\small 2}}{{\small 3}}}$
\\
lepton multiplet & : & ${\small L=}\left( {\small 1},{\small 2}\right) _{%
{\small -1}}$ & , & ${\small N=}\left( {\small 1,1}\right) _{{\small 0}}$ & ,
& ${\small E=}\left( {\small 1,1}\right) _{{\small 2}}$ \\
Higgs multiplet & : & ${\small H}_{u}{\small =}\left( {\small 1},{\small 2}%
\right) _{{\small -1}}$ & , & ${\small H}_{d}{\small =}\left( {\small 1},%
{\small 2}\right) _{+{\small 1}}$ & , &  \\ \hline
\end{tabular}
\label{ne}
\end{equation}%
In the brane set up, matter fields in the bi-fundamental representations
live at the brane intersections. This is the case of the superfields $Q$, $%
U^{c}$, $D^{c}$, $L$, $H_{u}$ and $H_{d}$; but not for the two superfields%
\textrm{\footnote{%
Implementation of the right handed leptons that are charged under the U$%
_{Y}\left( 1\right) $ requires adding a fourth brane stack \cite{A2}.}} $N$
and $E$ of eqs(\ref{ne}). For these superfields, the corresponding quiver
gauge graph requires rather four nodes; for more details see \textrm{\cite%
{22-23}}, see also\emph{\ }figure (\ref{F1}) for a brane representation.

\begin{figure}[tbph]
\begin{center}
\hspace{0cm} \includegraphics[width=8cm]{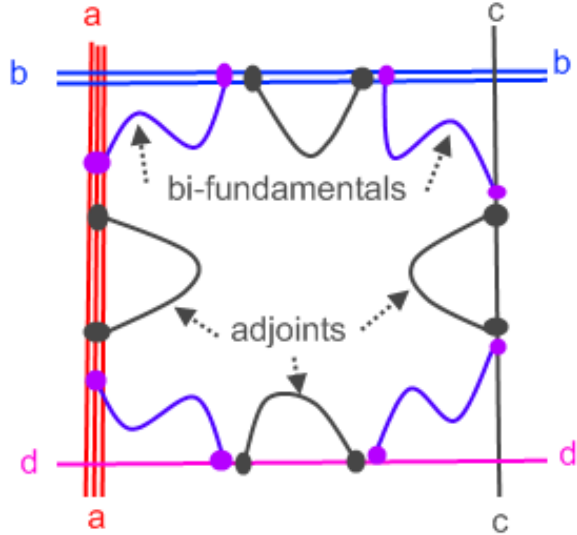}
\end{center}
\par
\vspace{-0.5 cm}
\caption{Quiver diagram of four stacks of D-branes of $U\left( 3\right)
\times U\left( 2\right) \times U\left( 1\right) \times U\left( 1\right) $
gauge model in type II superstrings compactification on Calabi-Yau threefold
orbifolds.}
\label{st}
\end{figure}

\ \ \ \newline
The gauge group of the \emph{MSQSM} is $U\left( 3\right) \times USp\left(
2\right) \times U\left( 1\right) $ with non anomalous hypercharge%
\begin{equation}
Q_{Y}=\frac{1}{2}Q_{{\small U}_{a}\left( {\small 1}\right) }-\frac{1}{3}Q_{%
{\small U}_{b}\left( {\small 1}\right) }
\end{equation}%
In \textrm{\cite{H2}}, a supersymmetric version of the minimal quiver
standard model has been constructed in F-theory on local CY4-folds by
partially Higgsing the brane probe theory of a del Pezzo $dP_{5}$ Calabi-Yau
singularity. This extension, which will be implicitly described in section
5, involves orientifolding ideas as a way to solve the problem of
engineering the leptonic right handed sector and anomaly cancelation in the
hypercharge sector. \newline
We end this section by describing rapidly the four nodes quiver gauge model
extending the three nodes one of figure (\ref{F1}). In the language of
intersecting D5-branes in type IIB superstrings on local Calabi-Yau
threefold orbifolds, the quiver gauge theory involves four stack of D5-
branes and an orientifold as depicted in the figure (\ref{st}) and table (%
\ref{tbb}).

\ \ \newline
The embedding of the \emph{MSSM} in type IIB superstring may be achieved in
this unoriented quiver gauge theories of at least four stacks of
intersecting D- branes leaving at the fix points of the orbifold action in
the type II superstring compactification to 4D space time. Under this
orbifolding, the particles content of the \emph{MSSM} is engineered by using
both
\begin{equation}
\begin{tabular}{llll}
$\left( N_{a},\bar{N}_{b}\right) $ & , & $\left( N_{a},N_{b}\right) $ & ,%
\end{tabular}%
\end{equation}%
bi-fundamental representations of the gauge group. This possibility is
familiar from type II orientifold models in which the world sheet of the
string is modded by some operation $\Omega R$ with $\Omega $ being the world
sheet parity operation and $R$ some geometrical action. Bi-fundamental
representations of type $\left( N_{a},\bar{N}_{b}\right) $ appear from open
strings stretched between branes $a$ and $b$ whereas those of type $\left(
N_{a},N_{b}\right) $ appear from those going between the branes $a$ to the
branes $b^{\ast }$; the mirror of the branes $b$ under $\Omega R$. Inclusion
of these representations in the string theoretic realization is crucial for
tadpole cancelation. Following \textrm{\cite{A1, A2}}, the spectrum of the
unoriented quiver gauge theory is given by%
\begin{equation}
\begin{tabular}{l|l|l|l|l|l|l|l}
intersection & matter & repres & Q$_{a}$ & Q$_{b}$ & Q$_{c}$ & Q$_{d}$ & $%
\frac{Y}{2}$ \\ \hline\hline
$ab$ & $Q_{L}$ & $\left( 3,2\right) $ & ${\small +1}$ & ${\small -1}$ & $\ \
{\small 0}$ & $\ \ {\small 0}$ & ${\small +}\frac{1}{6}$ \\
$ab^{\ast }$ & $q_{L}$ & $2\left( 3,2\right) $ & ${\small +1}$ & ${\small +1}
$ & $\ \ {\small 0}$ & $\ \ {\small 0}$ & ${\small +}\frac{{\small 1}}{%
{\small 6}}$ \\
$ac$ & $U_{R}$ & $3\left( \overline{3},1\right) $ & ${\small -1}$ & $\ \
{\small 0}$ & ${\small +1}$ & $\ \ {\small 0}$ & ${\small -}\frac{{\small 2}%
}{{\small 3}}$ \\
$ac^{\ast }$ & $D_{R}$ & $3\left( \overline{3},1\right) $ & ${\small -1}$ & $%
\ \ {\small 0}$ & ${\small -1}$ & $\ \ {\small 0}$ & ${\small +}\frac{%
{\small 1}}{{\small 3}}$ \\
$bd^{\ast }$ & $L$ & $3\left( 1,2\right) $ & $\ \ {\small 0}$ & ${\small -1}$
& $\ \ {\small 0}$ & ${\small -1}$ & ${\small -}\frac{{\small 1}}{{\small 2}}
$ \\
$cd$ & $E_{R}$ & $3\left( 1,1\right) $ & $\ \ {\small 0}$ & $\ \ {\small 0}$
& ${\small -1}$ & ${\small -1}$ & ${\small +1}$ \\
$cd^{\ast }$ & $N_{R}$ & $3\left( 1,1\right) $ & $\ \ {\small 0}$ & $\ \
{\small 0}$ & ${\small -1}$ & ${\small +1}$ & $\ \ \ {\small 0}$ \\ \hline
\end{tabular}
\label{tbb}
\end{equation}%
where the hypercharge $\frac{Y}{2}=\frac{1}{6}Q_{a}-\frac{1}{2}Q_{c}-\frac{1%
}{2}Q_{d}$.

\section{Non abelian gauge theory on 7- brane}

In this section, we describe some basic tools on brane physics to be used
later on when we study the 4D $\mathcal{N}=1$ supersymmetric GUT- type
models along the line of the BHV proposal \textrm{\cite{H1,H2,H3}}. \newline
In the first subsection, we review briefly Vafa's twelve dimensional
F-theory as it is the framework for building the \emph{4D} $\mathcal{N}=1$
supersymmetric GUT models. In the second subsection, we consider some useful
aspects on the geometry of the elliptic Calabi-Yau 4-folds,%
\begin{equation}
\begin{tabular}{lll}
$\mathbb{E}$ & $\rightarrow $ & $X_{4}$ \\
&  & $\downarrow \pi $ \\
&  & $B_{3}$%
\end{tabular}%
\end{equation}%
where $\mathbb{E}$ stands for the elliptic curve fibered on the complex
three dimension base $B_{3}$. For later use, we will particularly focus on
the following local geometry of the base
\begin{equation}
\begin{tabular}{lll}
$\mathbf{\Sigma }_{0}$ & $\rightarrow $ & $B_{3}$ \\
&  & $\downarrow \pi $ \\
&  & $S$%
\end{tabular}%
\end{equation}%
where the base $S$ is a complex surface and $\mathbf{\Sigma }_{0}\sim
\mathbb{P}^{1}$ a genus zero complex curves which locally may be thought of
as the complex line $\mathbb{C}$. So, the resulting local Calabi-Yau four-
folds $X_{4}$ reduces to the form (\ref{F}). Notice that although the
complex base $S$ could be a generic surface, we shall think about it as:%
\newline
(\textbf{i}) a del Pezzo surface $dP_{n}$ with H$_{2}$ homology as in eq(\ref%
{L}),\newline
(\textbf{ii}) a complex tetrahedral surface $\mathcal{T}$ of fig(\ref{by})
or its $\mathcal{T}_{n}$ blown ups studied in \textrm{\cite{SA}}.\newline
The second issue constitutes the basis of our contribution in the embedding
of GUT-like models building in the F-theory set up. \newline
With this picture in mind, we study the engineering of ADE\ gauge symmetries
in the fiber Y of eq(\ref{F}) with locus in the complex two codimension
surface $S$. \newline
In the third subsection, we study the colliding of the singularities in the
fiber as well as the enhancement of the gauge invariance at specific loci in
the complex surface $S$. As we will see later on, these collisions have a
nice realization in the complex tetrahedral geometry where gauge invariance
in the bulk gets enhanced once on the edges and twice at the vertices of the
tetrahedron of the figure (\ref{by}); thanks to toric geometry.\newline
To make direct contact with the usual \emph{4D} formulation of gauge theory
in \emph{SGUT} models building, we shall often use field theoretical method
to interpret geometric quantities in the compact real eight dimensional
manifold $X_{4}$.

\subsection{F-theory on elliptic CY manifolds}

We begin by noting that there are two main related approaches to introduce
Vafa's twelve dimensional F-theory: \newline
\textbf{(1)} in terms of strongly coupled \emph{10D} type IIB superstring,
or \newline
\textbf{(2) }by using superstrings dualities in lower space time dimensions.%
\newline
Besides its merit to incorporate F-theory as a part of a unifying picture
including the five superstring theories, the duality based manner for
defining F- theory has also the advantage to offer a way to engineer \emph{%
non abelian} gauge symmetries in terms of geometric singularities in the
internal manifolds. Before going into technical details, let us start by
reviewing rapidly these two constructions.

\subsubsection{\emph{10D} Type IIB set up\emph{\ }}

In type II superstring set up, the existence of twelve dimensional F-theory
underlying \emph{10D} type IIB superstring may be motivated by looking for a
link similar to the one existing between Witten's eleven dimensional
M-theory and \emph{10D} type IIA superstring \textrm{\cite{MT}}. In this
view, it has been observed in a seminal work by Vafa \textrm{\cite{FT}} that
the \emph{10D} type IIB superstring theory has indeed a remarkable
underlying \emph{12D} F- theory description with non constant dilaton and
axion. Recall that type IIB has, amongst others, the following features%
\newline
(\textbf{a}) a constant profile coupling constant $g_{{\small s}}$ over the
entire \emph{10D} spacetime,\newline
(\textbf{b}) a strong/weak self duality captured by the $SL\left( 2,\mathbb{Z%
}\right) $ symmetry, and \newline
(\textbf{c}) a NS-NS and R-R massless bosonic spectrum
\begin{equation}
\begin{tabular}{lllllllll}
NS-NS & :\qquad &  & $G_{MN}$ & $,$ & $B_{MN}$ & , & $\varphi $ & , \\
R-R & : &  & $\tilde{B}_{MN}$ & , & $\tilde{D}_{MNPQ}^{+}$ & , & $\tilde{%
\varphi}$ & .%
\end{tabular}
\label{da}
\end{equation}%
Moreover the dilaton and the axion vevs $\varphi $ and $\tilde{\varphi}$ of
eq(\ref{da}) are interpreted in terms of the complex structure modulus $\tau
_{{\small IIB}}\equiv \tau =\tilde{\varphi}+ie^{-\varphi }$ of an elliptic
curve $\mathbb{E}$ with the modular transformation
\begin{equation}
\begin{tabular}{llll}
$\tau \rightarrow \tau ^{\prime }=\frac{n_{1}\tau +n_{2}}{n_{3}\tau +n_{4}}$
& , & $\left(
\begin{array}{cc}
n_{1} & n_{2} \\
n_{3} & n_{4}%
\end{array}%
\right) \in SL\left( 2,Z\right) $ & .%
\end{tabular}%
\end{equation}%
By thinking about this 2-torus $\mathbb{T}^{2}$ as a universe \emph{geometric%
} entity with coordinates $x^{{\small 11}}\equiv x^{{\small 11}}+R_{{\small 1%
}}$ and $x^{{\small 12}}\equiv x^{{\small 12}}+R_{{\small 2}}$, one ends
with a $\left( 10+2\right) $ dimensional space time. \newline
To practically handle this complex elliptic curve $\mathbb{E}\sim \mathbb{T}%
^{2}$, it useful to embed it in the complex space $\mathbb{C}^{2}$ with a
local holomorphic coordinates $\left( \mathrm{u,v}\right) $. In this
embedding, the complex elliptic curve\textrm{\footnote{%
Generally, a complex elliptic curve is a nonsingular cubic curve in the $%
\left( u,v\right) $- complex plane with algebraic equation $%
\sum_{n,m=0}^{3}a_{nm}u^{n}v^{m}=0$ where $a_{nm}$ are some constants. This
complex cubic can be simplified however, by an appropriate change of
variables and brought to the usual Weierstrass form $v^{2}=u^{3}+au+b$ with
discriminant $\Delta =-16\left( 4a^{3}+27b^{3}\right) $. In our formulation,
we have kept the expression of the cubic quite general in order to give a
unified description of the ADE geometries.}} $\mathbb{E}$ may be naively
defined by the typical complex algebraic cubic,%
\begin{equation}
\mathbb{E}:\qquad \mathrm{v}^{2}=d\mathrm{u}^{3}+e\mathrm{u}^{2}+f\mathrm{u}%
+g\qquad ,\qquad d\neq 0\qquad ,  \label{el}
\end{equation}%
where $d,$ $e,$ $f$ and $g$ are some complex constants introduced for later
use. \newline
Recall in passing that in the \emph{Weierstrass form} of the complex
elliptic curve $\mathbb{E}$, we have $d=1$ and $e=0$; but here we have used
the equivalent form (\ref{el}) since later on the coefficients $d,$ $e,$ $f$
and $g$ will be promoted to holomorphic sections of a some canonical bundle
in the base of the CY4- folds. This promotion is needed in the engineering
of elliptic fibrations of CY4- folds and in the implementation of gauge ADE\
symmetries in the game. \newline
Twelve dimensional F- theory defines then a non perturbative vacua of type
IIB superstring theory with non constant dilaton and axion and may be
thought of as its strong string coupling limit ($\tau _{IIB}\rightarrow
\infty $); but with no local on shell dynamics along the two extra compact
directions $\left( x^{{\small 11}},x^{{\small 12}}\right) $. From this view,
\emph{10D} type IIB superstring theory may be seen as the compactification
of F-theory on $\mathbb{T}^{2}$
\begin{equation}
\begin{tabular}{llll}
{\small F-Theory}/$\mathbb{T}^{2}$ & $\leftrightarrow $ & {\small 10D Type
IIB} & .%
\end{tabular}
\label{ft}
\end{equation}%
Notice in passing that in ten dimensions, we also have a duality between
F-theory on a cylinder and $SO\left( 32\right) $ type I/Heterotic
superstrings. There, the modulus of the cylinder $\mathbb{S}^{1}\times
\mathbb{S}^{1}/\mathbb{Z}_{2}$ is identified with the type I/Heterotic
coupling constants \textrm{\cite{FS}}.

\subsubsection{Duality\ \ in lower dimensions}

Twelve dimensional F-theory may be nicely defined in terms of superstrings
dualities at various space time dimensions where more physical features are
expected. In eight space time dimensions, F-theory on elliptic K3 is dual%
\textrm{\footnote{%
Notice that the geometry of the compactification must be of a special type
for this duality to hold. In F-theory GUT models, it is precisely those
models that are not dual to heterotic superstring that are important as they
allow gauge breaking of the GUT group through the so called hyperflux method.%
}} to the \emph{10D} heterotic superstring on 2- torus $\mathbb{T}^{2}$,%
\begin{equation}
\begin{tabular}{llll}
$10D\text{ Heterotic superstring/}\mathbb{T}^{2}$ & $\longleftrightarrow $ &
F-theory on K3 & ,%
\end{tabular}
\label{tk}
\end{equation}%
where topologically $K3\sim \mathbb{E}\times \mathbb{P}^{1}$. \newline
This duality relation can be used to build other dualities in lower space
time dimensions by using the \textrm{adiabatic argument}. By further
compactifying (\ref{tk}) on a real two sphere $\mathbb{S}^{2}\sim \mathbb{P}%
^{1}$ reducing then the space time dimension to six, we get a duality
between F-theory on Calabi-Yau three-folds with elliptic K3 fibration and
the Heterotic superstring on elliptic K3,%
\begin{equation}
\begin{tabular}{llll}
$10D\text{ Heterotic superstring/}$K3 & $\longleftrightarrow $ & F-theory on
CY3 & ,%
\end{tabular}%
\end{equation}%
where topologically $CY3$ $\sim $ $K3\times \mathbb{P}^{1}$ or more
explicitly $\mathbb{E}\times \mathbb{P}^{1}\times \mathbb{P}^{1}$.\ \newline
In \emph{4D} space time, F-theory on Calabi-Yau four- folds is dual to the
Heterotic superstring on Calabi-Yau three-folds,%
\begin{equation}
\begin{tabular}{llll}
${\small 10D}\text{ {\small Heterotic superstring}/}${\small CY3} & $%
\longleftrightarrow $ & {\small F-theory on $CY_{\text{4}}$} & .%
\end{tabular}
\label{cc}
\end{equation}%
As we see, these duality based definitions of F- theory are related and they
can be used to build other dualities in various space dimensions by
implementing type I, type II superstrings, eleven dimension M- and twelve
dimension F- theories. These dualities turn out be crucial in the
engineering of \emph{non abelian} gauge symmetries, the bi-fundamental
matter and Yukawa couplings.

\subsection{Engineering non abelian gauge symmetries}

Before coming to the engineering non abelian gauge symmetries in F-Theory on
CY4- folds, let us start by recalling basic results that are helpful for the
understanding the links between the geometry of CY4- folds and \emph{4D}\
space time non abelian gauge invariance.

\textbf{(1)} \emph{Gauge fields in heterotic superstring}\newline
In building GUT models extending \emph{MSSM}, one needs, amongst others
\emph{4D} non abelian gauge fields $\mathcal{A}_{\mu }$; that is operator
fields with the non commutativity property $\left[ \mathcal{A}_{\mu },%
\mathcal{A}_{\nu }\right] \neq 0$. As it is well known, this non
commutativity feature is solved by taking the hermitian gauge field $%
\mathcal{A}_{\mu }$ in the adjoint representation of an ADE gauge group G as
given below,%
\begin{equation}
\mathcal{A}_{\mu }=\sum_{a=1}^{\dim G}T_{a}\mathcal{A}_{\mu }^{a}\qquad
,\qquad \mathcal{F}_{\mu \nu }=\partial _{\lbrack \mu }A_{\nu ]}+\left[
\mathcal{A}_{\mu },\mathcal{A}_{\nu }\right] ,
\end{equation}%
with
\begin{equation}
\left[ \mathcal{A}_{\mu },\mathcal{A}_{\nu }\right] =\sum_{a,b=1}^{\dim G}%
\mathcal{A}_{\mu }^{a}\mathcal{A}_{\nu }^{b}\left[ T_{a},T_{b}\right]
=\sum_{a,b=1}^{\dim G}C_{ab}^{c}\mathcal{A}_{\mu }^{a}\mathcal{A}_{\nu
}^{b}T_{c},  \label{tac}
\end{equation}%
where the $T_{a}$'s are the generators of $G$ and $C_{ab}^{c}$ its constant
structures,%
\begin{equation}
\left[ T_{a},T_{b}\right] =\sum_{c=1}^{\dim G}C_{ab}^{c}T_{c}.
\end{equation}%
These \emph{4D} massless gauge fields $\mathcal{A}_{\mu }^{a}$ together with
matter $\phi ^{a}$ in adjoint representations, which mediate the gauge
interactions between the elementary particles, have a nice origin in
quantized superstring theory.\newline
In the ten dimensional $E_{8}\times E_{8}$ or $SO\left( 32\right) $
heterotic superstrings with \emph{10D} massless bosonic fields

\begin{equation}
\begin{tabular}{llllll}
\hline\hline
$G_{MN}$ & $,$ & $B_{MN}$ & , & $\varphi $ &  \\
$\mathcal{A}_{M}=\sum_{a=1}^{\dim G}T_{a}\mathcal{A}_{M}^{a}$ & , & $%
G=E_{8}\times E_{8}$ & , & $G=SO\left( 32\right) $ &  \\ \hline
\end{tabular}%
\end{equation}

\ \ \ \newline
non abelian gauge fields $\mathcal{A}_{M}^{a}$ appear naturally in the
massless spectrum. Compactification down to lower space time dimensions,
\textrm{with some Wilson lines switched on to break partially gauge
invariance}, still have non abelian gauge fields $\mathcal{A}_{\mu }^{a}$.
It is this property which made first heterotic superstring much popular and
was behind the early days in building superstring inspired semi-realistic
\emph{MSSM} and \emph{GUT} models \textrm{\cite{BIN, HET}} by using
heterotic superstring compactifications down to \emph{4D}.

\textbf{(2)} \emph{Non abelian gauge fields in F-theory}\newline
In the \emph{10D} type IIB closed superstring with chiral $\mathcal{N}=2$
supersymmetry, which according to (\ref{ft}) may be also viewed as the
perturbative regime of \emph{12D} F-theory on $\mathbb{T}^{2}$, the massless
bosonic fields are as in eq(\ref{da}). As we see, there is no non abelian
gauge fields $\mathcal{A}_{M}^{a}$ in the massless spectrum of the theory.
But this is not a problem since \emph{10D} type IIB closed superstring has
Dp-branes with $p=1,3,5,7$. On a stack of $r$ Dp- branes live $r$ abelian $%
\left( p+1\right) $ dimensional gauge fields $\mathcal{A}_{M}^{I}$ belonging
to the spectrum of the quantized open superstrings that end on these branes.
These fields can be put altogether like $\mathcal{A}_{M}^{\left( {\small abel%
}\right) }=\sum_{I=1}^{r}\mathcal{A}_{M}^{I}H_{I}$ with the property%
\begin{equation}
\left[ \mathcal{A}_{M}^{\left( {\small abel}\right) },\mathcal{A}%
_{N}^{\left( {\small abel}\right) }\right] =\sum_{I,J=1}^{r}\mathcal{A}%
_{M}^{I}\mathcal{A}_{N}^{J}\left[ H_{I},H_{J}\right] =0,
\end{equation}
In fact, one should think about $\sum_{I=1}^{r}\mathcal{A}_{M}^{I}H_{I}$ as
the commuting part of a more general non abelian expansion involving as well
the gauge fields associated with step operators of Lie algebras $\mathcal{A}%
_{M}^{\pm \alpha }$ of the strings stretching between the D- branes. Indeed
for coincident branes, the gauge fields $\mathcal{A}_{M}^{\pm \alpha }$
become massless and one is left with a massless non abelian gauge field $%
\mathcal{A}_{M}^{a}\equiv \left( \mathcal{A}_{M}^{I},\mathcal{A}_{M}^{\pm
\alpha }\right) $ in the spectrum. Thanks to the extended solitonic objects
and open superstrings; these are exactly what is needed for engineering
\emph{non abelian} gauge symmetries in type II superstrings. \newline
By using the Heterotic string/F-theory duality (\ref{cc}), it is now clear
that gauge symmetries $G$ of the heterotic superstring on three- folds; with
\begin{equation}
G\subset E_{8}\times E_{8}\text{ \ \ \ or \ \ \ }G\subset SO\left( 32\right)
,
\end{equation}%
have a counterpart in the F-theory compactification on elliptically fibered
CY4- folds $X_{4}$. The origin of non abelian gauge fields in F-theory gauge
on CY4- folds is then due to the 7- \ brane wrapping 4-cycles in CY4- folds:

\begin{equation}
\begin{tabular}{lll}
$\text{Heterotic string/}${\small CY3} &  & {\small F-theory on $CY_{\text{4}%
}$} \\ \hline\hline
{\small gauge symmetry}\qquad $G$ & $\qquad \longleftrightarrow \qquad $ &
singularity \\
{\small gauge fields }$\mathcal{A}_{{\small M}}=\left( \mathcal{A}_{{\small M%
}}^{{\small I}},\mathcal{A}_{{\small M}}^{{\small \pm \alpha }}\right) $ & $%
\qquad \longleftrightarrow \qquad $ & coincident branes \ \ \ \ \ \  \\
\hline
\end{tabular}
\label{ab}
\end{equation}

\ \ \ \ \ \newline
In this table, the gauge fields $\mathcal{A}_{{\small M}}^{{\small I}}$ and $%
\mathcal{A}_{{\small M}}^{{\small \pm \alpha }}$ are respectively associated
with the Cartan Weyl basis generators $\left( H_{I},E_{\pm \alpha }\right) $
the of the Lie algebra \textbf{g }of the gauge symmetry G; i.e
\begin{equation}
\mathcal{A}_{{\small M}}=\sum_{{\small \alpha \in \Delta }}{\small E}_{%
{\small \pm \alpha }}\mathcal{A}_{{\small M}}^{{\small \pm \alpha }%
}+\sum_{I=1}^{r}{\small H}_{{\small I}}\mathcal{A}_{{\small M}}^{{\small I}%
}\qquad ,
\end{equation}%
with\textbf{\ }$\Delta =\Delta \left( \mathbf{g}\right) $ being the root
system of \textbf{g} and $r=r\left( \mathbf{g}\right) $\ is its rank. To
complete the table (\ref{ab}) for the case of F-theory GUT models, we still
need to study:\newline
\textbf{(a)} the engineering of non abelian gauge symmetry through the
implementation of the ADE geometric singularities in the elliptically K3
fiber of the local $CY4\sim K3\times S$,\newline
\textbf{(b)} the 4D effective gauge theory on the seven brane wrapping
compact 4- cycles in the base $S$ of the local CY4-fold.\newline
Below we study these two issues with some details.

\subsubsection{\emph{4D}\ gauge invariance in F-theory on CY4s}

In the F- theory set up of the duality (\ref{cc}), the \emph{4D}\ space time
gauge symmetry G has a very nice geometric interpretation. This invariance
is in fact captured by a Weierstrass ADE\ singularity living in the local
CY4-fold which may roughly be thought of as,%
\begin{equation}
\begin{tabular}{lll}
$\mathbb{E\times P}^{1}$ & $\rightarrow $ & $X_{4}$ \\
&  & $\downarrow {\small \pi }$ \\
&  & $S$%
\end{tabular}%
\end{equation}%
and described by the vanishing condition of the discriminant $\Delta _{E}$
of the elliptic curve $\mathrm{v}^{2}=\mathrm{u}^{3}+f\left( z\right)
\mathrm{u}+g\left( z\right) $. This condition reads as%
\begin{equation}
\Delta _{E}=\left( 16f^{3}+27g^{2}\right) =0,
\end{equation}%
and its solutions depend on the nature of the sections f$\left( z\right) $
and g$\left( z\right) $. \newline
To get more insight into the ways one deals with this condition for generic
elliptically fibered CY4- folds X$_{4}$, we start by recalling that elliptic
$X_{4}$ may be defined by the following complex four dimension elliptically
fibered hypersurface in $\mathbb{C}^{5}$,
\begin{equation}
\begin{tabular}{lll}
$\mathrm{v}^{2}=$ & $\ \ \ \mathcal{D}\left( {\small w}_{{\small 1}}{\small %
,w}_{{\small 2}}{\small ,w}_{{\small 3}}\right) \mathrm{u}^{3}$ $\ +$ $\
\mathcal{E}\left( {\small w}_{{\small 1}}{\small ,w}_{{\small 2}}{\small ,w}%
_{{\small 3}}\right) \mathrm{u}^{2}$ &  \\
& $+$ $\ \mathcal{F}\left( {\small w}_{{\small 1}}{\small ,w}_{{\small 2}}%
{\small ,w}_{{\small 3}}\right) \mathrm{u}$ $\ +$ $\ \mathcal{G}\left(
{\small w}_{{\small 1}}{\small ,w}_{{\small 2}}{\small ,w}_{{\small 3}%
}\right) $ & $.$%
\end{tabular}
\label{yw}
\end{equation}%
In this relation inspired from eq(\ref{el}), the holomorphic functions $%
\mathcal{D}\left( {\small w}\right) $, $\mathcal{E}\left( {\small w}\right) $%
, $\mathcal{F}\left( {\small w}\right) $ and $\mathcal{G}\left( {\small w}%
\right) $ are special sections on the base manifold $B_{3}$ of the elliptic
Calabi-Yau 4- fold
\begin{equation}
\begin{tabular}{lll}
$\mathbb{E}$ & $\rightarrow $ & $X_{4}$ \\
&  & $\downarrow \pi $ \\
&  & $B_{3}$%
\end{tabular}%
\end{equation}%
The complex variables $w=\left( w_{1},w_{2},w_{3}\right) $ are the
holomorphic coordinates of the base manifold $B_{3}$ and the $w$- dependence
in the holomorphic sections $\mathcal{D}\left( {\small w}\right) $, $%
\mathcal{E}\left( {\small w}\right) $, $\mathcal{F}\left( {\small w}\right) $
and $\mathcal{G}\left( {\small w}\right) $ are such that the various
monomials in eq(\ref{yw}) transform homogeneously under coordinates
transformations in the base.\newline
To explicitly exhibit the Weierstrass ADE\ singularity on the complex 3-
dimension base $B_{3}$ of the CY4- fold, it is interesting to factorize
these holomorphic sections $\mathcal{D}\left( {\small w}\right) $, $\mathcal{%
E}\left( {\small w}\right) $, $\mathcal{F}\left( {\small w}\right) $ and $%
\mathcal{G}\left( {\small w}\right) $ like%
\begin{equation}
\begin{tabular}{llll}
$\mathcal{D}\left( {\small w}\right) $ & $=$ & $\vartheta \left(
s_{1},s_{2}\right) \times \mathrm{d}\left( z\right) $ & , \\
$\mathcal{E}\left( {\small w}\right) $ & $=$ & $\vartheta \left(
s_{1},s_{2}\right) \times \mathrm{e}\left( z\right) $ & , \\
$\mathcal{F}\left( {\small w}\right) $ & $=$ & $\vartheta \left(
s_{1},s_{2}\right) \times \mathrm{f}\left( z\right) $ & , \\
$\mathcal{G}\left( {\small w}\right) $ & $=$ & $\vartheta \left(
s_{1},s_{2}\right) \times \mathrm{g}\left( z\right) $ & ,%
\end{tabular}%
\end{equation}%
where we have supposed $\vartheta \left( s_{1},s_{2}\right) \neq 0$ and
where $\left( s_{1},s_{2};z\right) $ are new local holomorphic coordinates
of $B_{3}$ related to the old complex coordinates $\left(
w_{1},w_{2},w_{3}\right) $ by some local analytic coordinate change,%
\begin{equation}
\begin{tabular}{llll}
$s_{1}$ & $=$ & $s_{1}\left( w_{1},w_{2},w_{3}\right) $ & , \\
$s_{2}$ & $=$ & $s_{2}\left( w_{1},w_{2},w_{3}\right) $ & , \\
$z$ & $=$ & $z\left( w_{1},w_{2},w_{3}\right) $ & .%
\end{tabular}
\label{ss}
\end{equation}%
In doing so, we have broken the $U\left( 3\right) $ structure group of the
tangent bundle $TB_{3}$ ( with $B_{3}\sim \Sigma _{0}$ $\mathbb{\times }$ $S$%
) down to $U_{R}\left( 1\right) \times U\left( 2\right) $ with $U_{R}\left(
1\right) $ and $U\left( 2\right) $ being respectively the structure group of
the tangent sub- bundles $T\Sigma _{0}$\ and $TS$. These complex structure
groups are contained in the R-symmetry groups of the compactification of
\emph{12D} F-theory down to \emph{4D},
\begin{equation}
\begin{tabular}{llll}
$U\left( 3\right) $ & $\subset $ & $SU\left( 4\right) \simeq SO_{R}\left(
6\right) $ & , \\
$U_{R}\left( 1\right) $ & $\simeq $ & $SO_{R}\left( 2\right) $ & , \\
$U\left( 2\right) $ & $\subset $ & $SU\left( 2\right) \times SU\left(
2\right) \simeq SO_{R}\left( 4\right) $ & .%
\end{tabular}
\label{ur}
\end{equation}%
Notice also the following features:\newline
(\textbf{1}) the complex holomorphic functions $\mathrm{d}\left( z\right) ,$
$\mathrm{e}\left( z\right) $, $\mathrm{f}\left( z\right) $ and $\mathrm{g}%
\left( z\right) $ are particular sections of the canonical bundle $\mathcal{K%
}_{\Sigma _{0}}$ on the curve $\Sigma _{0}$ in $B_{3}$. These holomorphic
functions, which will be specified later on for Weierstrass ADE
singularities; see table (\ref{tab1}), transform homogenously under the
change $z\rightarrow \varrho z$ as follows,%
\begin{equation}
\begin{tabular}{llll}
$\mathrm{d}\left( z\right) $ & $\rightarrow $ & $\varrho ^{n_{d}}\mathrm{d}%
\left( z\right) $ & , \\
$\mathrm{e}\left( z\right) $ & $\rightarrow $ & $\varrho ^{n_{e}}\mathrm{e}%
\left( z\right) $ & , \\
$\mathrm{f}\left( z\right) $ & $\rightarrow $ & $\varrho ^{n_{f}}\mathrm{f}%
\left( z\right) $ & , \\
$\mathrm{g}\left( z\right) $ & $\rightarrow $ & $\varrho ^{n_{g}}\mathrm{g}%
\left( z\right) $ & ,%
\end{tabular}%
\end{equation}%
where $n_{d}$, $n_{e}$, $n_{f}$ and $n_{g}$ are some positive integers.%
\newline
(\textbf{2}) In the coordinate frame $\left( u,v,z;s_{1},s_{2}\right) $, the
local CY4- fold is thought of as
\begin{equation}
\begin{tabular}{lll}
$Y_{2}$ & $\rightarrow $ & $X_{4}$ \\
&  & $\downarrow \pi $ \\
&  & $S$%
\end{tabular}
\label{ys}
\end{equation}%
where the local surface $Y_{2}$ is given by the elliptic curve $\mathbb{E}$
fibered on the complex line $\Sigma _{0}$ with coordinate $z$ ($Y_{2}\sim
\mathbb{E}$ $\mathbb{\times }$ $\Sigma _{0}$). In this realization, the
Calabi-Yau condition requires the two following:\newline
\textbf{(a)} the local coordinates $\left( u,v,z\right) $ have to transform
as sections of the canonical bundle over $S$. Under the \textquotedblleft
scaling\textquotedblright\ $s_{i}\rightarrow \sqrt{\lambda }s_{i}$, the
local coordinates $\left( u,v,z\right) $ transform like
\begin{equation}
\left( u,v,z\right) \qquad \rightarrow \qquad \left( \lambda ^{a}u,\lambda
^{b}v,\lambda ^{c}z\right) ,  \label{sc}
\end{equation}%
where $\lambda $ is a non zero complex number and the degrees a, b and c are
some integers to determine.\newline
\textbf{(b)} the holomorphic 2- form $\Omega ^{\left( 2,0\right) }=\frac{%
du\wedge dz}{v}$ over the local surface Y$_{2}$ should scale like $\Omega
^{\left( 2,0\right) }\rightarrow \lambda \Omega ^{\left( 2,0\right) }$. This
condition requires that the degrees a, b and c should be constrained as $%
a-b+c=1$.\newline
Substituting (\ref{ss}) back into (\ref{yw}) and setting $\mathrm{v}%
^{2}=\vartheta \mathrm{\tilde{v}}^{2}$, we can factorize it as follows%
\begin{equation}
\begin{tabular}{lll}
$\mathrm{\tilde{v}}^{2}=$ & $d\left( z\right) \mathrm{u}^{3}+e\left(
z\right) \mathrm{u}^{2}+f\left( z\right) \mathrm{u}+g\left( z\right) $ & ,
\\
$\vartheta =$ & $\vartheta \left( s_{1},s_{2}\right) $ & .%
\end{tabular}
\label{ell}
\end{equation}%
From these relations and following the analysis of \textrm{\cite{H1}}, one
can immediately read the Weierstrass form of the standard ADE singularities
by specifying the holomorphic sections $\mathrm{d}\left( z\right) $, $%
\mathrm{e}\left( z\right) $, $\mathrm{f}\left( z\right) $ and $\mathrm{g}%
\left( z\right) $ as given below,
\begin{equation}
\begin{tabular}{ll|llllllll}
{\small singularity} &  &  & $d$ &  & $e$ &  & $f$ &  & $g$ \\ \hline\hline
A$_{n}$ &  &  & 0 &  & 1 &  & 0 &  & z$^{n+1}$ \\
D$_{n}$ &  &  & 0 &  & z &  & 0 &  & z$^{n-1}$ \\
E$_{6}$ &  &  & 1 &  & 0 &  & 0 &  & z$^{4}$ \\
E$_{7}$ &  &  & 1 &  & 0 &  & z$^{3}$ &  & 0 \\
E$_{8}$ &  &  & 1 &  & 0 &  & 0 &  & z$^{5}$ \\ \hline
\end{tabular}
\label{tab1}
\end{equation}%
For later use we mainly need the engineering of the following singularities,
\begin{equation}
\begin{tabular}{llllll}
A$_{n}$ & : & $\tilde{v}^{2}=u^{2}+z^{n+1}$ & , & $n=4,5,6$ & , \\
D$_{n}$ & : & $\tilde{v}^{2}=zu^{2}+\alpha z^{n-1}$ & , & $n=5,6$ & , \\
E$_{6}$ &  & $\tilde{v}^{2}=u^{3}+z^{4}$ & , &  &
\end{tabular}%
\end{equation}%
in particular the A$_{4}$ geometry; its one -fold enhancements A$_{5}$ and D$%
_{5}$\ and the two- folds enhancements D$_{6}$ and E$_{6}$. These
enhancements, which are related to switching off Higgs vevs, have a
geometric realization in terms of colliding singularities. Below we give
some specific examples.

\subsubsection{Examples}

To fix the ideas on colliding singularities, we give below some illustrating
examples on the engineering of enhanced non abelian gauge symmetries by
colliding geometric singularities in the local CY4- folds. These examples
will be used later on when we consider gauge/brane interpretation as well as
the engineering of matter and Yukawa couplings.

$SU\left( 2\right) $\emph{\ gauge invariance}\newline
In the description we have been using so far, the Weierstrass\ singularity
capturing the $SU\left( 2\right) $ gauge invariance of the low energy
quantum field theory embedded in F-theory compactified on K3 fibered $CY4$-
folds, termed in Kodaira classification as $A_{1}$ \textrm{\cite{FS}}, is
given by,%
\begin{equation}
\begin{tabular}{lll}
$\mathrm{v}^{2}=$ & $\vartheta \left( \mathrm{u}^{2}+z^{2}\right) $ & , \\
$\vartheta =$ & $\vartheta \left( s_{1},s_{2}\right) $ & ,%
\end{tabular}%
\end{equation}%
where $\vartheta \left( s_{1},s_{2}\right) $ is a non zero holomorphic
function that live on the complex surface $S$ of eq(\ref{ys}). The locus of
this $A_{1}$ geometric singularity lives at,%
\begin{equation}
\left\{ P_{0}\right\} \times S,
\end{equation}%
with $P_{0}=\left( \mathrm{u,v,z}\right) =\left( 0,0,0\right) $ is the
singular point in the K3 fiber where the elliptic curve $\mathbb{E}$
degenerate. This means that the local CY4- folds (\ref{ys}) is given by the
local $A_{1}$ space
\begin{equation}
\mathrm{u}^{2}+\mathrm{\tilde{v}}^{2}+z^{2}=0
\end{equation}%
fibered on $S$. The fibration is captured by the relation $\mathrm{v}%
^{2}=-\vartheta \left( s_{1},s_{2}\right) \mathrm{\tilde{v}}^{2}$ and
extends directly to higher order A$_{n}$ geometries fibered on $S$.

\begin{figure}[tbph]
\begin{center}
\hspace{0cm} \includegraphics[width=10cm]{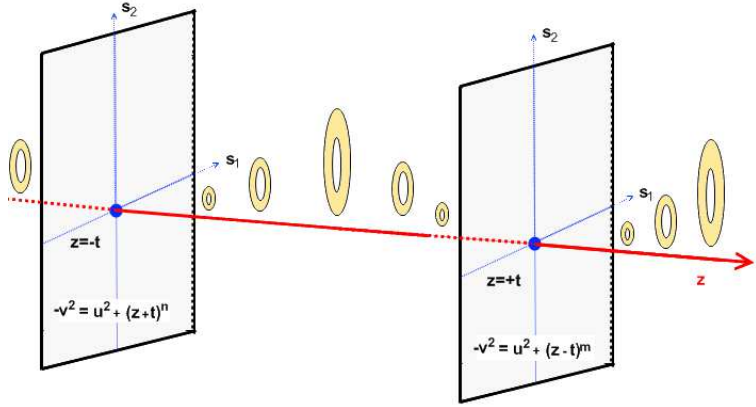}
\end{center}
\par
\vspace{-0.5 cm}
\caption{Geometric engineering of $SU\left( n\right) \times SU\left(
m\right) $ gauge invariance. The elliptic curve (yellow color) degenerate on
the surface $z=-t$ and $z=+t$. For $t=0$, the two singularities collide and
gauge symmetry gets enhanced to $SU\left( n+m\right) $.}
\label{mn}
\end{figure}

$SU\left( n\right) \times SU\left( m\right) $\emph{\ gauge invariance}%
\newline
From the preceding example, it is not difficult to see that the Weierstrass\
singularity capturing the semi simple $SU\left( n\right) \times SU\left(
m\right) \times U\left( 1\right) $ gauge invariance of the supersymmetric QFT%
$_{4}$ embedded in F- theory on the CY4- folds is then given by the
following algebraic relation%
\begin{equation}
\begin{tabular}{lll}
$\frac{\mathrm{v}^{2}}{\vartheta }=$ & $\mathrm{v}^{2}+\left( z+t\right)
^{n}\left( z-t\right) ^{m}$ & , \\
$\vartheta =$ & $\vartheta \left( s_{1},s_{2}\right) $ & .%
\end{tabular}
\label{nm}
\end{equation}%
The complex modulus $t$ is a section on $\mathcal{K}_{S}$ with same degree
as the variable $z$. This modulus may be physically thought of as a vevs of
a matter field $\phi $ in the adjoint representation of $SU\left( n+m\right)
$ with the following Cartan subalgebra value,
\begin{equation}
\left\langle \phi \right\rangle
=t\sum_{I=1}^{n-1}H_{I}-t\sum_{I=1}^{m-1}H_{n-1+I}.
\end{equation}%
Notice that the singularity $A_{n-1}$ lives at $\left\{ P_{1}\right\} \times
S$ with the point $P_{1}=\left( 0,0,-t\right) $ while the singularity $%
A_{m-1}$ lives at $\left\{ P_{2}\right\} \times S$ with $P_{2}=\left(
0,0,+t\right) $; see also the figure (\ref{mn}) for illustration.

\ \ \ \ \ \ \newline
The locus of the $A_{n-1}$ and $A_{m-1}$ singularity is then given by the
set $\left\{ P_{1},P_{2}\right\} \times S$. Notice moreover that in the case
where $t\rightarrow 0$, the two $A_{n-1}$ and $A_{m-1}$ singularities
collide and the gauge symmetry gets enhanced to \emph{SU}$\left( n+m\right) $
with singular geometry algebraic equation%
\begin{equation}
\begin{tabular}{lll}
$\mathrm{v}^{2}=$ & $\vartheta \left[ \mathrm{u}^{2}+z^{n+m}\right] $ & ,%
\end{tabular}%
\end{equation}%
with $\vartheta =\vartheta \left( s_{1},s_{2}\right) $ as before.

\emph{SO}$\left( 2n\right) $\emph{\ and E}$_{6}$ \emph{gauge invariances}%
\newline
The elliptic singularity capturing the \emph{SO}$\left( 2n\right) $\emph{\ }%
gauge symmetry is as follows
\begin{equation}
\begin{tabular}{lllll}
$\mathrm{v}^{2}=$ & $\left( \alpha ^{2}z^{n-1}-z\mathrm{u}^{2}\right)
\vartheta $ & , & $n\geq 4$ & , \\
$\vartheta =$ & $\vartheta \left( s_{1},s_{2}\right) $ &  &  & ,%
\end{tabular}%
\end{equation}%
where the modulus $\alpha $ is trick to handle gauge enhancements \textrm{%
\cite{H1}}. This singularity lives at $\left( \mathrm{u,v,z}\right) =\left(
0,0,0\right) $ whatever are the complex coordinates $\left(
s_{1},s_{2}\right) $. Then, the locus of the D$_{n}$ singularity is $\left\{
P_{0}\right\} \times S$ with $P_{0}=\left( 0,0,0\right) $. Notice that for $%
z\neq 0$; say $z=1$, the above relation describes an A$_{1}$ singularity at $%
\left( \mathrm{u,v,\alpha }\right) =\left( 0,0,0\right) $.\newline
Similarly, the elliptic singularity capturing the \emph{E}$_{6}$\emph{\ }%
gauge symmetry is
\begin{equation}
\begin{tabular}{lll}
$\mathrm{v}^{2}=$ & $\left( \mathrm{u}^{3}+z^{4}\right) \vartheta $ & ,%
\end{tabular}%
\end{equation}%
with $\vartheta =\vartheta \left( s_{1},s_{2}\right) $. This singularity
lives at $\left\{ \left( 0,0,0\right) \right\} \times S$. Aspects of
colliding of such kind of singularities will be detailed in the following
subsection.

\subsection{Colliding singularities: pure and hybrids}

Colliding singularities in the CY4- folds yields an enhancement of the gauge
invariance of the supersymmetric QFT$_{4}$ embedded in the F-theory
compactified on CY4. Generally speaking, we distinguish the following rough
classification:\newline
\textbf{(1)} \emph{Colliding singularities of same nature: pure colliding},%
\newline
(\textbf{2}) \emph{Colliding singularities of different types: hybrids}.%
\newline
Let us comment briefly this classification through some selected examples.

\subsubsection{Pure colliding}

This kind of gauge invariance enhancement concerns are the colliding of two
or several singularities of same type. Restricting the classification to the
ADE case, we distinguish for a given integer $l\geq 2$:\newline
\textbf{(a)} Unitary symmetry: $SU\left( n_{1}\right) \times ...\times
SU\left( n_{l}\right) $ with $n_{i}\geq 2,$\newline
\textbf{(b)} Orthogonal symmetry: $SO\left( 2n_{1}\right) \times ...\times
SO\left( 2n_{l}\right) $ with $n_{i}\geq 4,$\newline
\textbf{(c)} Exceptional symmetry $E_{s}^{\otimes l}$ with $s=6,7,8$.\newline
These collisions do not give necessary a singularity of same type as we will
show on the following examples. \newline
For the unitary series, the simplest example concerns colliding the \emph{A}$%
_{n}$ and $A_{m}$ singularities which lead to the enhancement,%
\begin{equation}
\emph{A}_{n}\times \emph{A}_{m}\rightarrow \emph{A}_{n+m},
\end{equation}%
whose algebraic relation is given by eq(\ref{nm}). In the case of three
singularities $\emph{A}_{n_{1}},$ $\emph{A}_{n_{2}}$ and $\emph{A}_{n_{3}}$,
the collision can be achieved in various ways and leads to the following
enhancement,%
\begin{equation}
\emph{A}_{n_{1}}\times \emph{A}_{n_{2}}\times \emph{A}_{n_{3}}\rightarrow
\left\{
\begin{array}{c}
\emph{A}_{n_{1}+n_{2}}\times \emph{A}_{n_{3}} \\
\emph{A}_{n_{1}}\times \emph{A}_{n_{2}+n_{3}} \\
\emph{A}_{n_{1}+n_{3}}\times \emph{A}_{n_{2}}%
\end{array}%
\right. \rightarrow \emph{A}_{n_{1}+n_{2}+n_{3}}
\end{equation}%
The colliding of $\emph{A}_{n}$ singularities is a commutative and
associative product. These collisions extend straightforwardly to the case
of colliding $l$ singular components A$_{n_{i}}$. We have several ways to do
these collisions; but with same result at the end:%
\begin{equation}
\emph{A}_{n_{1}}\times \emph{A}_{n_{2}}\times ...\times \emph{A}%
_{n_{l}}\rightarrow \left\{
\begin{array}{c}
\emph{A}_{n_{1}+n_{2}}\times ...\times \emph{A}_{n_{l}} \\
\emph{\vdots } \\
\emph{A}_{n_{1}+n_{l}}\times \emph{A}_{n_{2}}...%
\end{array}%
\right. \rightarrow \cdots \rightarrow \emph{A}_{n_{1}+n_{2}+\cdots +n_{l}}.
\end{equation}%
In the case where all singularities as well as their collision have all of
them the complex surface S as a locus in the local CY4- folds, then the
algebraic relation describing the colliding of these singularities reads as
follows
\begin{equation}
\begin{tabular}{lll}
$\frac{\mathrm{y}^{2}}{\vartheta }=$ & $\mathrm{x}^{2}+\dprod_{i=1}^{l}%
\left( z-t_{i}\right) ^{n_{i}}$ & , \\
$\vartheta =$ & $\vartheta \left( s_{1},s_{2}\right) $ & ,%
\end{tabular}%
\end{equation}%
where the complex moduli $t_{i}$ are sections on the canonical bundle $%
\mathcal{K}_{S}$ with same degree property as $z$.\newline
Regarding the orthogonal D- singularities, the colliding of two
singularities $D_{n}$ and $D_{m}$ gives an \emph{exotic} singularity which
is beyond the scope of the present study. These singularities may be
associated with the indefinite sector in the classification of Lie algebras
\textrm{\cite{AHL, AIT, EH, BS}}. The same thing is valid for the colliding
of the exceptional singularities.

\subsubsection{Hybrids}

Given several isolated singularities of type ADE, which are associated with
a semi simple gauge group invariance in the QFT$_{4}$ limit of F- theory on
CY4- folds, we can engineer enhanced singularities by performing collisions.
In addition to the pure colliding considered above, we also have, amongst
others, the following hybrids:\newline
$\alpha $) \emph{case} $\emph{A}_{n}\times \emph{D}_{m}\rightarrow \emph{D}%
_{n+m+1}$\newline
$\beta $) \emph{case} $\emph{A}_{n}\times \emph{E}_{6}\rightarrow $\emph{\
Exotic singularity, }$n>2$\newline
$\gamma $) \emph{case} $\emph{D}_{n}\times \emph{E}_{6}\rightarrow $\emph{\
Exotic singularity.}\newline
This analysis extends to the case of colliding more than two singularities.
For the case of three singularities, we have\newline
$\alpha $) \emph{case} $\emph{A}_{n}\times \emph{A}_{m}\times \emph{D}%
_{k}\rightarrow \emph{D}_{n+m+k+1}$\newline
$\beta $) \emph{case} $\emph{A}_{n}\times \emph{A}_{m}\times \emph{E}%
_{6}\rightarrow $\emph{\ Exotic singularity,}\newline
$\gamma $) \emph{case} $\emph{A}_{n}\times \emph{D}_{m}\times \emph{E}%
_{6}\rightarrow $\emph{\ Exotic singularity.}\newline
More hybrids such as the triangular geometries $T_{n,m,r}$ as those
considered in the geometric engineering of superconformal QFT$_{4}$ embedded
in type II compactification on CY3- folds \textrm{\cite{I3, BS}} as
technical details regarding these hybrids will be reported elsewhere.

\section{Seven brane wrapping 4-cycles}

In the \emph{4D} $\mathcal{N}=1$ supersymmetric QFT limit of F-theory on
local Calabi Yau four- folds $X_{4}$, there is a close relation between the
degeneracy loci of the elliptic curve $\mathbb{E}$ in $X_{4}$ and the seven
brane wrapping 4- cycles. The space time region
\begin{equation}
\begin{tabular}{ll}
$\left( x^{0},x^{1},x^{2},x^{3};s_{1},s_{2},\bar{s}_{1},\bar{s}_{2};z,\bar{z}%
,x^{{\small 11}},x^{{\small 12}}\right) $ &
\end{tabular}
\label{11}
\end{equation}%
of the 12- dimensional F-theory where the elliptic\textrm{\footnote{%
In (\ref{11}) the variables $\left( x^{{\small 11}},x^{{\small 12}}\right) $
are the two real compact coordinates of the extra 2-torus used in F-Theory
and which as been realized in terms of the algebraic curve $%
v=du^{3}+eu^{2}+fu+g.$}} curve $\mathbb{E}$ (\ref{ell}) degenerates,
corresponds precisely to the world volume $\mathcal{V}_{8}$ of the seven
brane,
\begin{equation}
\text{{\scriptsize X}}^{M}=\left( x^{0},\text{ }x^{1},\text{ }x^{2},\text{ }%
x^{3}\text{ };\text{ }s_{1},\text{ }s_{2},\text{ }\bar{s}_{1},\text{ }\bar{s}%
_{2}\right)  \label{xs}
\end{equation}%
In this region, the $\left( u,v,z\right) $ coordinates of the K3 fiber of $%
X_{4}$, with $u=u\left( x^{{\small 11}},x^{{\small 12}}\right) ,$ $v=v\left(
x^{{\small 11}},x^{{\small 12}}\right) $, take particular values and one is
left with the local coordinates (\ref{xs}) which parameterize the world
volume of the seven brane. Notice that in addition to the non compact \emph{%
4D} space time $\mathbb{R}^{1,3}$ with the usual real coordinates $\left(
x^{0},x^{1},x^{2},x^{3}\right) $, the holomorphic coordinates $\left( s_{1},%
\text{ }s_{2}\right) $ of (\ref{xs}) parameterize the compact complex
surface $S$ sitting in the complex three dimensional base $B_{3}$. The
compact real four dimensional manifold $S$ is just the locus of the elliptic
singularity in the Calabi Yau 4- folds.\newline
In this section, we first consider the $\mathcal{N}=1$ supergravity in \emph{%
8D }space time \textrm{\cite{SS}}; then we analyze its reduction to \emph{4D}
supergravity with four supersymmetric charges by borrowing ideas and results
from the twisted topological field analysis of \textrm{\cite{H1}; }in
particular the solutions of BPS equations. After that, we use these results
to study the \emph{4D} $\mathcal{N}=1$ non abelian gauge invariance in the
seven brane wrapping $S$ .

\subsection{General on $\mathcal{N}=1$ supergravity in \emph{8D}}

We start by recalling that in curved eight dimensional space time lives a
\emph{8D }$\mathcal{N}=1$ supergravity theory describing the interacting
dynamics of the supergravity multiplet $\mathcal{G}_{sugra}^{\left(
8D\right) }$ coupled to superYang Mills $\mathcal{V}_{SYM}^{\left( 8D\right)
}$. This supersymmetric gauge theory may be viewed as the field theory limit
of compactified superstrings theory at Planck scale; in particular as the
limit of F- theory on $K3$ which is dual \ to \emph{10D} heterotic
superstring on $\mathbb{T}^{2}$. In this subsection, we first review briefly
general results on this supersymmetric gauge theory having sixteen conserved
supercharges. Then we consider the super Yang-Mills theory in the limit of
decoupled supergravity in connection with the philosophy of the F-theory GUT
models building and the gauge theory on the seven brane wrapping 4-cycles of
the Calabi Yau 4- folds.

\subsubsection{Fields spectrum}

The massless spectrum of the $\mathcal{N}=1$ supergravity in \emph{8D}
involves two super multiplets: the supergravity multiplet $\mathcal{G}%
_{sugra}^{\left( 8D\right) }$ and the Maxwell (super-Yang-Mills) multiplet $%
\mathcal{V}_{8D}$.\newline
The \emph{8D} $\mathcal{N}=1$ supergravity multiplet $\mathcal{G}%
_{sugra}^{\left( 8D\right) }$ has the following fields content:

\begin{equation}
\begin{tabular}{ll|llll}
{\small Bosonic fields} & \qquad & \qquad & \qquad & {\small Fermions} &  \\
\hline\hline
$e_{M}^{A}$ $,$ $\mathcal{B}_{{\small MN}}$ $,$ $\mathcal{G}_{{\small M}}^{%
{\small 1}}$ $,$ $\mathcal{G}_{{\small M}}^{{\small 2}}$ $,$ $\mathrm{\sigma
}$ & \qquad &  & \qquad & $\psi _{_{M}}$ $,$ $\chi $ &  \\ \hline
\end{tabular}
\label{gr}
\end{equation}

\ \ \ \ \ \ \ \ \ \ \ \ \ \ \ \newline
The bosonic sector consists of the graviton (eightbein) $e_{M}^{A}$ with
space time metric $G_{MN}=e_{M}^{A}e_{NA}$, the antisymmetric field $%
\mathcal{B}_{{\small MN}}$, two gravi-photons $\mathcal{G}_{{\small M}}^{%
{\small 1}}$, $\mathcal{G}_{{\small M}}^{{\small 2}}$; and a scalar field $%
\mathrm{\sigma }$: the \emph{8D} dilaton. The fermionic sector consists of
the \emph{8D} Rarita-Scwhinger field $\mathcal{\psi }_{_{M}}$ and a \emph{8D}
pseudo Majorana fermion $\chi $. This supermultiplet contains \emph{48 + 48}
on shell propagating degrees of freedom capturing the pure supergravity
dynamics. \newline
Th\textbf{e }\emph{8D} $\mathcal{N}=1$ Maxwell supermultiplet $\mathcal{V}_{%
{\small Max}}^{\left( \QTR{sl}{8D}\right) }$ has the following fields
content:

\begin{equation}
\begin{tabular}{l|lll|llll}
superfield &  & {\small Bosonic fields} & \qquad & \qquad & \qquad & {\small %
Fermions} &  \\ \hline\hline
$\mathcal{V}_{{\small Max}}^{\left( \QTR{sl}{8D}\right) }$ &  & $\mathcal{A}%
_{M}$ $,$ $\mathcal{\phi }_{1}$ $,$ $\mathcal{\phi }_{2}$ & \qquad &  &
\qquad & $\mathcal{\lambda }_{\hat{a}}$ &  \\ \hline
\end{tabular}
\label{ma}
\end{equation}

\ \ \ \ \ \ \ \ \ \ \ \ \newline
where $\mathcal{A}_{M}$ ( to be some times denoted as $\mathcal{A}%
_{M}^{\left( {\small 8D}\right) }$ ) is a \emph{8D} Maxwell field. The
spinor field $\lambda _{\hat{a}}$ is the \emph{8D} gaugino having real \emph{%
8} propagating degrees of freedom and the fields $\left( \mathcal{\phi }_{1},%
\mathcal{\phi }_{2}\right) $ are two real \emph{8D} scalars parameterizing
the $SO\left( 1,2\right) /SO\left( 2\right) $ coset manifold
\begin{equation}
SO\left( 1,2\right) /SO\left( 2\right) \sim SU\left( 1,1\right) /U\left(
1\right)
\end{equation}%
defining the interactions of these scalar fields. \newline
Notice that the two scalar fields $\mathcal{\phi }_{m}=\left( \mathcal{\phi }%
_{1},\mathcal{\phi }_{2}\right) $ are charged under the $U_{R}\left(
1\right) \simeq SO_{R}\left( 2\right) $ symmetry of eq(\ref{ur}) with the $%
SO_{R}\left( 2\right) $ appearing in the breaking of the \emph{10D} space
time group
\begin{equation}
SO\left( 1,9\right) \supset SO\left( 1,7\right) \times SO_{R}\left( 2\right)
.
\end{equation}%
This property can be immediately viewed by thinking about $\left( \mathcal{A}%
_{M}^{\left( {\small 8D}\right) },\mathcal{\phi }_{1},\mathcal{\phi }%
_{2}\right) $ as following from the reduction of the dimensional field $%
\mathcal{A}_{M}^{\left( {\small 10D}\right) }$. Reducing the flat space time
dimension $\mathbb{R}^{1,9}$ down to $\mathbb{R}^{1,9}\times \mathbb{C}$,
the real 1- form gauge connexion $\mathcal{A}^{\left( {\small 10D}\right)
}=\sum_{M=0}^{9}\mathcal{A}_{M}^{\left( 10D\right) }dx^{M}$ splits as
\begin{equation}
\mathcal{A}^{\left( {\small 10D}\right) }=\left( \sum_{M=0}^{7}\mathcal{A}%
_{M}^{\left( {\small 8D}\right) }dx^{M}\right) +\left( \phi dz+\bar{\phi}d%
\bar{z}\right) ,
\end{equation}%
where we have set
\begin{equation}
\phi =\mathcal{\phi }_{1}+i\mathcal{\phi }_{2}=\frac{1}{2}\left( \mathcal{A}%
_{8}^{\left( {\small 10D}\right) }+i\mathcal{A}_{9}^{\left( {\small 10D}%
\right) }\right)
\end{equation}%
and where $z=x^{8}+ix^{9}$ stands for the coordinate of the complex line $%
\mathbb{C}$. Under the change $z\rightarrow e^{i\theta }z$, then we should
have
\begin{equation}
\phi \rightarrow e^{-i\theta }\phi
\end{equation}%
showing that $\phi $ carries a $U_{R}\left( 1\right) $ charge $q_{\phi }=-1$.

\emph{Supergravity action}\newline
To write down the more general supergravity action, let us consider the case
of $n$ Maxwell multiplets $\mathcal{V}_{{\small Max}}^{\text{ }\QTR{sl}{a}%
}=\left( \mathcal{A}_{M}^{a},\mathcal{\phi }_{1}^{a},\mathcal{\phi }_{2}^{a};%
\mathcal{\lambda }_{\hat{a}}^{a}\right) $ which, in the case of F- theory on
K3, may be thought of as dealing with the gauge theory of n separated 7-
branes. Combining these Maxwell multiplets $\mathcal{V}_{{\small Max}}^{%
\text{ }\QTR{sl}{a}}$ with the gravity multiplet $\mathcal{G}%
_{sugra}^{\left( 8D\right) }$, we get, in addition to the fermions $\psi
_{_{M}},\chi ,\mathcal{\lambda }_{\hat{a}}^{a}$, the following bosonic fields%
\begin{equation}
\begin{tabular}{llllllll}
$e_{M}^{A},$ & $\mathcal{B}_{{\small MN}},$ & $\mathcal{A}_{M}^{\Lambda },$
& $\mathcal{\phi }^{a},$ & $\mathrm{\sigma }$, &  &  &
\end{tabular}%
\end{equation}%
where we have set $\mathcal{\phi }^{a}=\left( \mathcal{\phi }_{1}^{a},%
\mathcal{\phi }_{2}^{a}\right) $. In this relation, the scalars $\mathcal{%
\phi }^{a}$ with $a=1,...,n$ parameterize the Kahler manifold
\begin{equation}
SO\left( n,2\right) /SO\left( n\right) \times SO\left( 2\right)
\end{equation}%
fixing the interactions of the scalars. Following \textrm{\cite{FE, S2}},
this coset manifold is conveniently parameterized by the following typical
representative $\left( n+2\right) \times \left( n+2\right) $ orthogonal
matrix,%
\begin{equation}
\begin{tabular}{llllll}
$L_{\Lambda }^{\underline{\Upsilon }}$ & $=$ & $\left(
\begin{array}{cc}
0_{{\small 2\times 2}} & \left( \phi ^{a}\right) _{2{\small \times n}} \\
\left( \phi _{a}\right) _{{\small n\times 2}} & 0_{{\small 2\times 2}}%
\end{array}%
\right) _{\left( n+2\right) \times \left( n+2\right) }$ & , & $L_{\Lambda }^{%
\underline{\Upsilon }}L_{\Sigma }^{\underline{\Gamma }}\eta _{\underline{%
\Upsilon }\underline{\Gamma }}=\eta _{\Lambda \Sigma }$ & ,%
\end{tabular}%
\end{equation}%
where the metric $\eta _{\Lambda \Sigma }={\small diag}\left(
++...+--\right) $ of the $\mathbb{R}^{n,2}$ real space.\newline
Regarding the 8D gauge vector fields $\left( \mathcal{A}_{M}^{a},\mathcal{G}%
_{M}^{1},\mathcal{G}_{M}^{2}\right) $ involved in this theory, they may be
combined as $\mathcal{A}_{M}^{\Lambda }$, with $\Lambda =1,...,n+2$. These
Maxwell type gauge fields transform as a vector of $SO\left( n,2\right) $
while the gaugino partners transform as a vector under $SO\left( n\right) $.
To describe the interactions of the scalar fields, we also need the gauge
connection $L^{-1}\partial _{M}L$ which may be split as follows%
\begin{equation}
\begin{tabular}{lllll}
$L^{-1}\partial _{M}L$ & $=$ & $\left(
\begin{array}{cc}
Q_{Ma}^{b} & P_{Ma}^{j} \\
P_{Mi}^{b} & Q_{Mi}^{j}%
\end{array}%
\right) _{\left( n+2\right) \times \left( n+2\right) }$ & , &
\end{tabular}%
\end{equation}%
where $Q_{M}^{ab}$ and $Q_{M}^{ij}$ are respectively the $SO\left( n\right) $
and $SO\left( 2\right) $ gauge connections and where $P_{Ma}^{j}$ are the
Cartan-Maurer Form transforming homogeneously under the $SO\left( n\right)
\times SO\left( 2\right) $ gauge symmetry.\newline
Following \textrm{\cite{SS}}, the component field action of the $\mathcal{N}%
=1$ supergravity in \emph{8D} describing the interacting dynamics of the $%
\mathcal{G}_{sugra}^{\left( 8D\right) }$ and the n vector multiplets $%
\mathcal{V}_{{\small Max}}^{\left( \QTR{sl}{8D}\right) }$ reads as%
\begin{eqnarray}
\frac{\mathcal{L}_{0}}{\det e} &\simeq &\frac{1}{4}\mathcal{R}-\frac{1}{4}e^{%
\mathrm{\sigma }}\mathrm{\tau }_{{\small \Lambda \Sigma }}\mathcal{F}%
_{MN}^{\Lambda }\mathcal{F}^{MN\Sigma }-\frac{1}{12}e^{2\mathrm{\sigma }}%
\mathcal{G}_{MNQ}\mathcal{G}^{MNQ}  \notag \\
&&+\frac{3}{8}\partial _{M}\mathrm{\sigma }\partial ^{M}\mathrm{\sigma }+%
\frac{1}{4}P_{Ma}^{i}P_{i}^{Ma} \\
&&+\text{ fermionic terms + gauge couplings}  \notag
\end{eqnarray}%
with $\mathcal{F}_{MN}^{\Lambda }$, $\mathcal{G}_{MNQ}$ and $\mathrm{\tau }_{%
{\small \Lambda \Sigma }}$ given by
\begin{equation}
\begin{tabular}{llll}
$\mathcal{F}_{MN}^{\Lambda }$ & $=$ & $\partial _{M}\mathcal{A}_{N}^{\Lambda
}-\partial _{N}\mathcal{A}_{M}^{\Lambda }$ & , \\
$\mathcal{G}_{MNQ}$ & $=$ & $\partial _{M}\mathcal{B}_{NQ}-\eta _{\Lambda
\Sigma }\mathcal{F}_{MN}^{\Lambda }\mathcal{A}_{Q}^{\Sigma }+$ {\small %
cyclic permutation} & , \\
$\mathrm{\tau }_{{\small \Lambda \Sigma }}$ & $=$ & $L_{\Lambda }^{%
\underline{a}}L_{\Sigma }^{\underline{a}}+L_{\Lambda }^{\underline{i}%
}L_{\Sigma }^{\underline{i}}$ & .%
\end{tabular}%
\end{equation}

\subsubsection{SYM$_{8}$ in decoupling gravity limit}

The \emph{8D} $\mathcal{N}=1$ supergravity multiplet $\mathcal{G}_{8D}$ may
also couple non abelian superYang-Mills multiplets.\ The field content of
these $\mathcal{N}=1$ non abelian gauge supermultiplets is given by,

\begin{equation}
\begin{tabular}{ll|llll}
{\small Bosons} & \qquad & \qquad & \qquad & {\small Fermions} &  \\
\hline\hline
$\left.
\begin{array}{c}
A_{M}=\sum_{a=1}^{\dim G}T_{a}\mathcal{A}_{M}^{a}, \\
\phi =\sum_{a=1}^{\dim G}T^{a}\phi _{a}%
\end{array}%
\right. $ & \qquad &  & \qquad & $\lambda =\sum_{a=1}^{\dim G}T_{a}\lambda
^{a}$ &  \\ \hline
\end{tabular}%
\end{equation}

\ \ \ \ \ \ \ \ \ \ \newline
where now the \emph{8D} gauge fields are valued in the Lie algebra of the
gauge symmetry with matrix generators $\left\{ T_{a}\right\} $ as in eq(\ref%
{tac}).\newline
The component fields action describing the classical interacting dynamics
may be constructed perturbatively by using Noether method. This field action
reads as
\begin{equation}
\mathcal{S}_{\text{{\small sugra}}}^{\left( {\small 8D}\right)
}=\dint_{R^{1,7}}d^{8}x\text{ }\mathcal{L}_{\text{sugra}}^{\left( {\small 8D}%
\right) }\left( x\right) \text{\textsf{\ \qquad ,}}
\end{equation}%
with $\mathcal{L}_{\text{sugra}}^{\left( {\small 8D}\right) }$ describing
the lagrangian density of the \emph{8D} supergravity fields
\begin{equation}
\mathcal{L}_{\text{sugra}}^{\left( {\small 8D}\right) }=\mathcal{L}\left(
e_{M}^{A},\mathcal{B}_{{\small MN}},\mathcal{G}_{{\small M}}^{{\small 1}},%
\mathcal{G}_{{\small M}}^{{\small 2}},\mathrm{\varphi },\psi _{_{M}},\chi ;%
\mathcal{A}_{M},\phi ^{1},\phi ^{2},\lambda \right) .
\end{equation}%
It is given by the sum of the \emph{8D} Hilbert-Einstein supergravity term $%
\mathcal{L}_{HE}$ \ \emph{plus } $\mathcal{L}_{SYM-E}$ the \emph{8D}
superYang-Mills term coupled to supergravity.\newline
In the limit of decoupled supergravity, the dynamics of the gravity
supermultiplet (\ref{gr}) is freezed and the above action $\mathcal{S}_{%
{\small sugra}}^{\left( {\small 8D}\right) }$ reduces to the usual
supersymmetric Yang Mills theory $\mathcal{S}_{{\small SYM}}^{\left( {\small %
8D}\right) }=\dint_{R^{1,7}}d^{8}x\mathcal{L}_{{\small SYM}}^{\left( {\small %
8D}\right) }$ with,%
\begin{equation}
\begin{tabular}{llll}
$\mathcal{S}_{{\small SYM}}^{\left( {\small 8D}\right) }$ & $=$ & $%
\dint_{R^{1,7}}d^{8}xTr\left( -\frac{1}{8}\mathcal{F}_{MN}\mathcal{F}^{MN}+%
\frac{i}{2}\bar{\lambda}\Gamma ^{M}D_{M}\lambda +D_{M}\overline{\phi }%
D^{M}\phi \right) +...$ & ,%
\end{tabular}
\label{den}
\end{equation}%
where $\mathcal{F}_{MN}=\partial _{M}\mathcal{A}_{N}-\partial _{N}\mathcal{A}%
_{M}+\left[ \mathcal{A}_{M},\mathcal{A}_{N}\right] $ is the \emph{8D} field
strength valued in the Lie algebra of the gauge group.

\subsubsection{Reduction to 4D $\mathcal{N}=1$ supersymmetry}

In the supersymmetric \emph{QFT}$_{4}$ set up of the F-theory \emph{GUT}
models building, the starting point is precisely the $\mathcal{N}=1$
supersymmetric Yang-Mills lagrangian density $\mathcal{L}_{\text{{\small SYM}%
}}^{\left( {\small 8D}\right) }$ (\ref{den}). Since the seven brane wraps
4-cycles $S$ in the CY4- folds, the \emph{8D} fields $\Phi \left( x;s,\bar{s}%
\right) $ of the 7- brane bulk theory may be thought of as a collection of
\emph{4D} space time fields
\begin{equation}
\Phi _{\left\{ s_{1},s_{2}\right\} }=\Phi _{\left\{ s_{1},s_{2}\right\}
}\left( x\right)
\end{equation}%
labeled by points $s_{m}=\left( s_{1},s_{2}\right) \in S$. Then, to reduce
the above $\mathcal{N}=1$ \emph{8D} SYM$_{8}$ down to a $\mathcal{N}=1$
supersymmetry in \emph{4D }as required by the compactification of F-theory
on CY4-folds, one needs to compactify $\mathbb{R}^{1,7}$ as $\mathbb{R}%
^{1,3}\times S$. Under this compactification, the action $\mathcal{S}_{\text{%
{\small SYM}}}^{\left( {\small 8D}\right) }$ (\ref{den}) gets reduced down
to $\mathcal{S}^{\left( {\small 4D}\right) }$ as follows,
\begin{equation}
\mathcal{S}_{SYM}^{\left( {\small 4D}\right) }=\dint_{R^{1,3}}d^{4}x\mathcal{%
L}_{SYM}^{\left( {\small 4D}\right) },  \label{sl}
\end{equation}%
where $\mathcal{L}_{SYM}^{\left( {\small 4D}\right) }$ is a priori given by,%
\begin{equation}
\mathcal{L}_{SYM}^{\left( {\small 4D}\right) }=\int_{S}d^{2}sd^{2}\bar{s}%
\text{ }\mathcal{L}_{{\small SYM}}^{\left( {\small 8D}\right) }\left[ \phi
\left( x;s,\bar{s}\right) \right] .  \label{ls}
\end{equation}%
Notice that in performing the reduction from \emph{8D} down to \emph{4D},
one should worry about two main things: (1) supersymmetry and (2) chiral
matter representations.

\emph{Supersymmetry}\newline
In the flat \emph{8D} $\mathcal{N}=1$ supersymmetric Yang-Mills (\ref{den}),
there are sixteen conserved supersymmetries. This is too much since
compactification of F- theory on CY4- folds has only four conserved
supersymmetries. Thus the reduction from \emph{8D} down to \emph{4D }should
preserve $\frac{1}{4}$ of the original sixteen. Using\emph{\ }the
compactification $\mathbb{R}^{1,7}\rightarrow \mathbb{R}^{1,3}\times S$, the
$SO\left( 1,7\right) $ structure group gets broken down like
\begin{equation}
SO\left( 1,7\right) \rightarrow SO\left( 1,3\right) \times U\left( 2\right)
=SO\left( 1,3\right) \times SU\left( 2\right) \times U_{J}\left( 1\right) .
\end{equation}%
The mechanism to perform the reduction preserving four supersymmetric
charges has been studied in \textrm{\cite{H1}} and is based on the mapping%
\begin{equation}
U_{R}\left( 1\right) \times U_{J}\left( 1\right) \rightarrow
U_{J_{top}}\left( 1\right) \text{ ,}  \label{top}
\end{equation}%
with twist charge,
\begin{equation}
J_{top}=J+2R\equiv T,  \label{tr}
\end{equation}%
borrowed from topological field theory ideas.

\emph{Chiral matter}\newline
In \emph{4D} $\mathcal{N}=1$ supersymmetric gauge theory, chiral matter is
described by chiral superfields transforming in complex representations of
the gauge group. The reduction of $\mathcal{N}=1$ \emph{SYM}$_{8D}$ down to
a \emph{4D} $\mathcal{N}=1$ supersymmetric theory gives indeed chiral
matter; but only in the \emph{adjoint} representation. This property is
immediately seen by decomposing the $\mathcal{N}=1$ super Yang-Mills
miltiplets $\mathcal{V}_{{\small SYM}}^{\left( \QTR{sl}{8D}\right) }$. This
supermultiplet belongs to the adjoint representation of the gauge group and
decomposes as follows:\newline
\textbf{(a)} $\mathcal{N}=1$ super Yang-Mills miltiplets $\mathcal{V}_{%
{\small SYM}}^{\left( \QTR{sl}{4D}\right) }$,\newline
\textbf{(b)} three massless chiral multiplets $\Phi _{0},\Phi _{1},\Phi _{2}$
in the adjoint representation,\newline
\textbf{(c}) an infinite tower of massive KK\ type modes which may be
denoted as $\mathcal{V}_{\left[ \pm n\right] }^{\left( \QTR{sl}{4D}\right) }$%
, $\Phi _{0\left[ n\right] },\Phi _{1\left[ n\right] }$ and $\Phi _{2\left[ n%
\right] }$; see also next subsection for more details.\newline
As we see, there is no chiral superfield in complex representations of the
gauge group. This difficulty has been solved in wonderful manner in \textrm{%
\cite{H1}}, by considering local Calabi-Yau four-folds,
\begin{equation}
\begin{tabular}{llll}
Y$_{2}$ & $\rightarrow $ & \quad X$_{4}$ &  \\
&  & $\quad \downarrow \pi $ &  \\
&  & $\quad S$ & ,%
\end{tabular}
\label{yc}
\end{equation}%
where now the base surface $S=\cup _{a}\mathcal{C}_{a}$ with non trivial 4-
cycles intersections%
\begin{equation}
\mathcal{C}_{a}\cap \mathcal{C}_{a}=\Sigma _{ab}.
\end{equation}%
where $\Sigma _{ab}$ are real 2- cycles inside the CY4- folds. Each real 2-
cycle $\Sigma _{ab}$ defines the locus of intersecting seven branes where
precisely live chiral matter. In the next subsection, we give some explicit
details.

\subsection{$SU\left( N\right) $ invariance in seven brane}

In\ the F-theory set up, non abelian gauge invariance has a remarkable
geometric engineering in terms of seven branes wrapping compact 4- cycles $%
\mathcal{C}_{a}$ in the CY4- folds. On each 4-cycle the world volume of the
seven brane splits into two blocks: \newline
\textbf{(1)} the four non compact real $\left( 1+3\right) $ space time
dimensions viewed as the \emph{4D} space where lives the $\mathcal{N}=1$
supersymmetric GUT. \newline
\textbf{(2)} four compact directions wrapping the 4-cycle $\mathcal{C}_{a}$
a number of times; say $N_{a}$ times.\newline
In the case of $SU\left( N_{a}\right) $ gauge invariance, the fiber $Y$ of
the Calabi-Yau 4- folds (\ref{yc}) has a $A_{N_{a}-1}$ singularity described
by the following algebraic equation%
\begin{equation}
\frac{\mathrm{v}^{2}}{\vartheta }=\mathrm{u}^{2}+z^{N_{a}}\qquad ,\qquad
\vartheta =\vartheta \left( s_{1},s_{2}\right) ,  \label{an}
\end{equation}%
where the integer $N_{a}$ in the monomial $z^{N_{a}}$ captures the number of
times the seven brane wraps $\mathcal{C}_{a}$.

\subsubsection{QFT$_{8D}$ set up}

In the supersymmetric field theory analysis, the non abelian gauge theory in
the seven brane involves the following:\newline
First, a non abelian \emph{8D} $\mathcal{N}=1$ supersymmetric $SU\left(
N_{a}\right) $ Yang Mills multiplet
\begin{equation}
\left( \mathcal{A}_{{\small M}}^{\left( {\small 8D}\right) },\lambda _{\hat{a%
}}^{\left( {\small 8D}\right) },\phi ^{\left( {\small 8D}\right) }\right) ,
\end{equation}%
with flat space time gauge dynamics given by the action $\mathcal{S}_{\text{%
{\small SYM}}}^{\left( {\small 8D}\right) }$ (\ref{den}). Since $SU\left(
N_{a}\right) $ is an exact gauge symmetry for the gauge theory engineered
from
\begin{equation}
\begin{tabular}{llll}
Y$_{2}$ & $\rightarrow $ & X$_{4}$ &  \\
&  & $\downarrow \pi $ &  \\
&  & $\mathcal{C}_{a}$ & ,%
\end{tabular}%
\end{equation}%
the vev of the scalar $\phi ^{\left( {\small 8D}\right) }$ in the adjoint
representation of the $SU\left( N_{a}\right) $ gauge symmetry should vanish,
i.e%
\begin{equation}
\left\langle \phi ^{\left( {\small 8D}\right) }\right\rangle =0,  \label{fi}
\end{equation}%
otherwise $SU\left( N_{a}\right) $ gauge invariance would be broken down to
a subsymmetry.\newline
Second, being $SU\left( N_{a}\right) $ matrices in the adjoint
representation, the \emph{8D} gauge fields can be expanded as follows:%
\begin{eqnarray}
\phi ^{\left( {\small 8D}\right) } &=&\sum_{\alpha \in \Delta }E_{\pm \alpha
}\phi ^{\pm \alpha }+\sum_{I=1}^{N_{a}-1}H_{I}\phi ^{I},  \notag \\
\mathcal{A}_{{\small M}}^{\left( {\small 8D}\right) } &=&\sum_{\alpha \in
\Delta }E_{\pm \alpha }\mathcal{A}_{{\small M}}^{\pm \alpha
}+\sum_{I=1}^{N_{a}-1}H_{I}\mathcal{A}_{{\small M}}^{I},  \label{if} \\
\lambda _{\hat{a}}^{\left( {\small 8D}\right) } &=&\sum_{\alpha \in \Delta
}E_{\pm \alpha }\mathcal{\lambda }_{\hat{a}}^{\pm \alpha
}+\sum_{I=1}^{N_{a}-1}H_{I}\mathcal{\lambda }_{\hat{a}}^{I},  \notag
\end{eqnarray}%
where $\left\{ H_{I},E_{\pm \alpha }\right\} $ is the Cartan Weyl basis of $%
su\left( N_{a}\right) $\textbf{\ }and\textbf{\ }$\Delta $ its root system.
Putting the expansion of $\phi ^{\left( {\small 8D}\right) }$ back into (\ref%
{if}) and setting $\left\langle \phi ^{\pm \alpha }\right\rangle =0$, we have%
\begin{equation}
\left\langle \phi ^{\left( {\small 8D}\right) }\right\rangle
=\sum_{I=1}^{N_{a}-1}H_{I}\left\langle \phi ^{I}\right\rangle .
\end{equation}%
Giving a non zero vev to some of the $\phi ^{I}$; say $\left\langle \phi
^{I}\right\rangle =t_{I}\neq 0$ with $I=1,...,N_{0}$, the gauge symmetry
gets broken down to $SU\left( N_{a}-N_{0}\right) \times U^{N_{0}}\left(
1\right) $. On the level of the geometry of the local Calabi-Yau 4- folds,
this breaking corresponds to performing a complex deformation of the
singularity which reduce the degree of the $A_{N_{a}-1}$ singularity (\ref%
{an}) down to
\begin{equation}
\frac{\mathrm{v}^{2}}{\vartheta }=\mathrm{u}^{2}+\left( z-t_{1}\right)
\left( z-t_{2}\right) ...\left( z-t_{N_{0}}\right) ^{N_{a}-N_{0}}\qquad ,
\end{equation}%
where $\vartheta =\vartheta \left( s_{1},s_{2}\right) $.\newline
Next, seen that the seven brane wraps the compact 4-cycle $\mathcal{C}_{a}$,
the above \emph{8D} gauge fields depend on the local coordinates $\left(
x^{0},x^{1},x^{2},x^{3};s_{1},s_{2},\bar{s}_{1},\bar{s}_{2}\right) $; that
is:
\begin{equation}
\begin{tabular}{llllll}
$\phi ^{\left( {\small 8D}\right) }$ & $=$ & $\phi ^{\left( {\small 8D}%
\right) }\left( x^{0},x^{1},x^{2},x^{3};s_{1},s_{2},\bar{s}_{1},\bar{s}%
_{2}\right) $ & $\equiv $ & $\phi ^{\left( {\small 8D}\right) }\left( x;s,%
\bar{s}\right) $ & , \\
$\mathcal{A}_{{\small M}}^{\left( {\small 8D}\right) }$ & $=$ & $\mathcal{A}%
_{{\small M}}^{\left( {\small 8D}\right) }\left(
x^{0},x^{1},x^{2},x^{3};s_{1},s_{2},\bar{s}_{1},\bar{s}_{2}\right) $ & $%
\equiv $ & $\mathcal{A}_{{\small M}}^{\left( {\small 8D}\right) }\left( x;s,%
\bar{s}\right) $ & , \\
$\lambda _{\hat{a}}^{\left( {\small 8D}\right) }$ & $=$ & $\lambda _{\hat{a}%
}^{\left( {\small 8D}\right) }\left( x^{0},x^{1},x^{2},x^{3};s_{1},s_{2},%
\bar{s}_{1},\bar{s}_{2}\right) $ & $\equiv $ & $\lambda _{\hat{a}}^{\left(
{\small 8D}\right) }\left( x;s,\bar{s}\right) $ & .%
\end{tabular}%
\end{equation}%
However, since $\mathcal{C}_{a}$ is a compact Kahler manifold\textrm{%
\footnote{%
The 4D fields are determined by the zero modes of the Dirac operator on the
complex base surface. The chiral and antichiral spectrum is determined by
the bundle valued cohomology groups $H_{\bar{\partial}}^{0}\left(
S,R^{v}\right) ^{v}$ $\oplus $ $H_{\bar{\partial}}^{1}\left( S,R\right) $ $%
\oplus $ $H_{\bar{\partial}}^{2}\left( S,R^{v}\right) ^{v}$ and $H_{\bar{%
\partial}}^{0}\left( S,R\right) $ $\oplus $ $H_{\bar{\partial}}^{1}\left(
S,R^{v}\right) ^{v}$ $\oplus $ $H_{\bar{\partial}}^{2}\left( S,R\right) $
where R is the vector bundle on the base surface whose sections transform in
the representation R of the structure group.}}; these fields may be expanded
in harmonic series in terms of the representations of the $U\left( 2\right) $
structure group of $T\mathcal{C}_{a}$. For the case of the scalar fields $%
\phi ^{\left( {\small 8D}\right) }$ and $\bar{\phi}^{\left( {\small 8D}%
\right) }$, we have a priori the following expansion\textrm{\footnote{%
BPS conditions \textrm{\cite{H1}} require furthermore that these expansions
to be holomorphic in the complex coordinates of the complex surface $%
\mathcal{C}$.}}
\begin{equation}
\begin{tabular}{llll}
$\phi ^{\left( {\small 8D}\right) }$ & $=$ & $\phi _{0}^{\left( {\small 4D}%
\right) }\left( x\right) +\sum_{n>0}\phi _{\left[ n\right] }^{\left( {\small %
4D}\right) }\left( x\right) R_{\left[ n\right] }$ & , \\
$\bar{\phi}^{\left( {\small 8D}\right) }$ & $=$ & $\bar{\phi}_{0}^{\left(
{\small 4D}\right) }\left( x\right) +\sum_{n>0}\bar{\phi}_{\left[ n\right]
}^{\left( {\small 4D}\right) }\left( x\right) \bar{R}_{\left[ n\right] }$ & ,%
\end{tabular}
\label{ex}
\end{equation}%
where the zero mode of the expansion,%
\begin{equation}
\phi _{0}^{\left( {\small 4D}\right) }\left( x\right) \equiv \phi \left(
x\right) \qquad ,\qquad \bar{\phi}_{0}^{\left( {\small 4D}\right) }\left(
x\right) \equiv \bar{\phi}\left( x\right) ,
\end{equation}%
stand for the \emph{4D} scalar fields and $\phi _{\left[ n\right] }^{\left(
{\small 4D}\right) }\left( x\right) $ and $\bar{\phi}_{\left[ n\right]
}^{\left( {\small 4D}\right) }$ for the non zero modes associated with the
non trivial $U\left( 2\right) $ representations $R_{\left[ n\right] }=R_{%
\left[ n\right] }\left[ s,\bar{s}\right] $ and $R_{\left[ -n\right] }=\bar{R}%
_{\left[ n\right] }\left[ s,\bar{s}\right] $.\newline
Moreover, following \textrm{\cite{H1,H2}} the BPS conditions require the
field to be holomorphic on the $S$ so that the representations\ $R_{\left[ n%
\right] }$ are holomorphic and may be taken as,%
\begin{equation}
R_{\left[ n\right] }=\dsum_{k=-n}^{n}s_{1}^{n-k}s_{2}^{k}.
\end{equation}%
Similarly, we have for the $8D$ vector gauge field,
\begin{equation}
\mathcal{A}_{{\small M}}^{\left( {\small 8D}\right) }=\left( \mathcal{A}_{%
{\small \mu }}^{\left( {\small 8D}\right) },\mathcal{A}_{{\small i}}^{\left(
{\small 8D}\right) },\mathcal{A}_{{\small \bar{\imath}}}^{\left( {\small 8D}%
\right) }\right) ,
\end{equation}%
the following mode expansion
\begin{equation}
\begin{tabular}{ll}
$\mathcal{A}_{{\small \mu }}^{\left( {\small 8D}\right) }\left( x;s,\bar{s}%
\right) =\mathcal{A}_{{\small \mu }}\left( x\right) +\sum_{n>0}\left( R_{%
\left[ n\right] }\mathcal{A}_{{\small \mu }}^{\left[ n\right] }\left(
x\right) +\bar{R}_{\left[ n\right] }\mathcal{A}_{{\small \mu }}^{\left[ -n%
\right] }\left( x\right) \right) $ & ,%
\end{tabular}
\label{ep}
\end{equation}%
where the zero mode $\mathcal{A}_{{\small \mu }}\left( x\right) \equiv
\mathcal{A}_{{\small \mu }}^{\left[ 0\right] }\left( x\right) $ stands for
the massless \emph{4D} gauge field and $\mathcal{A}_{{\small \mu }}^{\left[
\pm n\right] }\left( x\right) $ for the higher modes. Analogous expansions
are valid as well for the four other components on the compact manifold
namely%
\begin{equation}
\begin{tabular}{llll}
$\mathcal{A}_{{\small i}}^{\left( {\small 8D}\right) }$ & $=$ & $\mathcal{A}%
_{{\small i}}\left( x\right) +\sum_{n>0}R_{\left[ n\right] }\mathcal{A}_{%
{\small i}}^{\left[ n\right] }\left( x\right) $ & , \\
$\mathcal{A}_{{\small \bar{\imath}}}^{\left( {\small 8D}\right) }$ & $=$ & $%
\mathcal{A}_{{\small \bar{\imath}}}\left( x\right) +\sum_{n>0}\bar{R}_{\left[
n\right] }\mathcal{A}_{{\small \bar{\imath}}}^{\left[ n\right] }\left(
x\right) $ & ,%
\end{tabular}
\label{pe}
\end{equation}%
where the zero modes $\mathcal{A}_{{\small i}}\left( x\right) $ and $%
\mathcal{A}_{{\small \bar{\imath}}}\left( x\right) $ are two $U\left(
2\right) $ doublets of 4D scalars while $\mathcal{A}_{{\small i}}^{\left[ n%
\right] }$ and $\mathcal{A}_{{\small \bar{\imath}}}^{\left[ n\right] }$
describe massive excitations.\newline
Regarding the \emph{8D}\ fermionic field $\lambda _{\hat{a}}^{\left( {\small %
8D}\right) }$, the reduction is a little bit more technical as it requires
splitting this $SO\left( 1,7\right) $ spinor in terms of representations of $%
SO\left( 1,3\right) \times U\left( 2\right) $. Let us treat this
decomposition separately as it is interesting as well for the reduction of
the sixteen original supersymmetries down to the four conserved supercharges
in $\mathcal{N}=1$ supersymmetric theory in \emph{4D} space time.

\subsubsection{Twisted gauge theory}

We begin by recalling that the $SO\left( 1,7\right) $ space time group of
the \emph{8D} flat space time $\mathbb{R}^{1,7}$ decomposes in the case of
the seven- brane wrapping a 4- cycle $\mathcal{C}_{a}$ in the Calabi-Yau 4-
folds like,
\begin{equation}
\begin{tabular}{llll}
$SO\left( 1,7\right) \times U_{R}\left( 1\right) $ & $\supset $ & $SO\left(
1,3\right) \times SO\left( 4\right) \times U_{R}\left( 1\right) $ & , \\
& $\supset $ & $SO\left( 1,3\right) \times U\left( 2\right) \times
U_{R}\left( 1\right) $ & ,%
\end{tabular}%
\end{equation}%
where $U\left( 2\right) =U_{J}\left( 1\right) \times SU\left( 2\right) $ is
just the structure group of the tangent bundle of $\mathcal{C}_{a}$ and
where $U_{R}\left( 1\right) $ is as in eq(\ref{ur}). \newline
To twist the gauge theory in the seven brane, we combine the $U_{R}\left(
1\right) $ charge and the $U_{J}\left( 1\right) $ as in eq(\ref{top}) and
then think about the compact symmetry group as%
\begin{equation}
\begin{tabular}{llll}
$U_{R}\left( 1\right) \times U\left( 2\right) $ & $=$ & $U_{R}\left(
1\right) \times U_{J}\left( 1\right) \times SU\left( 2\right) $ & , \\
& $\supset $ & $U_{T}\left( 1\right) \times SU\left( 2\right) =U_{T}\left(
2\right) $ & .%
\end{tabular}%
\end{equation}%
with $T=J+2R$ as in the relation (\ref{tr}). In other words, we have the
following chain of breakings of space time groups%
\begin{equation}
\begin{tabular}{llll}
$SO\left( 1,9\right) $ & $\supset $ & $SO\left( 1,7\right) \times
U_{R}\left( 1\right) $ & , \\
& $\supset $ & $SO\left( 1,3\right) \times U_{R}\left( 1\right) \times
SO\left( 4\right) $ & , \\
& $\supset $ & $SO\left( 1,3\right) \times U_{R}\left( 1\right) \times
U_{J}\left( 1\right) \times SU\left( 2\right) $ & , \\
& $\supset $ & $SO\left( 1,3\right) \times U_{T}\left( 1\right) \times
SU\left( 2\right) $ & .%
\end{tabular}%
\end{equation}%
The sixteen components of the $SO\left( 1,7\right) $ spinor decomposes in
terms of the representations of $SO\left( 1,3\right) \times SU\left(
2\right) \times U_{T}\left( 1\right) $ as follows:%
\begin{equation}
\begin{tabular}{lllllll}
$16$ & $=$ &  & $\left( 2,1\right) \otimes 1_{0}$ & $\oplus $ & $\left(
1,2\right) \otimes 1_{0}$ &  \\
&  & $\oplus $ & $\left( 1,2\right) \otimes 2_{-}$ & $\oplus $ & $\left(
1,2\right) \otimes 2_{+}$ &  \\
&  & $\oplus $ & $\left( 1,2\right) \otimes 1_{--}$ & $\oplus $ & $\left(
2,1\right) \otimes 1_{++}$ & .%
\end{tabular}%
\end{equation}%
Thus the gaugino $\lambda _{\hat{a}}^{\left( {\small 8D}\right) }$
decomposes into two $U\left( 2\right) $ singlets $\eta _{\alpha }$ and $\chi
_{\alpha \left[ mn\right] }$\ of \emph{4D} Weyl spinors as well as a doublet
$\bar{\psi}_{\dot{\alpha}m}$:
\begin{equation}
\begin{tabular}{llllllll}
$\left( 2,1\right) \otimes 1_{0}$ & $\equiv $ & $\eta _{\alpha }$ & , & $%
\left( 1,2\right) \otimes 1_{0}$ & $\equiv $ & $\bar{\eta}_{\dot{\alpha}}$ &
, \\
$\left( 1,2\right) \otimes 2_{-}$ & $\equiv $ & $\bar{\psi}_{\dot{\alpha}m}$
& , & $\left( 2,1\right) \otimes 2_{+}$ & $\equiv $ & $\psi _{\alpha \bar{m}%
} $ & , \\
$\left( 2,1\right) \otimes 1_{--}$ & $\equiv $ & $\chi _{\alpha \left[ mn%
\right] }$ & , & $\left( 2,1\right) \otimes 1_{++}$ & $\equiv $ & $\bar{\chi}%
_{\dot{\alpha}\left[ \bar{m}\bar{n}\right] }$ & .%
\end{tabular}%
\end{equation}%
Each of these \emph{4D} Weyl spinor fields has a harmonic expansion\textrm{%
\footnote{%
Here also BPS conditions requires holomorphic/antiholomorphic fields on the
complex surface.}} as in (\ref{ex},\ref{ep}) and combine with the bosonic
fields (\ref{ex},\ref{ep},\ref{pe}) to form $\mathcal{N}=1$ supermultiplets
in \emph{4D} space time. The bosonic modes $\mathcal{\phi }^{\left[ \pm n%
\right] },$ $\mathcal{A}_{{\small \mu }}^{\left[ n\right] }$, $\mathcal{A}_{%
{\small i}}^{\left[ n\right] }$, $\mathcal{A}_{{\small \bar{\imath}}}^{\left[
-n\right] }$ and the fermionic ones $\eta _{\alpha }^{\left[ n\right] }$, $%
\bar{\psi}_{\dot{\alpha}m}^{\left[ n\right] }$, $\chi _{\alpha \left[ ij%
\right] }^{\left[ n\right] }=\varepsilon _{ij}\chi _{\alpha }^{\left[ n%
\right] }$ together with their complex conjugates combine to form $\mathcal{N%
}=1$ supermultiplets valued in the $su\left( N_{a}\right) $ Lie algebra. For
the zero modes, we have

\begin{equation}
\begin{tabular}{llll}
gauge multiplets & : & $V=\left( \mathcal{A}_{{\small \mu }},\eta _{\alpha },%
\bar{\eta}_{\dot{\alpha}}\right) $ & , \\
&  &  &  \\
chiral matter multiplets & : & $\left\{
\begin{tabular}{lll}
$\Phi _{ij}^{--}$ & $=$ & $\varepsilon _{ij}\left( \mathcal{\phi }^{--},\chi
_{\alpha }^{--}\right) $ \\
$\Upsilon _{\bar{\imath}}^{+}$ & $=$ & $\left( \mathcal{A}_{\bar{\imath}%
}^{+},\psi _{\alpha \bar{\imath}}^{+}\right) $%
\end{tabular}%
\right. $ & ,%
\end{tabular}
\label{adm}
\end{equation}%
where the upper charges refer to the $U_{T}\left( 1\right) $ twisted charge $%
T=J+2R$. Similar superfield relations are valid for each excitation level.

\section{Engineering F-theory GUT model}

In the engineering of supersymmetric \emph{GUT} models in the framework of
F-theory compactification on local CY4-folds, one has to specify, amongst
others, the base surface $S$. A priori, one may imagine several kinds of
compact complex surfaces by considering hypersurfaces in higher dimensional
complex Kahler manifolds. Typical examples of compact complex surfaces $S$
which have been considered in F-Theory GUT literature are given by the del
Pezzo\emph{\ }surfaces dP$_{n}$\emph{\ }with $n=0,1,...,8$ obtained by
preforming up to eight blow ups in the projective plane $\mathbb{P}^{2}$
\textrm{\cite{DP, H1, DQ, DR}}.\newline
Later on we develop a class of models based on toric manifold involving the
complex tetrahedral surface of \textrm{figure (\ref{by})} and its blown ups
\textrm{\cite{SA}}. But before that, we want to discuss here the \emph{dP}$%
_{n}$ based GUT model; as a front matter towards the study of the local
tetrahedron model. \newline
We take this opportuinity to study a realization of $SU\left( 5\right) $ GUT
model by using five intersecting 7-branes wrapping 4- cycles in the del
Pezzo $dP_{8}$ as illustrated by figure (\ref{7}).

\subsection{Del Pezzo surfaces $dP_{k}$}

Here, we give some useful tools on del Pezzo surfaces; these are needed for
the engineering of the corresponding $SU\left( 5\right) $ GUT model based on
$dP_{k}$ with $5\leq k\leq 8$.

\subsubsection{2- cycle homology of $dP_{k}$}

The $dP_{k}$ del Pezzo surfaces with $k\leq 8$ are defined as blow ups of
the complex projective space $\mathbb{P}^{2}$ at $k$ points. Taking into
account the overall size $r_{0}$ of the $\mathbb{P}^{2}$, a surface $dP_{k}$
has then real $\left( k+1\right) $ dimensional Kahler moduli,
\begin{equation}
\begin{tabular}{llllllll}
$r_{0}$ & , & $r_{1}$ & , & $\ldots $ & , & $r_{k}$ & ,%
\end{tabular}%
\end{equation}%
corresponding to the volume of each blown up cycle \textrm{\cite{H1, DR, SA}}%
. The $dP_{k}$s possess as well a moduli space of complex structures with
complex dimension $\left( 2k-8\right) $ where the eight gauge fixed
parameters are associated with the $GL\left( 3\right) $ symmetry of $\mathbb{%
P}^{2}$. As such, only surfaces with $5\leq k\leq 8$ admit a moduli space of
complex structures.\newline
The real 2-cycle homology group $\mathbb{H}_{2}\left( dP_{k},Z\right) $ is $%
\left( k+1\right) $ dimensional and is generated by $\left\{
H,E_{1},...,E_{k}\right\} $ where $H$ denotes the hyperplane class inherited
from $\mathbb{P}^{2}$ and the $E_{i}$ denote the exceptional divisors
associated with the blow ups. These generators have the intersection pairing
\begin{equation}
\begin{tabular}{llllll}
$H^{2}=1$ & , & $H.E_{i}=0$ & , & $E_{i}.E_{j}=-\delta _{ij}\quad ,\quad
i,j=1,...,k$ & ,%
\end{tabular}
\label{he}
\end{equation}%
so that the signature $\eta $ of the $\mathbb{H}_{2}\left( dP_{k},Z\right) $
group is given by ${\small diag}\left( +-...-\right) $.\newline
The first three blow ups giving $dP_{1},$ $dP_{2}$ and $dP_{3}$ complex
surfaces are of toric types while the remaining five others namely $%
dP_{4},...,$ $dP_{8}$ are non toric. These projective surfaces have the
typical toric fibration
\begin{equation*}
dP_{k}=\mathbb{T}^{2}\times \emph{B}_{2}^{\left( k\right) }\qquad ,\qquad
k=1,2,3,
\end{equation*}%
with real base $\emph{B}_{2}^{\left( k\right) }$ nicely represented by toric
diagrams $\Delta _{2}^{\left( k\right) }$ encoding the toric data of the
fibration%
\begin{equation}
\begin{tabular}{lll|ll|ll|ll}
{\small surface S} & {\small :} & ${\small dP}_{0}{\small =P}^{2}$ &  & $%
{\small dP}_{1}$ &  & ${\small dP}_{2}$ &  & ${\small dP}_{3}$ \\
\hline\hline
{\small blow ups} & {\small :} & ${\small k=0}$ &  & ${\small k=1}$ &  & $%
{\small k=2}$ &  & ${\small k=3}$ \\
{\small toric graph }$\Delta _{2}^{\left( k\right) }$ & {\small :} & {\small %
triangle} &  & {\small square} &  & {\small pentagon} &  & {\small hexagon}
\\
{\small generators} & {\small :} & ${\small H}$ &  & ${\small H}$ ${\small ,}
$ ${\small E}_{1}$ &  & ${\small H}$ ${\small ,}$ ${\small E}_{1}$ ${\small ,%
}$ ${\small E}_{2}$ &  & ${\small H}$ ${\small ,}$ ${\small E}_{1}$ ${\small %
,}$ ${\small E}_{2}$ ${\small ,}$ ${\small E}_{2}$ \\ \hline
\end{tabular}%
\end{equation}%
In terms of these basic classes of curves, one defines all the tools needed
for the present study; in particular the three following:\newline
\textbf{(1)} the generic classes $\left[ \Sigma _{a}\right] $ of holomorphic
curves in $dP_{k}$ given by the following linear combinations,
\begin{equation}
\Sigma _{a}=n_{a}H-\sum_{i=1}^{k}m_{ai}E_{i},  \label{ca}
\end{equation}%
with $n_{a}$ and $m_{a}$ are integers. The self- intersection numbers $%
\Sigma _{a}^{2}\equiv \Sigma _{a}\cdot \Sigma _{a}$ following from eqs(\ref%
{ca}) and (\ref{he}) are then given by
\begin{equation}
\Sigma _{a}^{2}=n_{a}^{2}-\sum_{i=1}^{k}m_{ai}^{2}.  \label{c}
\end{equation}%
\textbf{(2)} The canonical class $\Omega _{k}$ of the projective $dP_{k}$
surface, which is given by \emph{minus} the first Chern class $c_{1}\left(
dP_{k}\right) $ of the tangent bundle, reads as,
\begin{equation}
\Omega _{k}=-\left( 3H-\sum_{i=1}^{k}E_{i}\right) ,
\end{equation}%
and has a self intersection number $\Omega _{k}^{2}=9-k$ whose positivity
requires $k<9$. Obviously $k=0$ corresponds just to the case where there is
no blow up; i.e $dP_{0}=\mathbb{P}^{2}$ the complex projective plane.\
\newline
\textbf{(3)} the degree $d_{\Sigma }$ of a generic complex curve class $%
\Sigma =nH-\sum_{i=1}^{k}m_{i}E_{i}$ in $dP_{k}$ is given by the
intersection number between the class $\Sigma $ with the anticanonical class
$\left( -\Omega _{k}\right) $,
\begin{equation}
d_{\Sigma }=-\text{ }\left( \Sigma \cdot \Omega _{k}\right)
=3n-\sum_{i=1}^{k}m_{i}.
\end{equation}%
Positivity of this integer $d_{\Sigma }$ puts a constraint equation on the
allowed values of the $n$ and \ $m_{i}$ integers which should be like,%
\begin{equation}
\sum_{i=1}^{k}m_{i}\leq 3n.  \label{ccs}
\end{equation}%
Notice that there is a remarkable relation between the self intersection
number $\Sigma ^{2}$ (\ref{c}) of the classes of holomorphic curves and
their degrees $d_{\Sigma }$. This relation, which is known as the \emph{%
adjunction formula}, is given by
\begin{equation}
\Sigma ^{2}=2g-2+d_{\Sigma },
\end{equation}%
and allows to define the genus $g$ of the curve class $\Sigma $ as
\begin{equation}
g=1+\frac{n\left( n-3\right) }{2}-\sum_{i=1}^{k}\frac{m_{i}\left(
m_{i}-1\right) }{2}.
\end{equation}%
For instance, taking $\Sigma =3H$; that is $n=3$ and $m_{i}=0$, then the
genus $g_{3H}$ of this curve is equal to $1$ and so the curve $3H$ is in the
same class of the real 2- torus. In general, fixing the genus $g$ to a given
positive integer puts then a second constraint equation on $n$ and $m_{i}$
integers; the first constraint is as in (\ref{ccs}). For the example of
rational curves with $g=0$, we have
\begin{equation}
\Sigma ^{2}=d_{\Sigma }-2
\end{equation}%
giving a relation between the degree $d_{\Sigma }$ of the curve $\Sigma $
and its self intersection. For $d_{\Sigma }=0$, we have a rational curve
with self intersection $\Sigma ^{2}=-2$ while for $d_{\Sigma }=1$ we have a
self intersection $\Sigma ^{2}=-1$. To get the general expression of genus $%
g=0$ curves, one has to solve the constraint equation
\begin{equation}
\sum_{i=1}^{k}m_{i}\left( m_{i}-1\right) =2+n\left( n-3\right) ,
\end{equation}%
by taking into account the condition (\ref{ccs}). For $k=1$, this relation
reduces to $m\left( m-1\right) =2+n\left( n-3\right) $, its leading
solutions $n=1,$ $m=0$ and $n=0,$ $m=-1$ give just the classes $H$ and $E$
respectively with degrees $d_{H}=3$ and $d_{E}=1$. Typical solutions for
this constraint equation are given by the generic class $\Sigma
_{n,n-1}=nH-\left( n-1\right) E$ which is more convenient to rewrite it as
follows $\Sigma _{n,n-1}=H+\left( n-1\right) \left( H-E\right) $.

\subsubsection{Link with exceptional Lie algebras}

Del Pezzo surfaces $dP_{k}$ have also a remarkable link with the exceptional
Lie algebras. Decomposing the $\mathbb{H}_{2}$ homology group as,
\begin{equation}
\begin{tabular}{llll}
$\mathbb{H}_{2}\left( dP_{k},Z\right) _{k\geq 3}$ & $=$ & $\left\langle
\Omega _{k}\right\rangle \oplus \mathcal{L}_{k}$ & , \\
$\Omega _{k}$ & $=$ & $-3H+E_{i}+\cdots +E_{k}$ & , \\
$\mathcal{L}_{k}$ & $=$ & $\left\langle \Omega _{k}\right\rangle ^{\bot }$ &
,%
\end{tabular}
\label{h}
\end{equation}%
the sublattice $\mathcal{L}_{k}=\left\langle \alpha _{1},...,\alpha
_{k}\right\rangle $, orthogonal to\ $\Omega _{k}$, is identified with the
root space of the corresponding Lie algebra $E_{k}$. The generators $\alpha
_{i}$ of the lattice $\mathcal{L}_{k}$ are:
\begin{equation}
\begin{tabular}{ll}
$\alpha _{{\small 1}}=E_{{\small 1}}-E_{{\small 2}}$ & , \\
$\ \ \ \ \ \ \ \ \ \ \ \ \ \vdots $ &  \\
$\alpha _{{\small k-1}}=E_{{\small k-1}}-E_{{\small k}}$ & , \\
$\alpha _{{\small k}}=H-E_{1}-E_{2}-E_{3}$ & ,%
\end{tabular}
\label{ha}
\end{equation}%
with product $\alpha _{i}.\alpha _{j}$ equal to minus the Cartan matrix $%
C_{ij}\left( E_{k}\right) $ of the Lie algebra\textrm{\footnote{%
Here E$_{3}$, E$_{4}$, and E$_{5}$ denote respectively $SU\left( 3\right)
\times SU\left( 2\right) $, SU$\left( 5\right) $ and SO$\left( 10\right) .$}}
E$_{k}$. For the particular case of $dP_{2}$, the corresponding Lie algebra
is $su\left( 2\right) $. The mapping between the exceptional curves and the
roots of the exceptional Lie algebras is given in the following table%
\begin{equation}
\text{%
\begin{tabular}{ll|ll}
dP$_{k}$ surfaces & exceptional curves & Lie algebras & simple roots \\
\hline\hline
$dP_{1}$ & \multicolumn{1}{|l|}{${\small E}_{1}$} & - & \multicolumn{1}{|l}{-
} \\
$dP_{2}$ & \multicolumn{1}{|l|}{${\small E}_{1},${\small \ }${\small E}_{2}$}
& $su\left( 2\right) $ & \multicolumn{1}{|l}{${\small \alpha }_{{\small 1}}$}
\\
$dP_{3}$ & \multicolumn{1}{|l|}{${\small E}_{1},${\small \ }${\small E}_{2},$%
{\small \ }${\small E}_{3}$} & $su\left( 3\right) \times su\left( 2\right) $
& \multicolumn{1}{|l}{${\small \alpha }_{{\small 1}},${\small \ }${\small %
\alpha }_{{\small 2}},${\small \ }${\small \alpha }_{{\small 3}}$} \\
$dP_{4}$ & \multicolumn{1}{|l|}{${\small E}_{1},${\small \ }${\small E}_{2},$%
{\small \ }${\small E}_{3},{\small E}_{4}$} & $su\left( 5\right) $ &
\multicolumn{1}{|l}{${\small \alpha }_{{\small 1}},${\small \ }${\small %
\alpha }_{{\small 2}},${\small \ }${\small \alpha }_{{\small 3}},$ ${\small %
\alpha }_{{\small 4}}$} \\
$dP_{5}$ & \multicolumn{1}{|l|}{${\small E}_{1},${\small \ }${\small E}_{2},$%
{\small \ }${\small E}_{3},$ ${\small E}_{4},${\small \ }${\small E}_{5}$} &
$so\left( 10\right) $ & \multicolumn{1}{|l}{${\small \alpha }_{{\small 1}},$%
{\small \ }${\small \alpha }_{{\small 2}},${\small \ }${\small \alpha }_{%
{\small 3}},${\small \ }${\small \alpha }_{{\small 4}},{\small \alpha }_{%
{\small 5}}$} \\
$dP_{6},dP_{7},dP_{8}$ & \multicolumn{1}{|l|}{${\small E}_{1},{\small \ E}%
_{2},{\small \ ...},$ ${\small E}_{k}$} & $E_{6},E_{7},E_{8}$ &
\multicolumn{1}{|l}{${\small \alpha }_{{\small 1}},${\small \ ..., }${\small %
\alpha }_{{\small k}},$ ${\small k=6,7,8}$} \\ \hline
\end{tabular}%
}  \label{exp}
\end{equation}%
Notice that one can also use eqs(\ref{h},\ref{ha}) to express the generators
H and $\left\langle E_{i}\right\rangle _{1\leq i\leq k}$ in terms of the
anticanonical class $\Omega _{k}$ and the roots of the exceptional Lie
algebra. For the case of the del Pezzo $dP_{5}$, we have the following
useful relations%
\begin{equation}
\begin{tabular}{llll}
$\left(
\begin{array}{c}
H \\
E_{1} \\
E_{2} \\
E_{3} \\
E_{4} \\
E_{5}%
\end{array}%
\right) $ & $=$ & $\frac{-1}{4}\left(
\begin{array}{cccccc}
3 & 2 & 4 & 6 & 3 & 5 \\
1 & -2 & 0 & 2 & 1 & 3 \\
1 & 2 & 0 & 2 & 1 & 3 \\
1 & 2 & 4 & 2 & 1 & 3 \\
1 & 2 & 4 & 6 & 1 & 3 \\
1 & 2 & 4 & 6 & 5 & 3%
\end{array}%
\right) \left(
\begin{array}{c}
\Omega _{5} \\
\alpha _{1} \\
\alpha _{2} \\
\alpha _{3} \\
\alpha _{4} \\
\alpha _{5}%
\end{array}%
\right) $ & ,%
\end{tabular}%
\end{equation}%
from which we read the following classes of 2- cycles curves:
\begin{equation}
\begin{tabular}{llll}
$H$ & $=$ & $\frac{-1}{4}\left( 3\Omega _{5}+2\alpha _{1}+4\alpha
_{2}+6\alpha _{3}+3\alpha _{4}+5\alpha _{5}\right) $ & , \\
$H-E_{1}-E_{3}$ & $=$ & $\frac{-1}{4}\left( \Omega _{5}+2\alpha _{1}+4\alpha
_{2}+2\alpha _{3}+2\alpha _{4}-\alpha _{5}\right) $ & , \\
$2H-E_{1}-E_{5}$ & $=$ & $-$ $\Omega _{5}-\alpha _{1}-\alpha _{2}-\alpha
_{3}-\alpha _{5}$ & .%
\end{tabular}%
\end{equation}

\subsection{GUT model based on $dP_{8}$}

\begin{figure}[tbph]
\begin{center}
\hspace{0cm} \includegraphics[width=10cm]{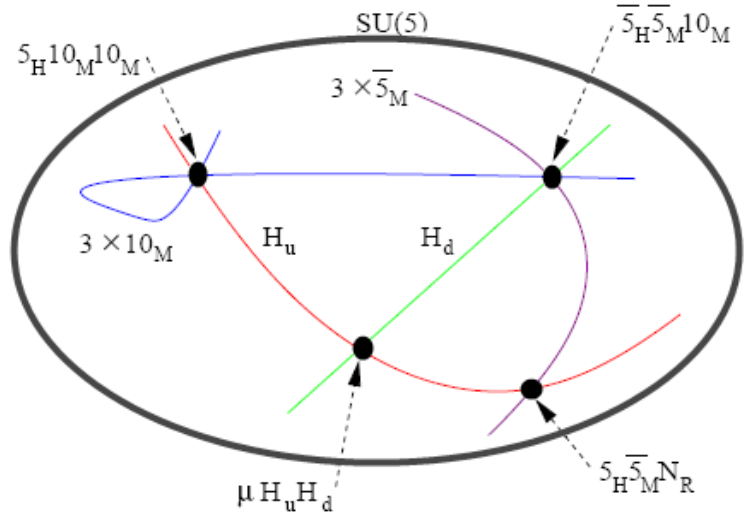}
\end{center}
\par
\vspace{-0.5 cm}
\caption{This figure is taken from ref \protect\cite{H2}: It represents the
various matter curves and Higgs ones in the $SU\left( 5\right) $ GUT model
based on del Pezzo surface $dP_{5}$. 4D Yukawa couplings live at the
intersection of the curves.}
\label{M2}
\end{figure}

In \textrm{\cite{H1, H2}, a }semi-realistic supersymmetric F- theory GUT
model based on\ del Pezzo surfaces $dP_{k}$ surfaces, $k\geq 5$, has been
constructed. The bulk gauge symmetry in the F- theory GUT model is broken
down to $SU_{C}\left( 3\right) \times SU_{L}\left( 2\right) \times
U_{Y}\left( 1\right) $ via an internal hypercharge flux in one to one
correspondence with the roots of underlying exceptional Lie algebras (\ref%
{exp}). Following these seminal works, the chiral matter of the \emph{MSSM}
localize on complex curves $\Sigma _{M}$ in the base surface $S$ of the CY4-
folds while Yukawa couplings localize at specific points $\mathcal{P}%
_{\gamma }$ in $S$. On the matter curves $\Sigma _{M}$, the bulk gauge
invariance G$_{r}$ gets enhanced to a rank r+1 symmetry G$_{r+1}$ while at
the points $P_{\gamma }$ it gets enhanced to a rank $\left( r+2\right) $
invariance $G_{r+2}$. A typical example is given by the figure (\ref{M2})

\ \ \ \newline
In this subsection, we use this example to develop an explicit realisation
of seven brane wrapping cycles of the BHV theory for the case of the $%
SU\left( 5\right) $ GUT model based on del Pezzo $dP_{8}$.

\subsubsection{BHV- $SU\left( 5\right) $ GUT model versus seven branes}

\begin{figure}[tbph]
\begin{center}
\hspace{0cm} \includegraphics[width=10cm]{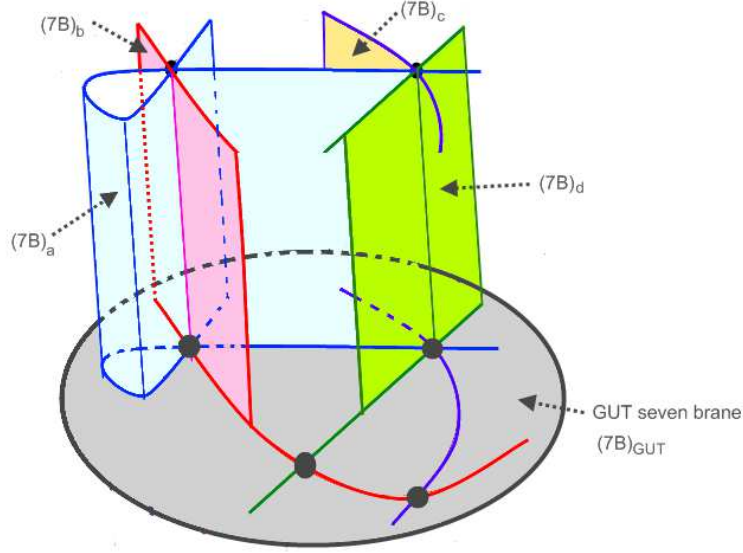}
\end{center}
\par
\vspace{-0.5 cm}
\caption{Brane representation of $SU\left( 5\right) $ GUT model. The
horizontal brane is the GUT brane; it intersects four other branes along
matter curves describing chiral matter.}
\label{7}
\end{figure}

In the $SU\left( 5\right) $ model, the chiral matter and Higgs superfields
as well as their Yukawa couplings localize on different curves in the base
of the local Calabi-Yau 4- folds. Matter and Higgs superfields in the $5$
and $\bar{5}$ representations of $SU\left( 5\right) $ localize on complex
curves $\Sigma _{i}^{\left( 5\right) }$ and $\Sigma _{i}^{\left( \bar{5}%
\right) }$ where the bulk $SU(5)$ singularity enhances to $SU(6)$ while
those in the $10$ and $\overline{10}$ representations localize on curves $%
\Sigma _{M}^{\left( 10\right) }$ and $\Sigma _{M}^{\left( \overline{10}%
\right) }$ where the bulk $SU(5)$ gets enhanced to $SO(10)$. Yukawa
couplings localize at four isolated points
\begin{equation}
\begin{tabular}{llll}
$P_{1},$ & $P_{2},$ & $P_{3},$ & $P_{4},$%
\end{tabular}%
\end{equation}%
in the base where the gauge symmetry gets enhanced either to $SU\left(
7\right) $, or $SO\left( 12\right) $ or $E_{6}$.\newline
To engineer the above typical $SU\left( 5\right) $ GUT model within the
framework of the BHV theory by using intersecting seven branes, we propose
the following:\newline
(\textbf{1}) We consider F- theory compactified on the local Calabi Yau
four- folds along the lines of BHV approach,%
\begin{equation}
\begin{tabular}{lll}
Y & $\rightarrow $ & X$_{4}$ \\
&  & $\downarrow \pi _{8}$ \\
&  & $dP_{8}$%
\end{tabular}
\label{P5}
\end{equation}%
with Kodaira type degenerating fiber Y \textrm{\cite{H1,H5}}.\newline
(\textbf{2}) We assume moreover that there are several singularities in the
fiber Y with different degeneracy types\ and different loci in $dP_{8}$. At
these loci live stacks of seven branes wrapping del Pezzo surfaces. These
seven brane stacks are as follows:\newline
(\textbf{a}) A bulk seven brane wrapping $dP_{4}\subset dP_{8}$ where the
fiber Y has an $SU\left( 5\right) $ singularity. We refer to this bulk seven
brane like $\left( 7B_{GUT}\right) _{SU\left( 5\right) }\equiv \left(
7B_{GUT}\right) _{5}$; it is given by the horizontal 7- brane depicted in
the figure (\ref{7})

\ \ \newline
(\textbf{b}) Together with this GUT seven brane, we have four more seven
branes intersecting the GUT brane along curves as shown on the figure (\ref%
{7}). \newline
To engineer these seven branes, we use the fact that $dP_{8}$ may be
obtained from the surface $dP_{4}$ by performing up to four more blow ups at
\emph{generic} points in $dP_{4}$. These blow ups generated by the
exceptional curves,
\begin{equation}
\begin{tabular}{llllllll}
$E_{5}$ & $,$ & $E_{7}$ & $,$ & $E_{8}$ & $,$ & $E_{9}$ & $,$%
\end{tabular}%
\end{equation}
together with the complex curves $\Sigma _{i}^{\left( 5\right) }$, $\Sigma
_{i}^{\left( \bar{5}\right) }$ and $\Sigma _{M}^{\left( 10\right) }$ and $%
\Sigma _{M}^{\left( \overline{10}\right) }$ of the figure (\ref{7}) allow to
determine the wrapping properties of the seven branes. We have:\newline
(\textbf{i}) a first seven brane wrapping the complex surface blown up of
the curve,%
\begin{equation}
\begin{tabular}{llll}
$E_{5}$ & $\rightarrow $ & $C_{a}$ &  \\
&  & $\downarrow \pi _{a}$ &  \\
&  & ${\small \Sigma }_{M}^{\left( 1\right) }$ &
\end{tabular}%
\end{equation}%
with base ${\small \Sigma }_{M}^{\left( 1\right) }$ given by the following
matter curve\textrm{\footnote{%
Notice that in \textrm{\cite{H2}}, the curve ${\small \Sigma }_{M}^{\left(
1\right) }$\ has been taken as $2H-E_{1}-E_{5}$. By performing the change $%
E_{4}\leftrightarrow E_{5}$, we get the same result.}} in $dP_{4}$,%
\begin{equation}
{\small \Sigma }_{M}^{\left( 1\right) }=2H-E_{1}-E_{4},
\end{equation}%
and where the fiber Y has a type I$_{1}$ geometry on $C_{a}$. On this seven
brane lives a Maxwell gauge supermultiplet with $U_{a}\left( 1\right) $
gauge invariance. We will refer below to this seven brane as $\left(
7B\right) _{a}$; see also the figure (\ref{7}). The non compact direction of
the $\left( 7B\right) _{a}$ brane fill the 4D space time while the four
compact ones wraps $C_{a}$.\newline
(\textbf{ii}) a second seven brane $\left( 7B\right) _{b}$ wrapping the
local 4- cycle
\begin{equation}
\begin{tabular}{llll}
$\left( E_{5}-nE_{6}\right) $ & $\rightarrow $ & $C_{b}$ &  \\
&  & $\downarrow \pi _{b}$ &  \\
&  & ${\small \Sigma }_{H_{d}}$ &
\end{tabular}%
\end{equation}%
with $n$ being an integer and the base ${\small \Sigma }_{H_{d}}$ same as
the BHV\ SU$\left( 5\right) $ model,
\begin{equation}
{\small \Sigma }_{H_{d}}=\left\langle H-E_{1}-E_{3}\right\rangle
\end{equation}%
and where Y has as well a type I$_{1}$ geometry.\newline
(\textbf{iii}) a third seven brane $\left( 7B\right) _{c}$ with a $%
U_{c}\left( 1\right) $ gauge symmetry wrapping%
\begin{equation}
\begin{tabular}{llll}
$\left( m_{1}E_{5}+m_{2}E_{6}-m_{3}E_{7}\right) $ & $\rightarrow $ & $C_{c}$
&  \\
&  & $\downarrow \pi _{c}$ &  \\
&  & ${\small \Sigma }_{H_{u}}$ &
\end{tabular}%
\end{equation}%
where ${\small \Sigma }_{H_{u}}=\left\langle H-E_{1}-E_{3}\right\rangle $
and where $m_{i}$ are integers which can be determined by solving the brane
intersection condition. \newline
(\textbf{iv}) a fourth seven brane $\left( 7B\right) _{d}$ with a$%
U_{d}\left( 1\right) $ gauge invariance wrapping%
\begin{equation*}
\begin{tabular}{llll}
$\left( k_{1}E_{5}+k_{2}E_{6}-k_{2}E_{7}-k_{3}E_{8}\right) $ & $\rightarrow $
& $C_{d}$ &  \\
&  & $\downarrow \pi _{d}$ &  \\
&  & ${\small \Sigma }_{M}^{\left( 2\right) }$ &
\end{tabular}%
\end{equation*}%
with ${\small \Sigma }_{M}^{\left( 2\right) }=H$ and where the $k_{i}$s are
integers.\newline
These branes intersect with the GUT branes along matter curves where the
gauge singularity gets enhanced either to $SU\left( 6\right) $ or $SO\left(
10\right) $. But, there are also branes intersections at four isolated
points $P_{\gamma }$ in the GUT branes as shown on the figure (\ref{7}). At
these points, the gauge symmetry gets enhanced to one of the following rank
six gauge groups,%
\begin{equation}
\begin{tabular}{llllll}
$SU\left( 7\right) $ & $,$ & $SO\left( 12\right) $ & $,$ & $E_{6}$ & ,%
\end{tabular}%
\end{equation}%
with the following typical breakings,%
\begin{equation}
\begin{tabular}{llllll}
$SU\left( 7\right) $ & $\rightarrow $ & $SU\left( 6\right) \times
U_{1}\left( 1\right) $ & $\rightarrow $ & $SU\left( 5\right) \times
U_{1}\left( 1\right) \times U_{2}\left( 1\right) $ & , \\
$SO\left( 12\right) $ & $\rightarrow $ & $SO\left( 10\right) \times
U_{1}^{\prime }\left( 1\right) $ & $\rightarrow $ & $SU\left( 5\right)
\times U_{1}^{\prime }\left( 1\right) \times U_{2}^{\prime }\left( 1\right) $
& , \\
$E_{6}$ & $\rightarrow $ & $SO\left( 10\right) \times U_{1}^{\prime \prime
}\left( 1\right) $ & $\rightarrow $ & $SU\left( 5\right) \times
U_{1}^{\prime \prime }\left( 1\right) \times U_{2}^{\prime \prime }\left(
1\right) $ & .%
\end{tabular}
\label{uu}
\end{equation}%
The decomposition of the adjoint representations of these groups namely the
\underline{$48$} of the $SU\left( 7\right) $ group, the \underline{$66$} of
the $SO\left( 12\right) $ symmetry and the \underline{$78$}\ for $E_{6}$,
give the bi- fundamental matters that localize on the curves $\Sigma
_{M}^{\left( 1\right) }$, $\Sigma _{M}^{\left( 2\right) }$ and $\Sigma
_{M}^{\left( 3\right) }$ for each group $G_{6}$. Below, we give some details
on the Yukawa tri-couplings that are invariant under these groups.\

\emph{Yukawa couplings at SU}$\left( 7\right) $ \emph{point}\newline
The $SU\left( 7\right) $ point is an isolated singular point in the surface $%
S$ where three matter curves $\Sigma _{1}$, $\Sigma _{2}$ and $\Sigma _{3}$\
meet. The geometric engineering of the $SU\left( 7\right) $ point in $S$ is
obtained by starting from a $SU\left( 7\right) $ singularity in the fiber Y
of the Calabi-Yau four-folds and switching on a $U_{1}\left( 1\right) \times
U_{2}\left( 1\right) $ bundle. The $U_{1}\left( 1\right) \times U_{2}\left(
1\right) $ fluxes give vevs to adjoint matter in the bulk theory
\begin{equation}
\left\langle \phi \right\rangle =t_{1}H_{1}+t_{2}H_{2}  \label{car}
\end{equation}%
with $H_{1}$ and $H_{2}$ being two Cartan generators of $SU\left( 7\right)
/SU\left( 5\right) $ and induces a geometric deformation in the fiber,
\begin{equation}
v^{2}=u^{2}+z^{5}\left( z-t_{1}\right) \left( z-t_{2}\right)  \label{5S}
\end{equation}%
where $t_{1}$ and $t_{2}$ are two complex moduli. This geometric deformation
induces as well a deformation in the base surface leading to rotation of the
branes.\newline
Notice that for $t_{1}=0$; but $t_{2}\neq 0$ and $t_{2}=0$; but $t_{1}\neq 0$
the $SU\left( 5\right) $ singularity (\ref{5S}) gets enhanced to $SU\left(
6\right) $ while for $t_{1}=t_{2}\neq 0$, it gets enhanced to $SU\left(
5\right) \times SU\left( 2\right) $. Notice also that for the particular
case $t_{1}=t_{2}=0$; that is when the $U_{1}\left( 1\right) \times
U_{2}\left( 1\right) $ fluxes are switched off; these singularities gets
further enhanced to the $SU\left( 7\right) $ singularity $v^{2}=u^{2}+z^{7}$.%
\newline
To get matters at brane intersections, we decompose the adjoint
representation $\underline{48}$ of $SU\left( 7\right) $\ in terms of
representations of $SU\left( 5\right) \times U_{1}\left( 1\right) \times
U_{2}\left( 1\right) $ namely\textrm{\footnote{%
To get the decomposition (\ref{add}), we have solved the traceless condition
of the fundamental representation of $SU\left( 7\right) $ in terms of $%
SU\left( 5\right) \times U^{2}\left( 1\right) $ as $%
7=5_{-1,-1}+1_{6,0}+1_{-1,5}$}}%
\begin{equation}
\begin{tabular}{llll}
$48$ & $=$ & $1_{0,0}$ $\oplus $ $1_{0,0}$ $\oplus $ $24_{0,0}$ &  \\
&  & $\oplus $ $\left( 5_{0,-6}\text{ }\oplus \text{ }\bar{5}_{0,6}\right) $
$\oplus $ $\left( 5_{-7,-1}\text{ }\oplus \text{ }\bar{5}_{7,1}\right) $ $%
\oplus $ $\left( 1_{7,-5}\text{ }\oplus \text{ }\bar{1}_{-7,5}\right) $ & .%
\end{tabular}
\label{add}
\end{equation}%
In addition to the usual uncharged adjoints, we have \ moreover the
following bi-fundamentals:\newline
(\textbf{a}) Four matter fields in the fundamental representations of $%
SU\left( 5\right) $: \newline
(\textbf{i}) two matter fields with charges $\left( 0,\mp 6\right) $; one in
the $5_{0,-6}$ and the other in the conjugate representation $\bar{5}_{0,6}$%
. Matter in these representations localize on the curve $\Sigma _{1}$
associated to $\pm 6t_{2}=0$. \newline
(\textbf{ii}) two more matter fields with charges $\pm \left( 7,1\right) $;
one in the $5_{-7,-1}$ representation and the other in the conjugate $\bar{5}%
_{7,1}$. They localize on the curve $\Sigma _{2}$ associated to $\pm \left(
7t_{1}+t_{2}\right) =0$.\newline
(\textbf{b}) Two $SU\left( 5\right) $ matter singlets with charges $\left(
7,-5\right) $ and $\left( -7,5\right) $ localizing on the curve $\Sigma _{3}$
associated to $\pm \left( 7t_{1}-5t_{2}\right) =0$.\newline
The $SU\left( 5\right) \times U_{1}\left( 1\right) \times U_{2}\left(
1\right) $ gauge invariant Yukawa tri- couplings is given by the following
fields overlapping:
\begin{equation}
\begin{tabular}{ll}
$5_{-7,-1}\otimes \bar{5}_{0,6}\otimes 1_{7,-5}$ & , \\
$5_{0,-6}\otimes \bar{5}_{7,1}\otimes \bar{1}_{-7,5}$ & .%
\end{tabular}
\label{y7}
\end{equation}%
Upon using the following fields identification
\begin{equation}
\begin{tabular}{llllll}
$5_{H_{u}}=5_{-7,-1}$ & , & $5_{M}=\bar{5}_{0,6}$ & , & $1_{X}=1_{7,-5}$ & ,%
\end{tabular}%
\end{equation}%
the three fields overlapping engineer the Yukawa coupling term $%
5_{H_{u}}\times 5_{M}\times 1_{X}$ originating then from points $P_{{\small %
SU}\left( 7\right) }$ in the base surface $S$ where the $SU(5)$ singularity
gets enhanced to a $SU\left( 7\right) $ singularity. \newline
This analysis extends directly to the $SO\left( 12\right) $ and $E_{6}$
gauge symmetries. Let us give some brief details.\ \

\emph{Yukawa couplings at }$SO\left( 12\right) $\emph{\ point}\newline
First, recall that the $SO\left( 12\right) $ singularity $%
v^{2}=u^{2}z+\alpha ^{2}z^{5}$ may be broken down to $SU(5)$ by using two
non zero vevs $t_{1}^{\prime }$ and $t_{2}^{\prime }$ like,%
\begin{equation}
\left\langle \phi ^{\prime }\right\rangle =t_{1}^{\prime }H_{1}^{\prime
}+t_{2}^{\prime }H_{2}^{\prime }
\end{equation}%
with $t_{1}^{\prime }$ and $t_{2}^{\prime }$ captured by two local Cartan $%
H_{1}^{\prime }$ and $H_{2}^{\prime }$ generators of $so\left( 12\right) $
Lie algebra. Under a one parameter deformation by $\left\langle \phi
^{\prime }\right\rangle =t_{1}^{\prime }H_{1}^{\prime }$ ($t_{2}^{\prime }=0$%
), we can either break the $SO\left( 12\right) $ singularity down to $%
SO\left( 10\right) $ or down to $SU\left( 5\right) \times SU\left( 2\right) $%
. By switching on the second deformation ($t_{2}^{\prime }\neq 0$), we can
break further the above singularity down to $SU(5)$ described by the
following relation%
\begin{equation}
v^{2}=\left( u-t_{1}^{\prime }\right) \left( u-t_{2}^{\prime }\right)
z+\alpha ^{2}z^{5}.
\end{equation}%
Under the $SO\left( 12\right) $ gauge symmetry breaking down to $SO\left(
10\right) \times U^{\prime }\left( 1\right) $, the adjoint representation
\underline{$66$} decomposes\textrm{\footnote{%
To get the decomposition of the adjoint of $SO\left( 12\right) $ in terms of
representations of $SO\left( 10\right) $ $\times $ $U\left( 1\right) $, we
have used the splitting $12$ $=$ $10_{0}$ $\oplus $ $1_{2}$ $\oplus $ $%
\overline{1}_{-2}$. To get the decomposition in terms of $SU\left( 5\right) $
$\times $ $U^{2}\left( 1\right) $ representations, we have used as well the
splitting $12$ $=$ $5_{0,2}$ $\oplus $ $\overline{5}_{0,-2}$ $\oplus $ $%
1_{2,0}$ $\oplus $ $\overline{1}_{-2,0}.$}} as $1_{0}+45_{0}+10_{2}+10_{-2}$
and by switching on the second flux, the $SO\left( 10\right) \times
U^{\prime }\left( 1\right) $ representation break further down to
representations of $SU\left( 5\right) \times U_{1}^{\prime }\left( 1\right)
\times U_{2}^{\prime }\left( 1\right) $ as given below,\emph{\ }%
\begin{equation}
\begin{tabular}{llll}
$66$ & $=$ & $1_{0,0}+1_{0,0}+24_{0,0}$ &  \\
&  & $\left( 5_{2,2}+\bar{5}_{-2,-2}\right) +\left( 5_{-2,2}+\bar{5}%
_{2,-2}\right) +10_{0,4}+\overline{10}_{0,-4}$ & .%
\end{tabular}
\label{66}
\end{equation}%
This decomposition involves two kinds of bi-fundamental matters. (\textbf{a}%
) Matter in $5_{2,2}$ and $5_{-2,2}$ representations which localize on the
curves in the $2\left( t_{1}^{\prime }\pm t_{2}^{\prime }\right) =0$ and (%
\textbf{b}) matter in the $10_{4,0}$ localizes on $\pm 4t_{1}^{\prime }=0$.%
\newline
The $SU\left( 5\right) \times U_{1}^{\prime }\left( 1\right) \times
U_{2}^{\prime }\left( 1\right) $ gauge invariant Yukawa couplings one can
write down by the combination of three matter fields is as follows:
\begin{equation}
\begin{tabular}{ll}
$\bar{5}_{+2,-2}\otimes \bar{5}_{-2,-2}\otimes 10_{0,+4}$ & , \\
$5_{-2,+2}\otimes 5_{+2,+2}\otimes \overline{10}_{0,-4}$ & .%
\end{tabular}
\label{y12}
\end{equation}%
Upon using the following fields identification
\begin{equation*}
\begin{tabular}{llllll}
$5_{H_{d}}=\bar{5}_{2,-2}$ & , & $5_{M}=\bar{5}_{-2,-2}$ & , & $%
10_{M}=10_{0,+4}$ & ,%
\end{tabular}%
\end{equation*}%
the three fields overlapping engineer the Yukawa coupling term $%
5_{H_{d}}\times 5_{M}\times 10_{M}$ originating then from points $P_{{\small %
SO}\left( 12\right) }$ in the base surface $S$ where the $SU(5)$ singularity
gets enhanced to a $SO\left( 12\right) $ singularity.

\emph{Yukawa couplings at }$E_{6}$ \emph{point}\newline
In the same manner, under the breaking of the $E_{6}$ gauge symmetry down to
$SO\left( 10\right) \times U\left( 1\right) $, the adjoint representation
\underline{$78$} decomposes as $1_{0}+45_{0}+16_{-3}+\overline{16}_{3}$ and
by a further breaking down to $SU\left( 5\right) \times U_{1}^{\prime \prime
}\left( 1\right) \times U_{2}^{\prime \prime }\left( 1\right) $ we get:%
\begin{equation}
\begin{tabular}{llll}
$78$ & $=$ & $1_{0,0}+1_{0,0}+24_{0,0}+$ &  \\
&  & $1_{5,3}+1_{-5,-3}+5_{-3,3}+\bar{5}_{3,-3}+10_{-1,-3}+\overline{10}%
_{1,3}+10_{4,0}+\overline{10}_{-4,0}$ & ,%
\end{tabular}%
\end{equation}%
where matter in the $5_{3,-3}$ and $\bar{5}_{3,-3}$ localizes on the curve $%
\left( t_{1}^{\prime \prime }-t_{2}^{\prime \prime }\right) =0$ and matter
in the $10_{-1,-3}$ and $10_{4,0}$ as well as their conjugates $\overline{10}%
_{1,3}$ and $\overline{10}_{-4,0}$ localize on the curves $\left(
t_{1}^{\prime \prime }+3t_{2}^{\prime \prime }\right) =0$ and $t_{1}^{\prime
\prime }=0$.\newline
The $SU\left( 5\right) \times U_{1}^{\prime \prime }\left( 1\right) \times
U_{2}^{\prime \prime }\left( 1\right) $ gauge invariant Yukawa couplings at
the E$_{6}$ point is given by the following three matter fields
interactions:
\begin{equation}
\begin{tabular}{lll}
& $5_{-3,3}\otimes 10_{-1,-3}\otimes 10_{4,0}$ & , \\
& $\bar{5}_{3,-3}\otimes \overline{10}_{1,3}\otimes \overline{10}_{-4,0}$ & .%
\end{tabular}
\label{en}
\end{equation}%
By using the fields identification
\begin{equation*}
\begin{tabular}{llllll}
$5_{H_{u}}=5_{-3,3}$ & , & $10_{M}=10_{-1,-3}$ & , & $10_{M}=10_{4,0}$ & ,%
\end{tabular}%
\end{equation*}%
the three overlapping (\ref{en}) engineer the Yukawa coupling term $%
5_{H_{u}}\times 5_{M}\times 10_{M}$ originating then from points in the base
surface S where the $SU(5)$ singularity gets enhanced to a $E_{6}$.

We end this study by giving more explicit expressions of the complex curves
on which matter localize. Following \textrm{\cite{H1,H2}} and using
fractional bundle idea, the configuration of the matter curves that engineer
a quasi-realistic F-theory $SU\left( 5\right) $ GUT model based on $dP_{8}$
are as follows:\newline
\textbf{(i)} the Higgs up $5_{H_{u}}$ and the Higgs down $\bar{5}_{H_{d}}$
are placed on two distinct matter curves $\Sigma _{H}^{\left( u\right) }$
and $\Sigma _{H}^{\left( d\right) }$ which intersect at a point in $dP_{8}$.%
\newline
\textbf{(ii)} the three generations of the fields in the$10_{M}$ are placed
on one self -intersecting $\mathbb{P}^{1}$ \newline
\textbf{(iii)} the three generations of the fields in the $\bar{5}_{M}$ are
placed on one smooth $\mathbb{P}^{1}$ which does not self-intersect.\newline
The matter content of this supersymmetric $SU\left( 5\right) $ model and the
corresponding fractional bundle assignments are collected in the following
table\textrm{\footnote{%
In \cite{H2}, this matter field configuration in terms of curves in dP$_{8}$
was named Model II.}, see also footnote 9}:

\begin{equation}
\begin{tabular}{l|l|l|l|l|l}
Model II\  & curve & class & $g_{\Sigma }$ & $\mathcal{L}_{\Sigma }$ & $%
\mathcal{L}_{\Sigma }^{m}$ \\ \hline\hline
${\small 1\times 5}_{H}$ & ${\small \Sigma }_{H}^{\left( u\right) }$ & $%
{\small H-E}_{1}{\small -E}_{3}$ & $0$ & $\mathcal{L}_{\Sigma _{H}^{\left(
u\right) }}^{1/5}\left( {\small 1}\right) $ & $\mathcal{L}_{\Sigma
_{H}^{\left( u\right) }}^{2/5}\left( {\small 1}\right) $ \\
${\small 1\times \bar{5}}_{H}$ & ${\small \Sigma }_{H}^{\left( d\right) }$ &
${\small H-E}_{1}{\small -E}_{3}$ & $0$ & $\mathcal{L}_{\Sigma _{H}^{\left(
d\right) }}^{1/5}\left( {\tiny -}{\small 1}\right) $ & $\mathcal{L}_{\Sigma
_{H}^{\left( d\right) }}^{2/5}\left( {\tiny -}{\small 1}\right) $ \\
${\small 3\times 10}_{M}$ & ${\small \Sigma }_{M}^{\left( 1\right) }$ & $%
{\small 2H-E}_{1}{\small -E}_{4}$ & $0$ & $\mathcal{L}_{\Sigma _{M}^{\left(
1\right) }}$ & $\mathcal{L}_{\Sigma _{M}^{\left( 1\right) }}\left( {\small 3}%
\right) $ \\
${\small 3\times \bar{5}}_{M}$ & ${\small \Sigma }_{M}^{\left( 2\right) }$ &
${\small H}$ & $0$ & $\mathcal{L}_{\Sigma _{M}^{\left( 2\right) }}$ & $%
\mathcal{L}_{\Sigma _{M}^{\left( 2\right) }}\left( {\small 3}\right) $ \\
\hline
\end{tabular}
\label{conf}
\end{equation}%
where $g_{\Sigma }$\ stands for the genus of the matter curves. The
geometrical figure representing the various matter curves in this $SU\left(
5\right) $ model are depicted in figure (\ref{M2}).\newline
The $\mathcal{N}=1$ chiral superpotential $W_{SU\left( 5\right) }$ capturing
the intersections of the various matter and Higgs curves is given by%
\begin{equation}
\begin{tabular}{lllll}
$W_{SU\left( 5\right) }=$ & $\sum_{i,j}\lambda _{ij}^{\left( d\right) }$ $\
\bar{5}_{H}\otimes \bar{5}_{M}^{\left( i\right) }\otimes 10_{M}^{\left(
j\right) }$ & $+$ & $\sum_{i,j}\lambda _{ij}^{\left( u\right) }$ $\
5_{H}\otimes 10_{M}^{\left( i\right) }\otimes 10_{M}^{\left( j\right) }$ &
\\
& $+$ $\sum_{i,a}\lambda _{ia}^{\left( u\right) }$ $\ 5_{H}\otimes \bar{5}%
_{M}^{\left( i\right) }\otimes N_{R}^{\left( a\right) }$ & $+$ & $\lambda
_{ud}^{\left( \phi \right) }$ $\ \Phi \otimes 5_{H}\otimes \bar{5}_{H}$ & ,%
\end{tabular}%
\end{equation}%
where the moduli $\lambda _{xy}^{\left( z\right) }$ stand for Yukawa
coupling constants. Notice that the interaction term $5_{H}\otimes \bar{5}%
_{M}^{\left( i\right) }\otimes N_{R}^{\left( a\right) }$ leads to a two-fold
enhancement in rank to an $SU\left( 7\right) $ singularity so that the
singlet $N_{R}^{\left( a\right) }$ may be identified with the right-handed
neutrinos. The interaction term $\Phi \otimes 5_{H}\otimes \bar{5}_{H}$ \
with vev $\left\langle \Phi \right\rangle $ determines the supersymmetric $%
\mu $- term \textrm{\cite{H2,H4}}.

\section{Quiver\ GUT models on tetrahedron}

In this section, we set up the basis for constructing a class of quiver GUT\
like models embedded in F- theory on CY4- folds by using the toric geometry
of the complex tetrahedral base surface. The key idea behind this
construction stems from thinking about the abelian gauge factors appearing
in eqs(\ref{uu}) as given by the toric symmetry
\begin{equation}
U\left( 1\right) \times U\left( 1\right)
\end{equation}
of the complex tetrahedral base surface $\mathcal{T}$. Denoting by $\left(
s_{1},s_{2}\right) $ the local holomorphic coordinates of $\mathcal{T}$,
this toric symmetry is given by,%
\begin{equation}
\begin{tabular}{llll}
$s_{1}$ & $\rightarrow $ & $e^{i\theta _{1}}s_{1}$ & , \\
$s_{2}$ & $\rightarrow $ & $e^{i\theta _{2}}s_{2}$ & ,%
\end{tabular}%
\end{equation}%
with $\theta _{1}$ and $\theta _{2}$ being the gauge parameters. This
abelian group action has degeneracy loci on the edges $\Sigma _{ab}$ and at
the vertices $P_{abc}$ of the tetrahedron (\ref{by}).\newline
In this section is organized into three parts, we study in the two first
ones the geometry of local Calabi-Yau four-folds based on $\mathcal{T}$ as a
matter to get a more insight of such a particular geometry. In the third
subsection, we construct three quiver $SU\left( 5\right) $ GUT- like models
embedded in F- theory on the tetrahedron based CY4s. GUT-like models
building using blown ups of tetrahedron will be considered in the next
section.

\subsection{4-cycles in CY4- folds}

We begin by recalling that real 4- cycles in Calabi-Yau 4- folds play an
important role in the engineering of F-theory GUT models. The seven brane
living at the elliptic singularity of the Calabi-Yau four folds has four non
compact directions filling the \emph{4D}\ Minkowski space time and four
compact directions that wrap compact real 4-cycles in the base of $X_{4}$.
Generally speaking, the Calabi-Yau 4- folds has an elliptic curve $\mathbb{E}
$ fibered on a complex three dimension base B$_{3}$,
\begin{equation}
\begin{tabular}{lll}
$\mathbb{E}$ & $\longrightarrow $ & $X_{4}$ \\
&  & $\downarrow \pi _{_{B}}$ \\
&  & $B_{3}$%
\end{tabular}%
\end{equation}%
but it is locally handled as a ADE\ geometry fibered on a complex surface $S$%
. Indeed, defining the elliptic fiber $\mathbb{E}$ by a cubic in the complex
plane with coordinates as usual like\textrm{\footnote{%
In the Weierstrass form of the elliptic curve, we have $d=1$ and $e=0$.}} $%
\mathrm{v}^{2}=d\mathrm{u}^{3}+e\mathrm{u}^{2}+f\mathrm{u}+g$ \textrm{, an
explicit }expression of $X_{4}$ is obtained by fibering the cubic on the
base $B_{3}$; i.e,%
\begin{equation}
\begin{tabular}{lll}
$\mathrm{v}^{2}=$ & $\ \ \ \mathcal{D}\left( {\small w}_{{\small 1}}{\small %
,w}_{{\small 2}}{\small ,w}_{{\small 3}}\right) \mathrm{u}^{3}$ $\ +$ $\
\mathcal{E}\left( {\small w}_{{\small 1}}{\small ,w}_{{\small 2}}{\small ,w}%
_{{\small 3}}\right) \mathrm{u}^{2}$ &  \\
& $+$ $\ \mathcal{F}\left( {\small w}_{{\small 1}}{\small ,w}_{{\small 2}}%
{\small ,w}_{{\small 3}}\right) \mathrm{u}$ $\ +$ $\ \mathcal{G}\left(
{\small w}_{{\small 1}}{\small ,w}_{{\small 2}}{\small ,w}_{{\small 3}%
}\right) $ & $.$%
\end{tabular}
\label{DE}
\end{equation}%
The complex variables $\left( {\small w}_{{\small 1}}{\small ,w}_{{\small 2}}%
{\small ,w}_{{\small 3}}\right) $ are local holomorphic coordinates
parameterizing the complex three dimension base $B_{3}$ while $\mathcal{D}%
\left( {\small w}\right) $, $\mathcal{E}\left( {\small w}\right) $, $%
\mathcal{F}\left( {\small w}\right) $ and $\mathcal{H}\left( {\small w}%
\right) $ are tri- holomorphic functions whose explicit expressions depend
on the type of the ADE singularity living in the CY4- folds.

\subsubsection{Factorization}

By breaking the $U\left( 3\right) $ group structure of the tangent bundle of
the complex three dimension base $TB_{3}$ down to the subgroup $U\left(
2\right) \times U\left( 1\right) $, we can locally split $TB_{3}$ like,%
\begin{equation}
TB_{3}\rightarrow TS\oplus \left( TS\right) ^{\perp },  \label{tb}
\end{equation}%
where $TS$ the tangent bundle of $S$ with group structure $U\left( 2\right) $
and where $\left( TS\right) ^{\perp }$ is the normal codimension one bundle
in $TB_{3}$. Under this decomposition, the fibration (\ref{DE}) can be
reduced down to the simple form%
\begin{equation}
\begin{tabular}{lll}
$\frac{\mathrm{v}^{2}}{\vartheta }=$ & $\ \ \ \mathrm{d}\left( z\right)
\mathrm{u}^{3}$ $\ +$ $\ \mathrm{e}\left( z\right) \mathrm{u}^{2}$ $+$ $\
\mathrm{f}\left( z\right) \mathrm{u}$ $\ +$ $\ \mathrm{g}\left( z\right) $ &
,%
\end{tabular}
\label{de}
\end{equation}%
where $\vartheta =\vartheta \left( s_{1},s_{2}\right) \neq 0$ is a
holomorphic function on the complex surface $S$. In this case, the local
CY4- folds is thought of as%
\begin{equation}
\begin{tabular}{lll}
$Y$ & $\longrightarrow $ & $X_{4}$ \\
&  & $\downarrow \pi _{s}$ \\
&  & $S$%
\end{tabular}%
\end{equation}%
where Y is an elliptic local K3 surface with a given ADE geometry; i.e $%
Y\sim E\times \Gamma $ with $\Gamma $\ being a projective line $P^{1}$\ or a
collection of intersecting $P^{1}$s. Comparing eq(\ref{DE}) to its
equivalent form (\ref{de}), we get the following relations,
\begin{equation}
\begin{tabular}{llll}
$\mathcal{D}\left( {\small w}_{{\small 1}}{\small ,w}_{{\small 2}}{\small ,w}%
_{{\small 3}}\right) $ & $=$ & $\vartheta \left( s_{1},s_{2}\right) \times
\mathrm{d}\left( z\right) $ & , \\
$\mathcal{E}\left( {\small w}_{{\small 1}}{\small ,w}_{{\small 2}}{\small ,w}%
_{{\small 3}}\right) $ & $=$ & $\vartheta \left( s_{1},s_{2}\right) \times
\mathrm{e}\left( z\right) $ & , \\
$\mathcal{F}\left( {\small w}_{{\small 1}}{\small ,w}_{{\small 2}}{\small ,w}%
_{{\small 3}}\right) $ & $=$ & $\vartheta \left( s_{1},s_{2}\right) \times
\mathrm{f}\left( z\right) $ & , \\
$\mathcal{G}\left( {\small w}_{{\small 1}}{\small ,w}_{{\small 2}}{\small ,w}%
_{{\small 3}}\right) $ & $=$ & $\vartheta \left( s_{1},s_{2}\right) \times
\mathrm{g}\left( z\right) $ & ,%
\end{tabular}
\label{fc}
\end{equation}%
where the holomorphic functions $\mathcal{D}\left( {\small w}\right) $, $%
\mathcal{E}\left( {\small w}\right) $, $\mathcal{F}\left( {\small w}\right) $
and $\mathcal{G}\left( {\small w}\right) $ get factorized in terms of
products of the holomorphic functions $\vartheta \left( s\right) $ on the
complex surface $S$ and the holomorphic functions $\mathrm{d}\left( z\right)
$, $\mathrm{e}\left( z\right) $, $\mathrm{f}\left( z\right) $ and $\mathrm{g}%
\left( z\right) $ on the normal line to the surface $S$ in the complex base B%
$_{3}$. \newline
In the case where the complex surface $S$ in the local CY4- folds has
several irreducible\textrm{\footnote{%
In the case where the base surface has several irreducible 4- cycles $S_{a}$%
, one has to specify the intersections $S_{a}\cap S_{b}$ as well as the
fibration of the ADE geometry; see below.}} compact components $S_{a}$ like,%
\begin{equation}
C_{4}=\dbigcup_{a=1}^{M}S_{a},  \label{69}
\end{equation}%
the factorizations (\ref{fc}) apply to each component $S_{a}$. \newline
Notice that the irreducible 4- cycle components $S_{a}$, describe as well
compact complex surfaces in the Calabi Yau 4- folds that are locally
parameterized by the complex coordinates $\left( s_{ma}\right) _{1\leq a\leq
n}$, i.e:
\begin{equation}
\begin{tabular}{llll}
$S_{1}$ & $=$ & $\left\{ s_{11},s_{21}\right\} $ & , \\
$S_{2}$ & $=$ & $\left\{ s_{12},s_{22}\right\} $ & , \\
$\vdots $ & $=$ & $\vdots $ & , \\
$S_{M}$ & $=$ & $\left\{ s_{1M},s_{1M}\right\} $ & .%
\end{tabular}%
\end{equation}%
Extending the factorizations (\ref{fc}) to each component $S_{a}$, we can
write,
\begin{equation}
\begin{tabular}{lll}
$\mathcal{D}$ & $=\vartheta \left( s_{1a},s_{2a}\right) \times \mathrm{d}%
\left( z_{a}\right) $ & , \\
$\mathcal{E}$ & $=\vartheta \left( s_{1a},s_{2a}\right) \times \mathrm{e}%
\left( z_{a}\right) $ & , \\
$\mathcal{F}$ & $=\vartheta \left( s_{1a},s_{2a}\right) \times \mathrm{f}%
_{i}\left( z_{a}\right) $ & , \\
$\mathcal{G}$ & $=\vartheta \left( s_{1a},s_{2a}\right) \times \mathrm{h}%
\left( z_{a}\right) $ & ,%
\end{tabular}%
\end{equation}%
with $z_{a}$ parameterizing the normal direction to $S_{a}$ in $B_{3}$.
Notice in passing that the geometry of the B$_{3}$ base of the Calabi-Yau 4-
folds is a little bit complicated. Because of cycles intersections, the
splitting (\ref{tb}) is not trivial. \newline
Focusing on the 4- cycles in the complex surface (\ref{69}), the irreducible
compact components $S_{a}$ have intersections captured by the following
typical relations,
\begin{equation}
\begin{tabular}{llll}
$S_{a}\cap S_{b}=\dbigcup_{\alpha =1}^{M^{\prime }}\mathcal{I}_{ab}^{\alpha
}\Sigma _{\alpha }$ & , & $\mathcal{I}_{ab}^{\alpha }=\mathcal{I}%
_{ba}^{\alpha }$ & , \\
&  &  &  \\
$\Sigma _{\alpha }\cap \Sigma _{\beta }=\dbigcup_{A=1}^{M^{\prime \prime }}%
\mathcal{J}_{\alpha \beta }^{A}P_{A}$ & , & $\mathcal{J}_{\alpha \beta }^{A}=%
\mathcal{J}_{\beta \alpha }^{A}$ & ,%
\end{tabular}%
\end{equation}%
where $\Sigma _{\alpha }$ and $P_{A}$ stand respectively for 2- and 0-
cycles in the local Calabi Yau 4- folds. The intersection numbers $\mathcal{I%
}_{ab}^{\alpha }$ and $\mathcal{J}_{\alpha \beta }^{A}$ fix also the manner
in which the $S_{a}$ 's are glued together. Moreover, the complex
coordinates $\left( s_{1a},s_{2a}\right) $ and $\left( s_{1b},s_{2b}\right) $%
\ of any two intersecting cycles $S_{a}$ and $S_{b}$ are obviously related
by holomorphic transition functions as usual.

\subsubsection{Toric surfaces and blown ups}

So far, we have been describing general geometric features of the base
surface of the local Calabi Yau 4- folds. A particular class of these
surfaces have been considered in the previous section; these are the del
Pezzo surfaces dP$_{n}$ with their remarkable links with:\newline
(\textbf{1}) the projective plane and its blown ups,\newline
(\textbf{2}) the finite dimensional exceptional Lie algebras $E_{n}$.
\newline
Here, we want to contribute to this direction by studding a particular class
of complex surfaces that may play the role of the base $S$ of the local
Calabi Yau 4- folds. This class of complex surfaces share basic features of
the projective plane
\begin{equation*}
\mathbb{P}^{2}=dP_{0}
\end{equation*}%
and the del Pezzos $dP_{n}$, but has also the property to allow more
possibilities. We will distinguish two kinds of surfaces:\newline
(\textbf{a}) complex tetrahedral surface $\mathcal{T}$ and its \emph{toric}
blown ups $\mathcal{T}_{n}^{toric}$,\newline
(\textbf{b}) \emph{non toric} blown ups $\mathcal{T}_{n}^{non\text{ }toric}$
of the tetrahedral surface $\mathcal{T}$.\newline
Below, we focus our attention on CY4- folds based on the complex tetrahedral
surface and its toric blow ups $\mathcal{T}_{n}^{toric}$.

\begin{figure}[tbph]
\begin{center}
\hspace{0cm} \includegraphics[width=10cm]{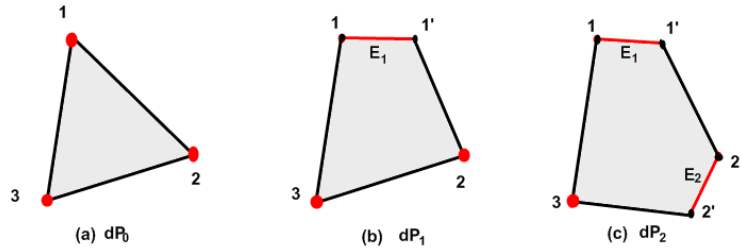}
\end{center}
\par
\vspace{-0.5 cm}
\caption{{\protect\small Toric graphs for dP}$_{0}${\protect\small \ , dP}$%
_{1}${\protect\small \ and dP}$_{2}${\protect\small .}}
\label{dpO}
\end{figure}

\emph{Toric surfaces}\newline
Toric surfaces $S$, which can be thought of as the fibration,
\begin{equation}
\begin{tabular}{lll}
$\mathbb{T}^{2}$ & $\rightarrow $ & $S$ \\
&  & $\downarrow \pi _{_{S}}$ \\
&  & $B_{S}$%
\end{tabular}%
\end{equation}%
with real two dimensional base $B_{S}$ and fiber $\mathbb{T}^{2}$, form a
particular generalization of the projective plane $dP_{0}$. These surfaces
have special features that are nicely engineered by using toric geometry
property encoded in a toric graph $\Delta _{S}$. The simplest toric surface
is obviously given by the compact $dP_{0}$; its toric graph is
\begin{equation}
\begin{tabular}{llll}
$\Delta _{dP_{0}}$ & $=$ & $\text{triangle }\left[ ABC\right] $ & .%
\end{tabular}
\label{tri}
\end{equation}%
Recall that $dP_{0}$ is defined as the projective plane in the non compact
complex three dimension space $\mathbb{C}^{3}$ like,%
\begin{equation}
dP_{0}=\left\{
\begin{array}{c}
\left( x_{1},x_{2},x_{3}\right) \equiv \left( \lambda x_{1},\lambda
x_{2},\lambda x_{3}\right) \\
\left( x_{1},x_{2},x_{3}\right) \neq \left( 0,0,0\right)%
\end{array}%
\right\}
\end{equation}%
with $\lambda $ a non zero complex constant. This compact surface has also a
nice supersymmetric linear sigma model representation given by%
\begin{equation}
\left\vert x_{1}\right\vert ^{2}+\left\vert x_{2}\right\vert ^{2}+\left\vert
x_{3}\right\vert ^{2}=r
\end{equation}%
with the gauge identification $x_{i}\equiv e^{i\theta }x_{i}$ and where $r$
is the Kahler parameter. Other complex surfaces directly related to $dP_{0}$
are given by the toric blown ups $dP_{1}$, $dP_{2}$ and $dP_{3}$ whose toric
graphs $\Delta _{dP_{1}}$, $\Delta _{dP_{2}}$ and $\Delta _{dP_{3}}$ are
depicted in the figure (\ref{dpO}).

\begin{figure}[tbph]
\begin{center}
\hspace{0cm} \includegraphics[width=5cm]{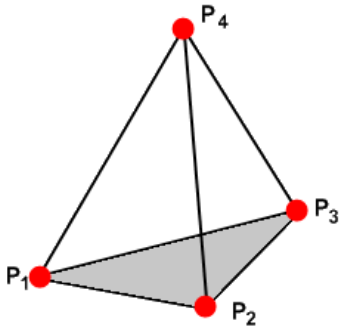}
\end{center}
\par
\vspace{-0.5 cm}
\caption{{\protect\small A toric complex surface given by the union of four
intersecting projective planes forming a toric tetrahedron. Each face of the
tetrahedron corresponds to an irreducible divisor }$S_{i}${\protect\small .
For instance }$S_{4}${\protect\small \ corresponds to the toric triangle }$%
\left[ P_{1}P_{2}P_{3}\right] ${\protect\small \ corresponding as well to a
toric representation of the projective plane.}}
\label{te}
\end{figure}

\ \ \ \ \newline
Complex two dimension toric surfaces may be also engineered by using
embedding in complex higher dimensional projective spaces $\mathbb{P}^{n}$
with $n\geq 3$ thought of as given by the fibration%
\begin{equation}
\begin{tabular}{lll}
$\mathbb{T}^{n}$ & $\rightarrow $ & $\mathbb{P}^{n}$ \\
&  & $\downarrow \pi _{_{\mathbb{P}^{n}}}$ \\
&  & $\Delta _{_{\mathbb{P}^{n}}}$%
\end{tabular}%
\end{equation}%
An interesting class of toric surfaces that we are interested in here is
given by the complex tetrahedral surface $\mathcal{T}$ and its toric blown
ups $\mathcal{T}_{n}^{toric}$. Let us consider first the \emph{non planar}
toric surface $\mathcal{T}$ with fibration%
\begin{equation}
\begin{tabular}{lll}
$\mathbb{T}^{2}$ & $\rightarrow $ & $\mathcal{T}$ \\
&  & $\downarrow \pi _{_{\mathcal{T}}}$ \\
&  & $\Delta _{_{\mathcal{T}}}$%
\end{tabular}%
\text{ }  \label{tet}
\end{equation}%
A nice way to define complex tetrahedral surface $\mathcal{T}$ is in terms
of divisors of the complex three dimensions projective space $\mathbb{P}^{3}$%
,%
\begin{equation}
\left\{
\begin{array}{c}
\left( x_{1},x_{2},x_{3},x_{4}\right) \equiv \left( \lambda x_{1},\lambda
x_{2},\lambda x_{3},\lambda x_{4}\right) \\
\left( x_{1},x_{2},x_{3},x_{4}\right) \neq \left( 0,0,0,0\right)%
\end{array}%
\right\}
\end{equation}%
with $\lambda \in \mathbb{C}^{\ast }$. Being a toric three- folds, the
complex space $\mathbb{P}^{3}$ may be also defined in terms of the
supersymmetric linear sigma model D- equation like,%
\begin{equation}
\mathbb{P}^{3}:\sum_{i=1}^{4}\left\vert x_{i}\right\vert ^{2}=R\quad ,\quad
x_{i}\equiv e^{i\theta }x_{i}\quad ,
\end{equation}%
where $R$ is the Kahler parameter of $\mathbb{P}^{3}$. Irreducible divisors $%
S_{a}$ in the space $\mathbb{P}^{3}$ are complex surfaces generated by the
equation $x_{a}=0$. There are four such divisors in $\mathbb{P}^{3}$ which
form altogether the complex tetrahedron depicted in figure (\ref{te}).

\ \ \newline
The complex tetrahedral surface $\mathcal{T}$ has some particular features
which we describe below.

(\textbf{a}) \emph{Link between }$\mathcal{T}$\emph{\ and }$\mathbb{P}^{2}$%
\newline
The complex tetrahedral surface $\mathcal{T}$ has the tetrahedron $\Delta _{%
\mathcal{T}}$ as a toric graph; it is then a natural extension of the
projective plane $\mathbb{P}^{2}=dP_{0}$ whose toric graph is a triangle (%
\ref{tri}). Since the tetrahedron $\Delta _{\mathcal{T}}$ has four
intersecting triangle faces; the non planar surface $\mathcal{T}$ involves
then four intersecting projective planes
\begin{equation}
\begin{tabular}{llllll}
$\mathbb{P}_{a}^{2}$ & $=$ & $dP_{0}^{\left( a\right) }$ & $,$ & $a=1,...,4$
& $.$%
\end{tabular}%
\end{equation}%
Using the link between the projective plane and the del Pezzo surfaces, we
may refer to the complex tetrahedral surface $\mathcal{T}$ as follows
\begin{equation}
\mathcal{T}_{{\small 0}}=dP_{k_{1},k_{2},k_{3},k_{4}}\qquad ,\qquad \left(
k_{1},k_{2},k_{3},k_{4}\right) =\left( 0,0,0,0\right) \text{,}
\end{equation}%
where the four integers $\left( k_{1},k_{2},k_{3},k_{4}\right) $ refer to
the number of blow ups of the faces of the non surface $\mathcal{T}$. Notice
that these blow ups form in fact just a particular family of a larger one. A
way to see this feature is to focus on toric singularities where tetrahedron
involves both at its edges and its vertices; for useful details see below
but for an explicit study regarding these blow ups see \textrm{\cite{SA}.}

(\textbf{b}) \emph{Tetrahedron and gauge enhancements}\newline
As a toric surface, the tetrahedron $\mathcal{T}$ \ $\sim $\ \ $\Delta _{_{%
\mathcal{T}}}\times \mathbb{T}^{2}$ has a natural $U^{2}\left( 1\right) $
symmetry on $\mathbb{T}^{2}$ with fix points on the following loci:\newline
\textbf{(i) }the six edges of the toric surface $\mathcal{T}$ where a
1-cycle of $\mathbb{T}^{2}$ shrinks to zero,\newline
\textbf{(ii)} its four vertices where 2- cycles shrink to zero.\newline
The $U\left( 1\right) \times U\left( 1\right) $ toric gauge symmetry of the
fiber of the toric surface $\mathcal{T}$ may be:

\begin{itemize}
\item interpreted in terms of two wrapped seven branes $\left( 7B\right)
_{1} $ and $\left( 7B\right) _{1}^{\prime }$,

\item used to engineer the enhancement of gauge symmetry along the edges and
at the vertices of the tetrahedral surface.
\end{itemize}

(\textbf{c})\emph{Blown ups of the tetrahedron}\newline
Mimicking the relation between the projective plane $dP_{0}$ and the del
Pezzo surfaces $dP_{k}$, and using the relation between the complex
tetrahedral surface $\mathcal{T}$ and the complex projective plane, we can
perform blow ups of the toric surface $\mathcal{T}$. Generally, we
distinguish two kinds of blow ups: toric blow ups and non toric ones \textrm{%
\cite{SA}}. Regarding the toric blow ups, one has to distinguish as well two
classes of blow ups:\newline
(\textbf{i}) blow ups by projective lines of the edges $\Sigma _{ab}$ of the
tetrahedron,\newline
(\textbf{ii}) blow ups of the vertices $P_{abc}$ by projective planes.%
\newline
Regarding the edges, the bow up at each point on a edge $\Sigma $ is done in
terms of projective line $\mathbb{P}^{1}$. As such the blow up of the full
edge $\Sigma \sim \mathbb{P}^{1}$ is given by a del Pezzo surface dP1:%
\begin{equation}
\begin{tabular}{llllll}
$\Sigma $ & $\rightarrow $ & $dP_{1}$ & $\sim $ & $\mathbb{P}^{1}\times
\mathbb{P}^{1}$ & .%
\end{tabular}%
\end{equation}
Concerning, the blow up of each vertex of the tetrahedron, it is done by a
projective plane $\mathbb{P}^{2}$; for illustration see figure (\ref{t1}).
\newline
To avoid technicalities, it is enough to notice that for each plane $\mathbb{%
P}_{a}^{2}$ associated with a given face of the complex tetrahedral surface $%
\mathcal{T}$, one may perform up to eight blow ups as given below,%
\begin{equation}
\begin{tabular}{llllll}
$\mathbb{P}_{a}^{2}=dP_{0}^{\left( a\right) }$ & $\longrightarrow $ & $%
dP_{k_{a}}^{\left( a\right) }$ & , & $k_{a}=1,...,8,$ $a=1,2,3,4$ & .%
\end{tabular}%
\end{equation}%
In doing so, we a priori get the following blown up surfaces of the
tetrahedron,%
\begin{equation}
\begin{tabular}{llllll}
$\mathcal{T}_{n}=dP_{k_{1},k_{2},k_{3},k_{4}}$ & $,$ & $%
n=k_{1}+k_{2}+k_{3}+k_{4}$ & $,$ & $k_{a}=1,...,8$ & .%
\end{tabular}%
\end{equation}%
Clearly, these complex surfaces give generalizations of the del Pezzo ones
which are recovered by setting three of these integers to zero to get $%
dP_{k_{1},0,0,0}$ by taking $k_{1}=k_{2}=k_{3}=0$. \ Explicit examples will
be given in section 7.

\subsection{More on tetrahedron geometry}

We first describe subspaces in the complex tetrahedral geometry where the
bulk gauge invariance in the GUT seven brane undergoes transitions. Then we
build explicitly the local Calabi Yau 4- folds based on the tetrahedral
surface $\mathcal{T}$.

\subsubsection{subspaces in tetrahedron}

In the complex tetrahedral geometry, the 4-cycle $\mathcal{C}_{4}$ is given
by the union of four intersecting components $S_{1},$ $S_{2}$, $S_{3}$\ and $%
S_{4}$
\begin{equation}
\mathcal{C}_{4}=S_{1}\dbigcup S_{2}\dbigcup S_{3}\dbigcup S_{4},  \label{C4}
\end{equation}%
where the compact toric surfaces $\left( S_{a}\right) _{1\leq a\leq 4}$ are
four intersecting complex projective surfaces $\mathbb{P}_{a}^{2}$ belonging
to four different planes of the complex three dimension space $\mathbb{P}%
^{3} $. We have the relations:%
\begin{equation}
\begin{tabular}{llll}
$S_{1}\cap S_{2}=\Sigma _{\left( 12\right) }$ & , & $S_{2}\cap S_{3}=\Sigma
_{\left( 23\right) }$ & , \\
$S_{1}\cap S_{3}=\Sigma _{\left( 13\right) }$ & , & $S_{2}\cap S_{4}=\Sigma
_{\left( 24\right) }$ & , \\
$S_{1}\cap S_{4}=\Sigma _{\left( 14\right) }$ & , & $S_{3}\cap S_{4}=\Sigma
_{\left( 34\right) }$ & .%
\end{tabular}
\label{cs}
\end{equation}%
Moreover, since the complex tetrahedral surface is toric, all the edges $%
\Sigma _{\left( ab\right) }$ are precisely given by projective lines $%
\mathbb{P}^{1}$. Furthermore, seen that the tetrahedron is compact, these
projective lines $\Sigma _{\left( ab\right) }$ intersect mutually at four
points $P_{A}$ in the base of the local Calabi-Yau 4- folds,%
\begin{equation}
\begin{tabular}{llllllll}
$\Sigma _{\left( 23\right) }\cap \Sigma _{\left( 24\right) }$ & $=$ & $%
\Sigma _{\left( 23\right) }\cap \Sigma _{\left( 34\right) }$ & $=$ & $\Sigma
_{\left( 24\right) }\cap \Sigma _{\left( 34\right) }$ & $=$ & $P_{1}$ & , \\
$\Sigma _{\left( 14\right) }\cap \Sigma _{\left( 34\right) }$ & $=$ & $%
\Sigma _{\left( 13\right) }\cap \Sigma _{\left( 34\right) }$ & $=$ & $\Sigma
_{\left( 13\right) }\cap \Sigma _{\left( 14\right) }$ & $=$ & $P_{2}$ & , \\
$\Sigma _{\left( 12\right) }\cap \Sigma _{\left( 24\right) }$ & $=$ & $%
\Sigma _{\left( 14\right) }\cap \Sigma _{\left( 24\right) }$ & $=$ & $\Sigma
_{\left( 12\right) }\cap \Sigma _{\left( 14\right) }$ & $=$ & $P_{3}$ & , \\
$\Sigma _{\left( 12\right) }\cap \Sigma _{\left( 23\right) }$ & $=$ & $%
\Sigma _{\left( 13\right) }\cap \Sigma _{\left( 23\right) }$ & $=$ & $\Sigma
_{\left( 12\right) }\cap \Sigma _{\left( 13\right) }$ & $=$ & $P_{4}$ & .%
\end{tabular}%
\end{equation}%
Using eq(\ref{cs}), these intersecting points may be also viewed as the
intersection of three faces as shown below
\begin{equation}
\begin{tabular}{llll}
$S_{2}\cap S_{3}\cap S_{4}$ & $=$ & $P_{1}$ & , \\
$S_{1}\cap S_{3}\cap S_{4}$ & $=$ & $P_{2}$ & , \\
$S_{1}\cap S_{2}\cap S_{4}$ & $=$ & $P_{3}$ & , \\
$S_{1}\cap S_{2}\cap S_{4}$ & $=$ & $P_{4}$ & .%
\end{tabular}
\label{C5}
\end{equation}%
Tetrahedron geometry has other remarkable properties; in particular each
face $S_{a}$ of the tetrahedron has a toric fibration%
\begin{equation}
\begin{tabular}{lll}
$\mathbb{T}_{a}^{2}$ & $\rightarrow $ & $S_{a}$ \\
&  & $\downarrow \pi _{_{S}}$ \\
&  & $\Delta _{S_{a}}$%
\end{tabular}%
\end{equation}%
with real two dimension base $\Delta _{S_{a}}$ represented by a triangular
toric graph and a fiber $\mathbb{T}_{a}^{2}=\mathbb{S}_{a}^{1}\times \mathbb{%
S}_{a}^{1},$ $a=1,2,3,4,$ where the $\mathbb{S}_{a}^{1}$s are associated
with the $U_{a}\left( 1\right) $ toric actions. Notice that these toric
fibers $\mathbb{T}_{a}^{2}$ are not the same for all the faces $S_{a}$; they
change from a $S_{a}$ to an other $S_{b}$; but intersect along a 1- cycle $%
\mathbb{S}_{ab}^{1}$. Thus, given two faces $S_{a}$ and $S_{b}$ with
intersecting curve,%
\begin{equation}
\Sigma _{\left( ab\right) }=S_{a}\cap S_{b},
\end{equation}%
we have the following,%
\begin{equation}
\begin{tabular}{ll|lll|lll|ll}
subspaces &  &  & 2d- base &  &  & 2d-fiber &  &  & toric action \\
\hline\hline
$S_{a}$ &  &  & $\Delta _{S_{a}}$ &  &  & $\mathbb{S}_{a}^{1}\times \mathbb{S%
}_{ab}^{1}$ &  &  & $U_{a}\left( 1\right) \times U_{ab}\left( 1\right) $ \\
$S_{b}$ &  &  & $\Delta _{S_{b}}$ &  &  & $\mathbb{S}_{b}^{1}\times \mathbb{S%
}_{ab}^{1}$ &  &  & $U_{b}\left( 1\right) \times U_{ab}\left( 1\right) $ \\
$\Sigma _{\left( ab\right) }$ &  &  & $\Delta _{\Sigma _{\left( ab\right) }}$
&  &  & $\mathbb{S}_{ab}^{1}$ &  &  & $U_{ab}\left( 1\right) $ \\ \hline
\end{tabular}%
\end{equation}%
The toric fibers $\mathbb{S}_{a}^{1}\times \mathbb{S}_{ab}^{1}$ degenerate
\emph{once} on the projective edges $\Sigma _{\left( ab\right) }$ and
degenerate \emph{twice} at the four vertices $P_{A}$. Notice that the
1-cycles $\mathbb{S}_{a}^{1}$ and $\mathbb{S}_{b}^{1}$ shrink to zero on $%
\Sigma _{\left( ab\right) }$
\begin{equation}
\begin{tabular}{lll}
$\mathbb{S}_{ab}^{1}$ & $\rightarrow $ & $\Sigma _{ab}$ \\
&  & $\downarrow \pi _{_{\Sigma }}$ \\
&  & $\Delta _{_{\Sigma }}$%
\end{tabular}%
\end{equation}%
Furthermore, the cycle $\mathbb{S}_{ab}^{1}$ shrinks down to zero at the
meeting point of the two curves $\Sigma _{\left( ab\right) }$ and $\Sigma
_{\left( ac\right) }$.\newline
With these tools at hand, we turn now to build the explicit expression of
the algebraic equation of the local elliptic Calabi-Yau 4- folds based on
the tetrahedron.

\subsubsection{Local tetrahedron}

In toric language, one may directly read the intersections in the base of
the elliptically K3 fibered Calabi-Yau four- folds,%
\begin{equation}
\begin{tabular}{llll}
$Y$ & $\longrightarrow $ & $X_{4}$ &  \\
&  & $\downarrow \pi _{_{_{\mathcal{T}}}}$ &  \\
&  & $\mathcal{T}$ &
\end{tabular}%
\end{equation}%
Using the irreducible $S_{a}$s, the complex tetrahedral surface may be
defined as
\begin{equation}
\begin{tabular}{llll}
$\mathcal{T}$ & $=$ & $S_{1}\dbigcup S_{2}\dbigcup S_{3}\dbigcup S_{4}$ & ,%
\end{tabular}
\label{ts}
\end{equation}%
with the following intersections,%
\begin{equation}
\begin{tabular}{llllll}
$S_{a}\cap S_{b}$ & $=$ & $\Sigma _{\left( ab\right) }$ & $,$ & $a<b=1,...,4$
& $,$ \\
$S_{a}\cap S_{b}\cap S_{c}$ & $=$ & $P_{\left( abc\right) }$ & $,$ & $a<b<c$
& $.$%
\end{tabular}%
\end{equation}%
Being a toric surface, the toric fibration of the tetrahedral surface $%
\mathcal{T}$ \ \ $\sim $ \ \ $\Delta _{_{\mathcal{T}}}\times \mathbb{T}_{_{%
\mathcal{T}}}^{2}$ \ is not homogeneous; it decomposes in terms of the toric
fibrations,
\begin{equation}
\begin{tabular}{llll}
$\mathcal{T}$ & $\sim $ & $\dbigcup_{a}\left( \Delta _{_{S_{a}}}\times
\mathbb{T}_{a}^{2}\right) $ & ,%
\end{tabular}
\label{tt}
\end{equation}%
with toric graph given by the figure (\ref{te}). From this toric graph, one
can directly read the toric data of each component $S_{a}$ and then those of
$\mathcal{T}$. \newline
In the toric graph picture of the complex base, the local Calabi-Yau four-
folds $X_{4}$ may be thought of as fibering on each point of $\Delta _{_{%
\mathcal{T}}}$ a complex three dimension fiber $Z$ given by the 2-torus $%
\mathbb{T}_{_{\mathcal{T}}}^{2}$ \emph{times} the the complex two dimension
fiber $Y$. Roughly, we have%
\begin{equation}
\begin{tabular}{llllll}
$X_{4}$ & $\sim $ & $\mathcal{T}\times Y$ & $\sim $ & $\Delta _{_{\mathcal{T}%
}}\times Z$ & ,%
\end{tabular}%
\end{equation}%
with $Z$ $\sim $ $\mathbb{T}_{_{\mathcal{T}}}^{2}\times Y$. \newline
Below, we construct the explicit expression for $X_{4}$ as a complex 4
dimension hypersurface in the complex space $\mathbb{C}^{5}$. First, we give
the algebraic equation of the complex base tetrahedron $\mathcal{T}$. Then,
we study the fiber singularity on the edges $\Sigma _{\left( ab\right) }$
and at the vertices $P_{\left( abc\right) }$ of the tetrahedron.

\emph{Base surface} $\mathcal{T}$\newline
Since the four irreducible components $S_{a}$ of the complex tetrahedron are
given by different projective planes in $\mathbb{C}^{4}$, we start by
introducing the projective coordinates of the complex three dimension
projective space $\mathbb{P}^{3}$,%
\begin{equation}
\left( x_{1},x_{2},x_{3},x_{4}\right) \equiv \left( \lambda x_{1},\lambda
x_{2},\lambda x_{3},\lambda x_{4}\right) ,  \label{ga}
\end{equation}%
with projective parameter $\lambda \in \mathbb{C}^{\ast }$ and $\left(
x_{1},x_{2},x_{3},x_{4}\right) \neq \left( 0,0,0,0,0\right) $. The
tetrahedron surface is engineered by thinking about the compact surfaces $%
S_{a}$ as the planar divisors of $\mathbb{P}^{3}$,%
\begin{equation}
x_{a}=\left\vert x_{a}\right\vert e^{i\varphi _{a}}=0\qquad ,\qquad
a=1,2,3,4.
\end{equation}%
As noticed earlier, this representation has an equivalent description in the
supersymmetric linear sigma model set up of toric manifolds. There, the
divisors $S_{a}$ are given by the standard D- term equations
\begin{equation}
S_{a}:\left( \sum_{b=1}^{4}\left\vert x_{b}\right\vert ^{2}\right)
_{x_{a}=0}=R\quad ,\quad x_{b}\equiv e^{i\theta }x_{b}\quad ,  \label{si}
\end{equation}%
where $R$ is the Kahler parameter of $\mathbb{P}^{3}$ and $e^{i\theta }$ is
the $U\left( 1\right) $ compact part of the gauge transformation (\ref{ga}).%
\newline
In the mirror complex holomorphic description, it is not difficult to see
that the complex algebraic equation describing the base manifold $\mathcal{T}
$ of the local Calabi-Yau 4- folds is given by the following complex two
dimension surface in the projective space $\mathbb{P}^{3}$,
\begin{equation}
\mathrm{\mu }\left( \dprod_{a=1}^{4}x_{a}\right) =\mathrm{\mu }%
x_{1}x_{2}x_{3}x_{4}=0.  \label{mu}
\end{equation}%
In this relation, $\mathrm{\mu }$\ is a complex number and the divisors $%
S_{a}$ are precisely given by the solutions of this relation. This equation
may be viewed as well as the large complex structure limit ($\mathrm{\mu
\rightarrow \infty }$) of the quartic%
\begin{equation}
\begin{tabular}{lll}
$\sum_{i=1}^{4}A_{i}x_{i}^{4}+\sum_{i=1}^{4}\left[ x_{i}^{3}\left(
\sum_{j\neq i}B_{ij}x_{j}\right) \right] +\sum_{i=1}^{4}\left[
x_{i}^{2}\left( \sum_{j\neq l\neq i}C_{ijl}x_{j}x_{l}\right) \right] $ &  &
\\
$+\sum_{i=1}^{4}\left[ x_{i}\left( \sum_{j\neq l\neq m\neq
i}D_{ijlm}x_{j}x_{l}x_{m}\right) \right] +\mathrm{\mu }x_{1}x_{2}x_{3}x_{4}$
& $=$ & $0$%
\end{tabular}%
\end{equation}%
where the $A_{i}$'s, $B_{ij}$'s, $C_{ijk}$'s and $D_{ijlm}$'s are complex
structures. In mirror geometry, the divisors $S_{a}$ are explicitly given by,%
\begin{equation}
\begin{tabular}{llllll}
$S_{1}$ & : & $\left\{ \left( x_{2},x_{3},x_{4}\right) \equiv \left( \lambda
x_{2},\lambda x_{3},\lambda x_{4}\right) \right\} $ & $\equiv $ & $\mathbb{P}%
_{1}^{2}$ & , \\
$S_{2}$ & : & $\left\{ \left( x_{1},x_{3},x_{4}\right) \equiv \left( \lambda
x_{1},\lambda x_{3},\lambda x_{4}\right) \right\} $ & $\equiv $ & $\mathbb{P}%
_{2}^{2}$ & , \\
$S_{3}$ & : & $\left\{ \left( x_{1},x_{2},x_{4}\right) \equiv \left( \lambda
x_{1},\lambda x_{2},\lambda x_{4}\right) \right\} $ & $\equiv $ & $\mathbb{P}%
_{3}^{2}$ & , \\
$S_{4}$ & : & $\left\{ \left( x_{1},x_{2},x_{3}\right) \equiv \left( \lambda
x_{1},\lambda x_{2},\lambda x_{3}\right) \right\} $ & $\equiv $ & $\mathbb{P}%
_{4}^{2}$ & ,%
\end{tabular}%
\end{equation}%
with the $\mathbb{C}^{\ast }$ action generated by the complex parameter $%
\lambda $ inherited from the projective action of the $\mathbb{P}^{3}$
space. Similarly, the intersections $S_{a}\cap S_{b}=\Sigma _{\left(
ab\right) }$ can be determined explicitly from above relations. These are
given by the following projective lines in $\mathbb{P}^{3}$,
\begin{equation}
\begin{tabular}{llllll}
$\Sigma _{\left( 12\right) }$ & $=$ & $\left\{ \left( x_{3},x_{4}\right)
\equiv \left( \lambda x_{3},\lambda x_{4}\right) \right\} $ & $\equiv $ & $%
\mathbb{P}_{1}^{1}$ & , \\
$\Sigma _{\left( 13\right) }$ & $=$ & $\left\{ \left( x_{2},x_{4}\right)
\equiv \left( \lambda x_{2},\lambda x_{4}\right) \right\} $ & $\equiv $ & $%
\mathbb{P}_{2}^{1}$ & , \\
$\Sigma _{\left( 14\right) }$ & $=$ & $\left\{ \left( x_{2},x_{3}\right)
\equiv \left( \lambda x_{2},\lambda x_{3}\right) \right\} $ & $\equiv $ & $%
\mathbb{P}_{3}^{1}$ & , \\
$\Sigma _{\left( 23\right) }$ & $=$ & $\left\{ \left( x_{1},x_{4}\right)
\equiv \left( \lambda x_{1},\lambda x_{4}\right) \right\} $ & $\equiv $ & $%
\mathbb{P}_{4}^{1}$ & , \\
$\Sigma _{\left( 24\right) }$ & $=$ & $\left\{ \left( x_{1},x_{3}\right)
\equiv \left( \lambda x_{1},\lambda x_{3}\right) \right\} $ & $\equiv $ & $%
\mathbb{P}_{5}^{1}$ & , \\
$\Sigma _{\left( 34\right) }$ & $=$ & $\left\{ \left( x_{1},x_{2}\right)
\equiv \left( \lambda x_{1},\lambda x_{2}\right) \right\} $ & $\equiv $ & $%
\mathbb{P}_{6}^{1}$ & . \\
&  &  &  &  &
\end{tabular}%
\end{equation}%
Their supersymmetric linear sigma model description may be directly deduced
from eqs(\ref{si}) by putting to zero two of the four variables. \newline
Finally, the intersections of these curves are given by points in $\mathcal{T%
}$. Up on making an appropriate choice of the $\mathbb{C}^{\ast }$ action,
these points may be taken as,%
\begin{equation}
\begin{tabular}{ll}
$P_{1}=\left( 1,0,0,0\right) $ & , \\
$P_{2}=\left( 0,1,0,0\right) $ & , \\
$P_{3}=\left( 0,0,1,0\right) $ & , \\
$P_{4}=\left( 0,0,0,1\right) $ & .%
\end{tabular}%
\end{equation}

\emph{Fibering Y over the base }$\mathcal{T}$\newline
Using the above analysis, we can now write down the explicit algebraic
relation defining the local Calabi Yau four- fold based on the tetrahedron $%
\mathcal{T}$. In the large complex structure limit $\mathrm{\mu \rightarrow
\infty }$, the local elliptic Calabi-Yau four- folds may defined by the
following algebraic relations,%
\begin{equation}
\mathrm{v}^{2}=\left( \dprod_{l=1}^{4}x_{l}\right) \times \mathrm{\tilde{v}}%
^{2}\qquad ,\qquad \left( \dprod_{l=1}^{4}x_{l}\right) =0\qquad ,  \label{vv}
\end{equation}%
with the singular term $\mathrm{\tilde{v}}^{2}$ given by,%
\begin{eqnarray}
\mathrm{\tilde{v}}^{2} &=&\text{ \ \ }\sum_{i=1}^{4}\frac{1}{x_{i}}\left[
\mathrm{d}_{i}\left( z\right) \mathrm{u}^{3}\ +\ \mathrm{e}_{i}\left(
z\right) \mathrm{u}^{2}+\ \mathrm{f}_{i}\left( z\right) \mathrm{u}\ +\
\mathrm{h}_{i}\left( z\right) \right]  \notag \\
&&+\sum_{i>j=1}^{4}\frac{1}{x_{i}x_{j}}\left[ \mathrm{d}_{\left( ij\right)
}\left( z\right) \mathrm{u}^{3}\ +\ \mathrm{e}_{\left( ij\right) }\left(
z\right) \mathrm{u}^{2}+\ \mathrm{f}_{\left( ij\right) }\left( z\right)
\mathrm{u}\ +\ \mathrm{h}_{\left( ij\right) }\left( z\right) \right]
\label{su} \\
&&+\sum_{i>j>k=1}^{4}\frac{1}{x_{i}x_{j}x_{k}}\left[ \mathrm{d}_{\left(
ijk\right) }\left( z\right) \mathrm{u}^{3}\ +\ \mathrm{e}_{\left( ijk\right)
}\left( z\right) \mathrm{u}^{2}+\ \mathrm{f}_{\left( ijk\right) }\left(
z\right) \mathrm{u}\ +\ \mathrm{h}_{\left( ijk\right) }\left( z\right) %
\right] .  \notag
\end{eqnarray}%
These holomorphic relations involve several terms which deserve some
comments.\newline
(\textbf{1}) Since $\Pi _{l=1}^{4}x_{l}=0$ as required by the defining
relation of the tetrahedron, then eq(\ref{vv}) is non trivial unless if $%
\mathrm{\tilde{v}}^{2}$ has poles so that the product $\mathrm{\tilde{v}}%
^{2}\times \Pi _{l=1}^{4}x_{l}$ make sense, that is%
\begin{equation}
\mathrm{\tilde{v}}^{2}\times \dprod_{l=1}^{4}x_{l}\text{ \ }\rightarrow
\text{ \ finite,}
\end{equation}%
in the limit $x_{k}\rightarrow 0$. \newline
(\textbf{2}) Since $\Pi _{l=1}^{4}x_{l}=0$ has simple, double and triple
zeros, then the poles in $\mathrm{\tilde{v}}^{2}$ should be of three kinds:
simple, double and triple,%
\begin{equation}
\mathrm{\tilde{v}}^{2}\sim \frac{1}{x^{3}}+\frac{1}{x^{2}}+\frac{1}{x}+\text{
...,}
\end{equation}%
where the dots stand for irrelevant regular terms.\newline
(\textbf{3}) the simple poles are located at $x_{a}=0$ and so are associated
with the divisors $S_{a}$. These simple poles correspond to the first terms
in eq(\ref{su}). Upon multiplication by $\Pi _{l=1}^{4}x_{l}$, we get cubic
monomials $x_{i}x_{j}x_{k}$. More explicitly, this term reads as,%
\begin{equation}
\begin{tabular}{ll}
$+$ $\ x_{2}x_{3}x_{4}\left[ \mathrm{d}_{1}\left( z\right) \mathrm{u}^{3}\
+\ \mathrm{e}_{1}\left( z\right) \mathrm{u}^{2}+\ \mathrm{f}_{1}\left(
z\right) \mathrm{u}\ +\ \mathrm{h}_{1}\left( z\right) \right] \delta \left(
x_{1}\right) $ &  \\
$+$ $\ x_{1}x_{3}x_{4}\left[ \mathrm{d}_{2}\left( z\right) \mathrm{u}^{3}\
+\ \mathrm{e}_{2}\left( z\right) \mathrm{u}^{2}+\ \mathrm{f}_{2}\left(
z\right) \mathrm{u}\ +\ \mathrm{h}_{2}\left( z\right) \right] \delta \left(
x_{2}\right) $ &  \\
$+\ \ x_{1}x_{2}x_{4}\left[ \mathrm{d}_{3}\left( z\right) \mathrm{u}^{3}\ +\
\mathrm{e}_{3}\left( z\right) \mathrm{u}^{2}+\ \mathrm{f}_{3}\left( z\right)
\mathrm{u}\ +\ \mathrm{h}_{3}\left( z\right) \right] \delta \left(
x_{3}\right) $ &  \\
$+\ \ x_{1}x_{2}x_{3}\left[ \mathrm{d}_{4}\left( z\right) \mathrm{u}^{3}\ +\
\mathrm{e}_{4}\left( z\right) \mathrm{u}^{2}+\ \mathrm{f}_{4}\left( z\right)
\mathrm{u}\ +\ \mathrm{h}_{4}\left( z\right) \right] \delta \left(
x_{4}\right) $ & ,%
\end{tabular}
\label{xx}
\end{equation}%
where we have added the Dirac delta function $\delta \left( x_{a}\right) $
to refer to the divisor $S_{a}$ in question. \newline
Furthermore, the extra term between brackets, namely
\begin{equation}
\mathrm{\tilde{v}}_{a}^{\prime 2}=\mathrm{d}_{a}\left( z\right) \mathrm{u}%
^{3}\ +\ \mathrm{e}_{a}\left( z\right) \mathrm{u}^{2}+\ \mathrm{f}_{a}\left(
z\right) \mathrm{u}\ +\ \mathrm{h}_{a}\left( z\right)  \label{vt}
\end{equation}%
where we have set
\begin{equation*}
\mathrm{\tilde{v}}_{a}^{\prime 2}=\frac{\left( \mathrm{\tilde{v}}%
^{2}x_{a}\right) }{\left( \Pi _{b=1}^{4}x_{b}\right) },
\end{equation*}%
captures the way the fiber Y degenerates on $S_{a}$ as a locus. In the $%
SU\left( 5\right) $ GUT type model, eq(\ref{vt}) takes the form%
\begin{equation}
\mathrm{\tilde{v}}_{a}^{\prime 2}=\mathrm{u}^{2}+z_{a}^{5}\left(
z_{a}-t_{a1}\right) \left( z-t_{a2}\right) \qquad ,\qquad a=1,2,3,4,
\end{equation}%
where $t_{a1}$ and $t_{a2}$ are vevs as in eq(\ref{car}). \newline
(\textbf{4}) the double poles are located at $x_{a}=x_{b}=0$ and are
associated with the complex curves $\Sigma _{\left( ab\right) }=S_{a}\cap
S_{b}$. These double poles correspond to the second term in eqs(\ref{su}).
Up on multiplying by $\left( \Pi _{b=1}^{4}x_{b}\right) $, one ends with
quadratic monomial $x_{a}x_{b}$ associated with the six matter curves $%
\Sigma _{\left( ab\right) }$.\newline
Moreover, the elliptic curves fibered on the matter curves $\Sigma _{\left(
ab\right) }$ namely%
\begin{equation}
\mathrm{\tilde{v}}_{ab}^{\prime 2}=\mathrm{d}_{\left( ab\right) }\left(
z\right) \mathrm{u}^{3}\ +\ \mathrm{e}_{\left( ab\right) }\left( z\right)
\mathrm{u}^{2}+\ \mathrm{f}_{\left( ab\right) }\left( z\right) \mathrm{u}\
+\ \mathrm{h}_{\left( ab\right) }\left( z\right) ,
\end{equation}%
with $\mathrm{\tilde{v}}_{ab}^{\prime 2}=\left( \mathrm{\tilde{v}}%
^{2}x_{a}x_{b}\right) /\left( \Pi _{c=1}^{4}x_{c}\right) $, capture the one-
fold enhanced gauge symmetry.\newline
(\textbf{5}) the triple poles located at $x_{a}=x_{b}=x_{c}=0$ are
associated with the vertices of the tetrahedron. Furthermore, the elliptic
curves fibered on the vertices of the tetrahedron
\begin{equation}
\mathrm{\tilde{v}}_{abc}^{\prime 2}=\mathrm{d}_{\left( abc\right) }\left(
z\right) \mathrm{u}^{3}\ +\ \mathrm{e}_{\left( abc\right) }\left( z\right)
\mathrm{u}^{2}+\ \mathrm{f}_{\left( abc\right) }\left( z\right) \mathrm{u}\
+\ \mathrm{h}_{\left( abc\right) }\left( z\right) ,
\end{equation}%
with $\mathrm{\tilde{v}}_{abc}^{\prime 2}=\left( \mathrm{\tilde{v}}%
^{2}x_{a}x_{b}x_{c}\right) /\left( \Pi _{l=1}^{4}x_{l}\right) $, capture the
two-fold enhanced gauge symmetry namely $SO\left( 12\right) $, $E_{6}$ and $%
SU\left( 7\right) $.

\subsection{$SU\left( 5\right) $ Quiver models}

In this subsection, we consider F-theory on local Calabi-Yau 4- folds based
on tetrahedron and we construct a class of three kinds of \emph{4D} $%
\mathcal{N}=1$ supersymmetric $SU\left( 5\right) $ quiver GUT- type models.
By using the $SU\left( 5\right) $ group as a gauge invariance on the
surfaces $S_{a}$ of the tetrahedron, we distinguish three models according
to the gauge enhanced symmetry that live at the vertices of the tetrahedron.
These \emph{unrealistic} models have respectively a $SU\left( 7\right) $, a $%
SO\left( 12\right) $ or a $E_{6}$ enhanced invariance.

\subsubsection{$SU\left( 7\right) $ vertex}

\begin{figure}[tbph]
\begin{center}
\hspace{0cm} \includegraphics[width=8cm]{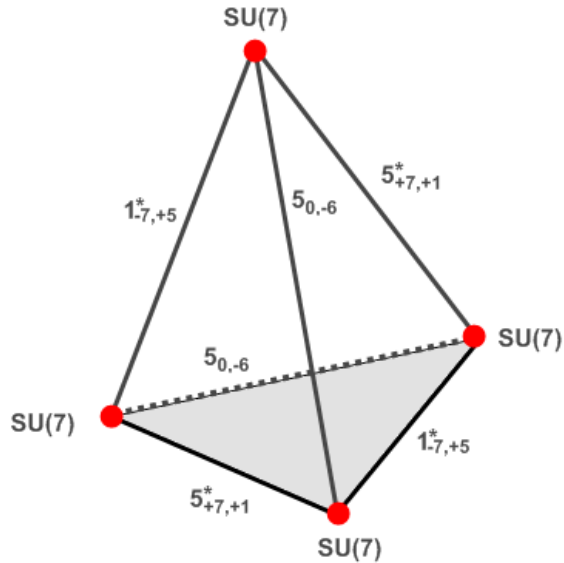}
\end{center}
\par
\vspace{-0.5 cm}
\caption{Quiver gauge diagram for $SU\left( 5\right) $ GUT- like model with $%
SU\left( 7\right) $ enhanced gauge symmetry at the vertices.}
\label{74}
\end{figure}

The quiver gauge diagram of the 4D $\mathcal{N}=1$ supersymmetric $SU\left(
5\right) $ GUT-type model with a $SU\left( 7\right) $ enhanced gauge
symmetry is depicted in figure (\ref{74}).

\ \ \ \newline
The $SU\left( 7\right) $ symmetry at the vertices of the tetrahedron breaks
down to subgroups on the edges and the surfaces. The simplest \emph{4D} $%
\mathcal{N}=1$ supersymmetric $SU\left( 5\right) $ GUT- type model one
engineers from the $SU\left( 7\right) $ singularity involves the Yukawa
couplings (\ref{y7}). The chiral superfields \ configuration of the model
reads as:%
\begin{equation}
\begin{tabular}{ll|ll|ll}
\multicolumn{2}{l}{chiral superfields} &  & $SU\left( 5\right) \times
U^{2}\left( 1\right) $ &  & number \\ \hline\hline
{\small Matter like } & $\Phi _{M}$ &  & $\bar{5}_{7,1}$ &  & 4 \\
{\small Higgs like }$\ \ $ & $\Phi _{H}$ &  & $5_{0,-6}$ &  & 4 \\
{\small Neutino like } & $\Phi _{N}$ &  & $\bar{1}_{-7,5}$ &  & 4 \\ \hline
\end{tabular}%
\end{equation}%
These superfields follow from the decomposition of the adjoint
representation $\underline{48}$ of the enhanced gauge symmetry $SU\left(
7\right) $ living at the vertices of the tetrahedron in terms of
representations $SU\left( 5\right) \times U^{2}\left( 1\right) $ group as
shown below,
\begin{equation}
\begin{tabular}{llll}
$48$ & $=$ & $1_{0,0}\oplus 1_{0,0}\oplus 24_{0,0}$ &  \\
&  & $\oplus \left( 5_{0,-6}\oplus \bar{5}_{0,6}\right) \oplus \left(
5_{-7,-1}\oplus \bar{5}_{7,1}\right) \oplus \left( 1_{7,-5}\oplus \bar{1}%
_{-7,5}\right) $ & .%
\end{tabular}%
\end{equation}%
From this decomposition, we see that one can build several tri-coupling
gauge invariant terms; These are given by the following tri- couplings%
\begin{equation}
\begin{tabular}{llll}
$W_{1}=5_{0,-6}\times 1_{0,0}\times \bar{5}_{0,6}$ & , & $%
W_{3}=5_{0,-6}\times \bar{1}_{-7,5}\times \bar{5}_{7,1}$ & , \\
$W_{2}=5_{-7,-1}\times 1_{0,0}\times \bar{5}_{7,1}$ & , & $W_{4}=\bar{5}%
_{0,6}\times 1_{7,-5}\times 5_{-7,-1}$ & .%
\end{tabular}%
\end{equation}%
Notice that the superpotentials $W_{1}$ and $W_{2}$ involve, in addition to
two chiral superfields transforming into conjugates bi-fundamentals, an
adjoint bulk matter singlet. The superpotentials $W_{3}$ and $W_{4}$\
involve however only chiral matter in the bi-fundamentals.\newline
A typical $\mathcal{N}=1$ chiral superpotential that involve a Higgs like
superfield H$_{u}$, matter in the $\bar{5}$ and neutrino like superfields
reads as follows%
\begin{equation}
\int d^{2}\theta \text{ }W_{3}=\sum_{a=1}^{4}\lambda _{a}\int d^{2}\theta
\text{ }\Phi _{H}^{a}\Phi _{M}^{a}\Phi _{N}^{a}
\end{equation}%
where the $\lambda _{a}$s are coupling constants. Notice that along the
matter curve in the $5_{0,-6}$ and $\bar{5}_{7,1}$\ representations, the
bulk $SU\left( 5\right) \times U^{2}\left( 1\right) $ gauge symmetry gets
enhanced to $SU\left( 6\right) \times U\left( 1\right) $ which gets further
enhanced to $SU\left( 7\right) $ at the vertices. Along the matter curve
associated with $\bar{1}_{-7,5}$, the $SU\left( 5\right) $ singularity on
the surface gets enhanced to $SU\left( 5\right) \times SU\left( 2\right) $.

\subsubsection{$SO\left( 12\right) $ enhanced singularity}

\begin{figure}[tbph]
\begin{center}
\hspace{0cm} \includegraphics[width=8cm]{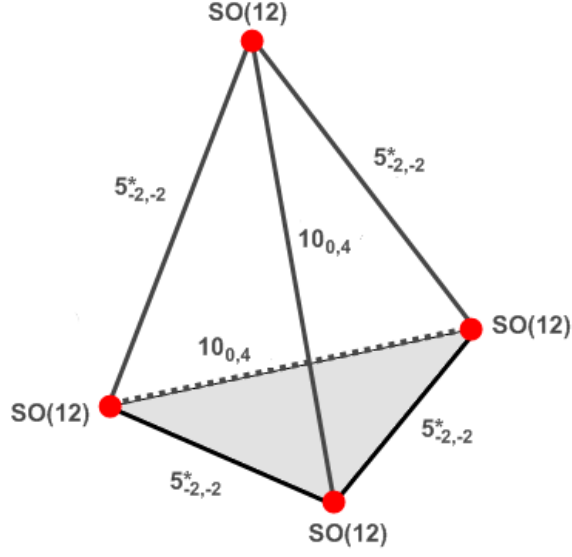}
\end{center}
\par
\vspace{-0.5 cm}
\caption{Quiver gauge diagram for SU$\left( 5\right) $ GUT- like model with $%
SO\left( 12\right) $ enhanced gauge symmetry at the vertices.}
\label{12}
\end{figure}

The quiver gauge diagram of the supersymmetric $SU\left( 5\right) $ GUT-type
model with an SO$\left( 12\right) $ enhanced singularity is depicted in
figure (\ref{12}). \newline
\ \ \newline
The chiral superfield configuration of this model reads as,%
\begin{equation}
\begin{tabular}{ll|ll|ll}
\multicolumn{2}{l}{chiral superfields} &  & $SU\left( 5\right) \times
U^{2}\left( 1\right) $ &  & number \\ \hline\hline
{\small Matter like } & $\Phi _{\bar{5}}$ &  & $\bar{5}_{-2,-2}$ &  & 4 \\
{\small Matter like } & $\Phi _{10}$ &  & $10_{0,4}$ &  & 4 \\
{\small Higgs like }$\ \ $ & $\Phi _{H}$ &  & $\bar{5}_{-2,-2}$ &  & 4 \\
\hline
\end{tabular}%
\end{equation}%
where now, we have both matter in the $\bar{5}$ and $10$ representations as
well as the Higgs $H_{d}$. These complex superfields follow from the
decomposition of the adjoint representation $\underline{66}$ of the two fold
enhanced $SO\left( 12\right) $ symmetry, living at the vertices of the
tetrahedron, in terms of representations $SU\left( 5\right) \times
U^{2}\left( 1\right) $
\begin{equation}
\begin{tabular}{llll}
$66$ & $=$ & $1_{0,0}+1_{0,0}+24_{0,0}$ &  \\
&  & $\left( 5_{2,2}+\bar{5}_{-2,-2}\right) +\left( 5_{-2,2}+\bar{5}%
_{2,-2}\right) +10_{0,4}+\overline{10}_{0,-4}$ & .%
\end{tabular}%
\end{equation}%
From this decomposition, we see that one can build several tri-coupling
gauge invariant terms; These are given by%
\begin{equation}
\begin{tabular}{llll}
$W_{1}^{\prime }=5_{2,2}\times 1_{0,0}\times \bar{5}_{-2,-2}$ & , & $%
W_{3}^{\prime }=\bar{5}_{-2,-2}\times \bar{5}_{-2,-2}\times 10_{0,4}$ & , \\
$W_{2}^{\prime }=10_{0,4}\times 1_{0,0}\times \overline{10}_{0,-4}$ & , & $%
W_{4}^{\prime }=5_{2,2}\times 5_{2,2}\times \overline{10}_{0,-4}$ & .%
\end{tabular}%
\end{equation}%
Like in the $SU\left( 7\right) $ case, here also the $SO\left( 12\right) $
gauge symmetry gets broken down to subgroups on the edges and the faces of
the tetrahedron. Moreover, the superpotentials $W_{1}^{\prime }$ and $%
W_{2}^{\prime }$ involve, in addition to two bi-fundamentals, an adjoint
singlet while $W_{3}^{\prime }$ and $W_{4}^{\prime }$ involve only matter in
the bi-fundamentals which is used to describe Yukawa couplings of GUT- like
models. The $\mathcal{N}=1$ chiral superpotential reads as follows%
\begin{equation}
\int d^{2}\theta \text{ }W_{3}^{\prime }=\sum_{a=1}^{4}\lambda _{a}^{\prime
}\int d^{2}\theta \text{ }\Phi _{H}^{a}\Phi _{\bar{5}}^{a}\Phi _{10}^{a}
\end{equation}%
where the $\lambda _{a}^{\prime }$ s are coupling constants. Notice that,
along the matter curves represented by the edges, the $SU\left( 5\right)
\times U\left( 1\right) $ gauge symmetry on the surface of the tetrahedron
gets enhanced to $SO\left( 10\right) \times U\left( 1\right) $ which in
turns gets further enhanced to $SO\left( 12\right) $ at the vertices.

\subsubsection{$E_{6}$ enhanced singularity}

\begin{figure}[tbph]
\begin{center}
\hspace{0cm} \includegraphics[width=8cm]{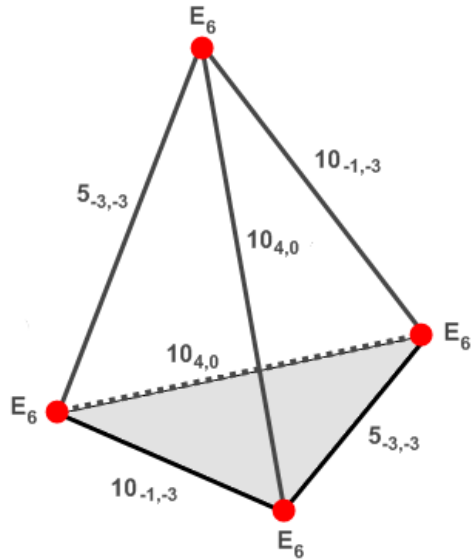}
\end{center}
\par
\vspace{-0.5 cm}
\caption{Quiver gauge diagram for SU$\left( 5\right) $ GUT- like model with $%
E_{6}$ enhanced gauge symmetry at the vertices \ and a $SU\left( 5\right) $
model involving $5\times 10\times 10$ tri-couplings.}
\label{E}
\end{figure}

The quiver gauge diagram of the supersymmetric $SU\left( 5\right) $ GUT-type
model with an $E_{6}$ enhanced singularity at the vertices of the
tetrahedron is depicted in the figure (\ref{E}),

\ \ \ \newline
The quiver gauge model has the following chiral superfield spectrum:%
\begin{equation}
\begin{tabular}{ll|ll|ll}
\multicolumn{2}{l}{chiral superfields} &  & $SU\left( 5\right) \times
U^{2}\left( 1\right) $ &  & number \\ \hline\hline
{\small Matter like } & $\Phi _{5}$ &  & $5_{-3,3}$ &  & 4 \\
{\small Matter like } & $\Phi _{10}$ &  & $10_{-1,-3}$ &  & 4 \\
{\small Matter like }$\ \ $ & $\Phi _{H}$ &  & $10_{4,0}$ &  & 4 \\ \hline
\end{tabular}%
\end{equation}%
These chiral superfields follow from the decomposition of the adjoint
representation $\underline{78}$ of the enhanced $E_{6}$ in terms of
representations $SU\left( 5\right) \times U^{2}\left( 1\right) $ namely,
\begin{equation}
\begin{tabular}{llll}
$78$ & $=$ & $1_{0,0}+1_{0,0}+24_{0,0}+$ &  \\
&  & $1_{5,3}+1_{-5,-3}+5_{-3,3}+\bar{5}_{3,-3}+10_{-1,-3}+\overline{10}%
_{1,3}+10_{4,0}+\overline{10}_{-4,0}$ & .%
\end{tabular}%
\end{equation}%
From this decomposition, we see that we can build several tri-coupling gauge
invariant terms; these are:%
\begin{equation}
\begin{tabular}{ll}
$W_{1}^{\prime \prime }=5_{-3,3}\times 1_{0,0}\times \bar{5}_{3,-3}$ & , \\
$W_{2}^{\prime \prime }=10_{-1,-3}\times 1_{0,0}\times \overline{10}_{1,3}$
& , \\
$W_{3}^{\prime \prime }=10_{4,0}\times 1_{0,0}\times \overline{10}_{-4,0}$ &
, \\
$W_{4}^{\prime \prime }=5_{-3,3}\times 10_{-1,-3}\times 10_{4,0}$ & , \\
$W_{5}^{\prime \prime }=\bar{5}_{3,-3}\times \overline{10}_{1,3}\times
\overline{10}_{-4,0}$ & .%
\end{tabular}%
\end{equation}%
Similarly as before, the superpotentials $W_{1}^{\prime \prime }$, $%
W_{2}^{\prime \prime }$ and $W_{3}^{\prime \prime }$ involve, besides two
bi-fundamentals, an adjoint singlet while $W_{4}^{\prime \prime }$ and $%
W_{5}^{\prime \prime }$ \ involve only matter in the bi-fundamentals.\newline
The $\mathcal{N}=1$ chiral superpotential describing the tri-coupling of the
matter in the bi-fundamentals is given by $W_{4}^{\prime \prime }$.
Moreover, along the matter curves in the tetrahedron, the $SU\left( 5\right)
\times U^{2}\left( 1\right) $ gauge symmetry on the surface of the
tetrahedron gets enhanced to $SO\left( 10\right) \times U\left( 1\right) $
which gets further enhanced to $E_{6}$ at the vertices.\newline
In the end of this section, notice that in these SU$\left( 5\right) $
GUT-type models based on tetrahedron, the gauge symmetry at the vertices is
of same nature. In what follows, we study other configurations where
different gauge symmetries live at the vertices of the tetrahedron. This
kind of quiver gauge models requires however performing blown ups of the
tetrahedron surface.

\section{GUT- like models on blown up Tetrahedron}

We start by giving further details on the blown up $\mathcal{T}_{n}$ on the
complex tetrahedral geometry $\mathcal{T}$; in particular the blown ups by
projective planes $\mathbb{P}^{2}$ at one and two vertices of $\Delta _{_{%
\mathcal{T}}}$. Then, we consider the building of $SU\left( 5\right) $ GUT-
type models that are embedded in F-theory on Calabi-Yau four-folds based on
these geometries.

\subsection{More on blown ups of tetrahedron}

Starting from the non planar tetrahedral surface $\mathcal{T}$ with its four
projective planar faces $S_{a}$, its six projective line edges $\Sigma
_{\left( ab\right) }$ and the four vertices $P_{\left( abc\right) }$, we can
perform blown ups of the tetrahedral surface $\mathcal{T}$ at a finite set
of points. Roughly, we distinguish:\newline
\textbf{(1)} blown ups at the four vertices of the tetrahedron $\Delta _{_{%
\mathcal{T}}}$,\newline
\textbf{(2)} blown ups at the edges of the tetrahedron\newline
(\textbf{3}) blown ups at a finite number of generic points of the
tetrahedron.\newline
In what follows, we will consider the first case of these blown ups and
illustrate the main idea by studying $SU\left( 5\right) $ GUT-type models
based on $\mathcal{T}_{1}$ and $\mathcal{T}_{2}$ geometries.

\subsubsection{Blown up at a vertex}

\begin{figure}[tbph]
\begin{center}
\hspace{0cm} \includegraphics[width=5cm]{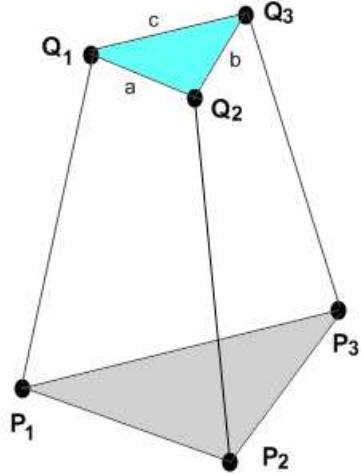}
\end{center}
\par
\vspace{-0.5 cm}
\caption{{\protect\small The toric graph representing a blown up of the
tetrahedral geometry. The vertex }$P_{4}${\protect\small \ has been replaced
by a projective plane }$\left[ {\protect\small Q}_{4}{\protect\small Q}%
_{5}Q_{6}\right] ${\protect\small .}}
\label{t1}
\end{figure}

Recall that the toric graph of the tetrahedron $\Delta _{_{\mathcal{T}}}$
has four vertices $P_{\left( abc\right) }$ where meet simultaneously\textrm{%
\footnote{%
Three projective planes meet as well at each vertex of the tetrahedron.}}
three projective lines $\Sigma _{\left( ab\right) }$, $\Sigma _{\left(
ac\right) }$ and $\Sigma _{\left( bc\right) }$. Starting from such a graph
and focusing on the fourth vertex $P_{4}$ of the figure (\ref{te}), the
blown up of this vertex $P_{4}$ by a projective plane amounts to replacing $%
P_{4}$ by a projective plane,
\begin{equation}
\begin{tabular}{llll}
$\text{point }P_{4}$ & $\longrightarrow $ & $\text{projective plane }\mathbb{%
P}^{2}$ & .%
\end{tabular}%
\text{ }
\end{equation}%
Since in toric geometry, a projective plane is described by a triangle, the
blown up of the vertex $P_{4}$ amounts to substitute this point by a
triangle $\left[ Q_{1}Q_{2}Q_{3}\right] $ as depicted in the figure (\ref{t1}%
).

\ \ \ \newline
The resulting toric geometry of the blown up tetrahedron at a vertex by a
projective plane, to which we shall refer below to as $\mathcal{T}_{1}$, has
five intersecting faces namely: \newline
\textbf{(1)} two complex projective planes with toric graphs given by the
triangles
\begin{equation}
\begin{tabular}{llll}
$\left[ P_{1}P_{2}P_{3}\right] $ & , & $\left[ Q_{1}Q_{2}Q_{3}\right] $ & ,%
\end{tabular}%
\end{equation}%
of the figure (\ref{t1}). As we see from this figure, these triangles have
no edge intersection.\newline
\textbf{(2)} three del Pezzo surfaces $dP_{1}$ with toric graphs given by
the quadrilaterals ,
\begin{equation}
\begin{tabular}{lllll}
$\left[ P_{1}P_{2}Q_{1}Q_{2}\right] $ & , & $\left[ P_{2}P_{3}Q_{2}Q_{3}%
\right] $ & , & $\left[ P_{1}P_{3}Q_{1}Q_{3}\right] $%
\end{tabular}%
\end{equation}%
Thinking about the three edges $\left[ Q_{1}Q_{2}\right] $, $\left[
Q_{2}Q_{3}\right] $, $\left[ Q_{1}Q_{3}\right] $ of the exceptional
triangles $\left[ Q_{1}Q_{2}Q_{3}\right] $ that generate the blown up of the
vertex $P_{4}$ as describing complex projective lines with the Kahler\textrm{%
\footnote{%
Notice that the projective plane has one Kahler parameter; it should not be
confused with the auxiliary parameters a, b and c. }} parameters,%
\begin{equation}
\begin{tabular}{llll}
$\left[ Q_{1}Q_{2}\right] $ & $\rightarrow $ & a & , \\
$\left[ Q_{2}Q_{3}\right] $ & $\rightarrow $ & b & , \\
$\left[ Q_{1}Q_{3}\right] $ & $\rightarrow $ & c & ,%
\end{tabular}%
\end{equation}%
and considering the singular limit of the geometry (\ref{t1}) where one or
two of these parameters are sent to zero, one recovers new "singular"
topologies of blown up of the tetrahedron $\Delta _{_{\mathcal{T}}}$. For
instance, putting $a=0$, and $b=c\neq 0$, the points $Q_{1}$ and $Q_{2}$
gets identified,
\begin{equation}
Q_{1}=Q_{2}\equiv Q_{0},
\end{equation}%
and so the triangle $\left[ Q_{1}Q_{2}Q_{3}\right] $ gets reduced to a
singular line%
\begin{equation}
\begin{tabular}{llll}
$\left[ Q_{1}Q_{2}Q_{3}\right] $ & $\rightarrow $ & $\left[ Q_{0}Q_{3}\right]
$ & .%
\end{tabular}%
\end{equation}%
Consequently, we get a degenerating blown up of the tetrahedron where the
vertex P$_{4}$ is replaced by the projective line $\left[ Q_{0}Q_{3}\right] $%
. The resulting geometry has three intersecting projective planes dP$_{0}$;
intersecting as well two del Pezzo surfaces dP$_{1}$. Notice that in the
special case where $a=b=c=0$, we recover obviously the standard tetrahedron $%
\Delta _{_{\mathcal{T}}}$.

\subsubsection{Blown up at two vertices}

\begin{figure}[tbph]
\begin{center}
\hspace{0cm} \includegraphics[width=6cm]{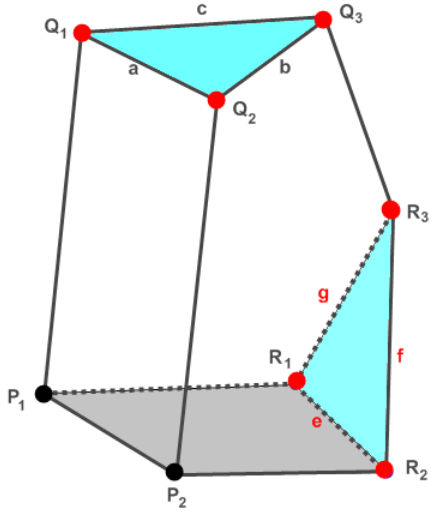}
\end{center}
\par
\vspace{-0.5 cm}
\caption{blown up of the tetrahedron at the two points $P_{3}$ and $P_{4}$
which have been replaced by the projective planes with toric graphs given by
the triangles {\protect\small \ }$\left[ {\protect\small R}_{1}%
{\protect\small R}_{2}R_{3}\right] $ and $\left[ {\protect\small Q}_{1}%
{\protect\small Q}_{2}Q_{3}\right] $ respectively{\protect\small .}}
\label{lbd}
\end{figure}

The blown up of the tetrahedron $\Delta _{_{\mathcal{T}}}$ at two vertices,
say P$_{3}$ and P$_{4}$ of the \textrm{figure (\ref{te}}), is achieved by
replacing these two points by projective planes. In toric graph language,
this amounts to replace P$_{3}$ and P$_{4}$ by the triangles,%
\begin{equation}
\begin{tabular}{llllllll}
P$_{3}$ & $\rightarrow $ & $\left[ R_{1}R_{2}R_{3}\right] $ & , & P$_{4}$ & $%
\rightarrow $ & $\left[ Q_{1}Q_{2}Q_{3}\right] $ & .%
\end{tabular}%
\end{equation}%
The toric graph of the two blown up tetrahedron is depicted in the figure (%
\ref{lbd}).

\ \ \ \newline
The obtained surface, denoted as $\mathcal{T}_{2}$, has six intersecting
faces namely: \newline
\textbf{(1)} two projective planes with toric graphs given by the triangles
of the figure (\ref{lbd}) namely $\left[ R_{1}R_{2}R_{3}\right] $ and $\left[
Q_{1}Q_{2}Q_{3}\right] .$\newline
\textbf{(2)} two del Pezzo surfaces $dP_{1}$ with toric graphs given by the
quadrilaterals $\left[ P_{1}P_{2}R_{1}R_{2}\right] $ and $\left[
P_{1}P_{2}Q_{1}Q_{2}\right] $.\newline
\textbf{(3)} two del Pezzo surfaces $dP_{2}$ with toric graphs given by the
pentagons $\left[ P_{1}R_{1}R_{3}Q_{3}Q_{1}\right] $ and $\left[
P_{2}R_{2}R_{3}Q_{3}Q_{2}\right] $.\newline
Similarly as in the previous case, one can recovers new singular topologies
of the blown tetrahedron (\ref{lbd}) by taking singular limits of the Kahler
parameters a, b, c , e, f and g. The case where $e=f=g=0$ leads to the
figure (\ref{t1}) and the case where all these parameters are set to zero
gives the standard tetrahedron.

\subsection{$SU\left( 5\right) $ GUT model on $\mathcal{T}_{1}$ and $%
\mathcal{T}_{2}$}

In this subsection, we engineer various unrealistic $SU\left( 5\right) $
GUT-type models that are embedded in consider F-theory on local elliptic K3
fibered Calabi Yau four- folds based on the surfaces $\mathcal{T}_{1}$ and $%
\mathcal{T}_{2}$. We first construct GUT -type models based on $\mathcal{T}%
_{1}$ and then we build other models based on $\mathcal{T}_{2}$.

\subsubsection{$SU\left( 5\right) $ GUT type models on $\mathcal{T}_{1}$}

\begin{figure}[tbph]
\begin{center}
\hspace{0cm} \includegraphics[width=12cm]{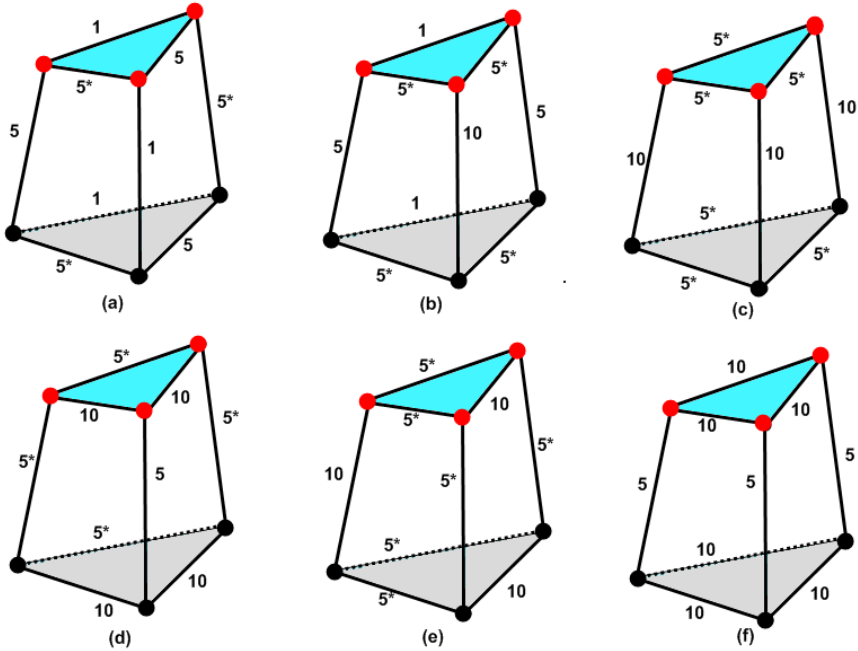}
\end{center}
\par
\vspace{-0.5 cm}
\caption{tetrahedron models: On top models involving chiral matter in
singlets, $5$, $5^{\ast }$, $10$. On bottom $SU\left( 5\right) $ models
involving $5\times 10\times 10$ and $5^{\ast }\times 5^{\ast }\times 10$
vertices.}
\label{sa}
\end{figure}

The toric graph of the complex surface $\mathcal{T}_{1}$ is given by the
figure (\ref{t1}); the fix points of the toric action encode data on the
seven brane intersections with the following features:\newline
(\textbf{1}) $\mathcal{T}_{1}$ has five faces where live the \emph{4D} $%
\mathcal{N}=1$ supersymmetric gauge theory with bulk gauge symmetry $%
SU\left( 5\right) \times U\left( 1\right) \times U\left( 1\right) $ where
the extra factor $U^{2}\left( 1\right) $ is the toric symmetry in the
2-torus in the toric surface $\mathcal{T}_{1}$.\newline
\textbf{(2)} $\mathcal{T}_{1}$ has nine edges where localize matter in the
fundamental and antisymmetric representations of the $SU\left( 5\right) $
gauge symmetry. On these curves, the rank of the gauge invariance gets
enhanced by one.\newline
\textbf{(3)} $\mathcal{T}_{1}$ has six vertices where live tri-fields Yukawa
couplings and where the gauge symmetry gets enhanced to $SU\left( 7\right) $%
, or $SO\left( 12\right) $ or also $E_{6}$.\newline
Now using the fact that at the vertices of the surface $\mathcal{T}_{1}$,
the tri- fields interactions should be gauge invariant under the gauge group
$SU\left( 5\right) \times U^{2}\left( 1\right) $, one can engineer various
gauge invariant configurations; in particular the ones depicted in the
figures (\ref{sa}),

To engineer SU$\left( 5\right) $ GUT- type gauge invariant models with
different Yukawa couplings, we use the following relations,

\begin{equation}
\begin{tabular}{lllll}
Yukawa tri-fields couplings &  &  &  & enhanced singularity at vertices \\
\hline\hline
$1\otimes 5\otimes \bar{5}$ &  & $\rightarrow $ &  & $\qquad SU\left(
7\right) $ \\
$\bar{5}\otimes \bar{5}\otimes 10$ &  & $\rightarrow $ &  & $\qquad SO\left(
12\right) $ \\
$5\otimes 5\otimes 10$ &  & $\rightarrow $ &  & $\qquad E_{6}$ \\ \hline
\end{tabular}%
\end{equation}

\ \ \ \newline
to choose the kind of the ADE singularity one has to put in the fiber over
each vertex of the base surface $\mathcal{T}_{1}$. Let us illustrate the
idea by describing the examples depicted in the figures (\ref{sa}).\newline
The six toric vertices of $\mathcal{T}_{1}$ of the figure (\ref{sa}-a)
involves the tri-coupling $1\otimes 5\otimes \bar{5}$ and so have a SU$%
\left( 7\right) $ enhanced singularity. In the figure (\ref{sa}-b), four
toric vertices have a SU$\left( 7\right) $ singularity and the two others
have a SO$\left( 12\right) $ one since the tri-couplings are given by
\begin{equation}
\bar{5}\otimes \bar{5}\otimes 10.
\end{equation}
The six toric vertices of the figure (\ref{sa}-c) have all of them an SO$%
\left( 12\right) $ enhanced singularity. Using the same philosophy, four
toric vertices of the figure (\ref{sa}-d) have an SO$\left( 12\right) $
enhanced singularity and the two others are of type
\begin{equation}
5\otimes 5\otimes 10
\end{equation}%
and so are associated with an E$_{6}$ singularity. Finally, all the six
toric vertices of the figure (\ref{sa}-f) are of E$_{6}$ type while the tri-
fields couplings given by the figure (\ref{sa}-e) are equivalent to those of
the figure (\ref{sa}-c).

\subsubsection{$SU\left( 5\right) $ GUT type models on $\mathcal{T}_{2}$}

The toric graph of the surface $\mathcal{T}_{2}$ is given by the figure (\ref%
{lbd}); it has:\newline
(\textbf{1}) Six toric faces where localize a \emph{4D} $\mathcal{N}=1$
supersymmetric gauge theory with $SU\left( 5\right) \times U^{2}\left(
1\right) $ gauge symmetry. These faces are given by del Pezzo surfaces of
different types:\newline
(\textbf{a}) two isolated $dP_{0}$'s; each one intersects a surface $dP_{1}$
and two surfaces $dP_{2}$,\newline
(\textbf{b}) two intersecting $dP_{1}$'s, each one of these $dP_{1}$'s
intersects a $dP_{1}$ and two $dP_{2}$\newline
(\textbf{c}) two intersecting dP$_{2}$'s, each one of these $dP_{2}$'s
intersects the two $dP_{0}$'s and the two $dP_{1}$'s.\newline
\textbf{(2)} Thirteen toric edges describing the intersections of the del
Pezzo surfaces. On these curves localize matter in the singlet, the
fundamentals and the antisymmetric representations of $SU\left( 5\right) $.
For the last representations, the gauge symmetry gets enhanced either to $%
SU\left( 6\right) \times U\left( 1\right) $ or to $SO\left( 10\right) \times
U\left( 1\right) $.\newline
\textbf{(3)} Eight toric vertices where live Yukawa couplings and the
enhanced gauge singularity. Each of these vertices is associated with the
intersection of three edges and it localizes tri- fields Yukawa coupling.

\emph{Model I} \newline
Now using the same approach as for the surface $\mathcal{T}_{1}$, we can
engineer various gauge invariant configurations.

\begin{figure}[tbph]
\begin{center}
\hspace{0cm} \includegraphics[width=8cm]{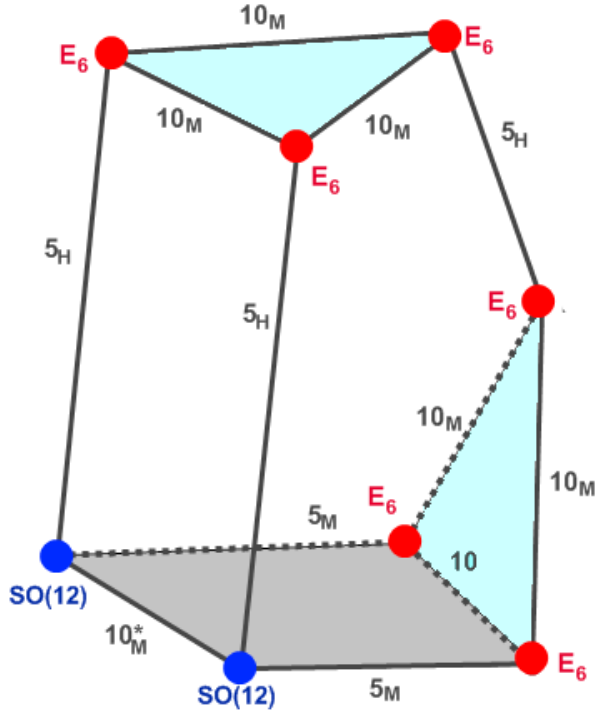}
\end{center}
\par
\vspace{-0.5 cm}
\caption{$SU\left( 5\right) $ GUT- type model based on the $\mathcal{T}_{2}$
geometry. This model has six vertices with an $E_{6}$ singularity and two
others with an $SO\left( 12\right) $ one.}
\label{trz}
\end{figure}

\ \ \newline
One of these configurations, depicted in the figure (\ref{trz}), involves
eight Yukawa couplings: six Yukawa couplings of type $5\otimes 5\otimes 10$
and two Yukawa couplings of type $5\otimes 5\otimes \overline{10}$.

\begin{equation}
\begin{tabular}{ll|lll|l}
{\small Yukawa couplings} &  &  & {\small Singularity} &  & {\small number
of vertices} \\ \hline\hline
\begin{tabular}{l}
$5\otimes 5\otimes 10$ \\
$\bar{5}\otimes \bar{5}\otimes \overline{10}$%
\end{tabular}
&  &  & $E_{6}$ &  &
\begin{tabular}{l}
$6$ \\
$0$%
\end{tabular}
\\ \hline
\begin{tabular}{l}
$5\otimes 5\otimes \overline{10}$ \\
$\bar{5}\otimes \bar{5}\otimes 10$%
\end{tabular}
&  &  & $SO\left( 12\right) $ &  &
\begin{tabular}{l}
$2$ \\
$0$%
\end{tabular}
\\ \hline
\begin{tabular}{l}
$5\otimes \bar{5}\otimes 1$%
\end{tabular}
&  &  & $SU\left( 7\right) $ &  & $%
\begin{tabular}{l}
$0$%
\end{tabular}%
$ \\ \hline
\end{tabular}%
\end{equation}
An equivalent configuration is also given by the conjugate representations.

\begin{figure}[tbph]
\begin{center}
\hspace{0cm} \includegraphics[width=8cm]{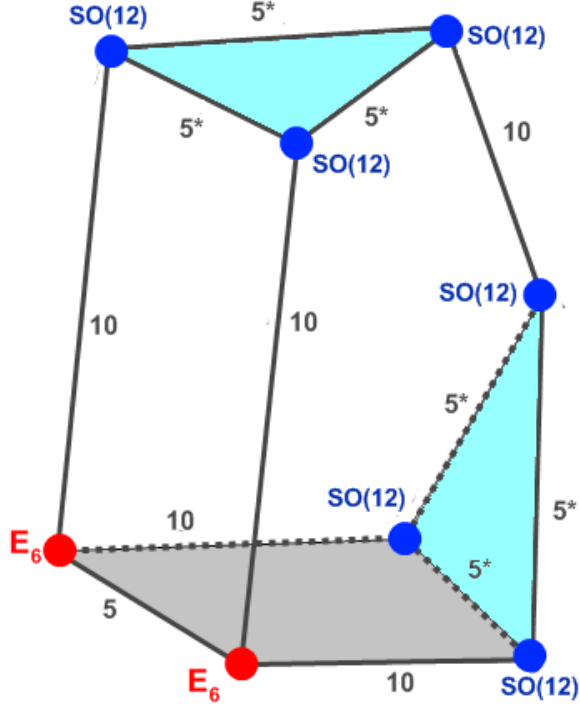}
\end{center}
\par
\vspace{-0.5 cm}
\caption{$SU\left( 5\right) $ GUT- like model based on the $\mathcal{T}_{2}$
geometry; it has five $E_{6}$ vertices and two $SO\left( 12\right) $
vertices.}
\label{tz}
\end{figure}

\emph{Model II}\ \newline
This SU$\left( 5\right) $ GUT- type model is the dual of the previous one.
This duality is in the sense that six of the eight vertices have an SO$%
\left( 12\right) $ singularity while the two remaining others have an E$_{6}$
one. We distinguish two class of models depending on the intersecting
surfaces and intersecting edges. We will refer to these models as IIa and
IIb:\newline
\emph{Model IIa}: In this model, the Yukawa couplings are depicted in the
figure (\ref{tz})

\begin{figure}[tbph]
\begin{center}
\hspace{0cm} \includegraphics[width=8cm]{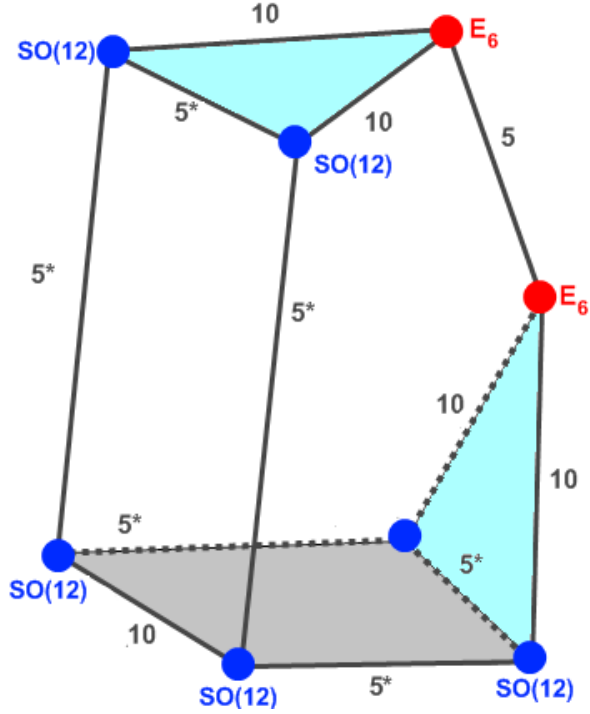}
\end{center}
\par
\vspace{-0.5 cm}
\caption{$SU\left( 5\right) $ GUT- like model based on the $\mathcal{T}_{2}$
geometry; it has two $E_{6}$ vertices and six $SO\left( 12\right) $
vertices. }
\label{tzz}
\end{figure}

\ \ \ \ \ \newline
From the figure (\ref{tz}), we see that the vertices are as in the following
table,
\begin{equation}
\begin{tabular}{ll|lll|l}
{\small Yukawa couplings} &  &  & {\small Singularity} &  & {\small number
of vertices} \\ \hline\hline
\begin{tabular}{l}
$5\otimes 5\otimes 10$ \\
$\bar{5}\otimes \bar{5}\otimes \overline{10}$%
\end{tabular}
&  &  & $E_{6}$ &  &
\begin{tabular}{l}
$2$ \\
$0$%
\end{tabular}
\\ \hline
\begin{tabular}{l}
$5\otimes 5\otimes \overline{10}$ \\
$\bar{5}\otimes \bar{5}\otimes 10$%
\end{tabular}
&  &  & $SO\left( 12\right) $ &  &
\begin{tabular}{l}
$6$ \\
$0$%
\end{tabular}
\\ \hline
\begin{tabular}{l}
$5\otimes \bar{5}\otimes 1$%
\end{tabular}
&  &  & $SU\left( 7\right) $ &  & $%
\begin{tabular}{l}
$0$%
\end{tabular}%
$ \\ \hline
\end{tabular}
\label{tab}
\end{equation}%
We learn also that the two $E_{6}$ and the six $SO\left( 12\right) $
vertices are given by the intersections of three del Pezzo surfaces as given
below,%
\begin{equation}
\begin{tabular}{llllllll}
$6\times E_{6}$ & : & $dP_{1}^{\left( 1\right) }$ & $\cap $ & $%
dP_{1}^{\left( 2\right) }$ & $\cap $ & $dP_{2}$ & , \\
$2\times SO\left( 12\right) $ & : & $dP_{0}$ & $\cap $ & $dP_{1}$ & $\cap $
& $dP_{2}$ & ,%
\end{tabular}%
\end{equation}%
where $dP_{1}^{\left( 1\right) }$ and $dP_{1}^{\left( 2\right) }$\ are the
two del Pezzo surfaces of the blown up surface $\mathcal{T}_{2}$.\newline
\emph{Model IIb}. \qquad In this model, the configuration of the Yukawa
couplings, depicted in the figure (\ref{tzz}), are as in the table (\ref{tab}%
)

\ \newline
The two vertices with $E_{6}$ singularity and the other six vertices with $%
SO\left( 12\right) $ singularity are given by the following intersections,%
\begin{equation}
\begin{tabular}{llllllll}
$2\times E_{6}$ & : & $dP_{0}$ & $\cap $ & $dP_{2}^{\left( 1\right) }$ & $%
\cap $ & $dP_{2}^{\left( 2\right) }$ & , \\
$6\times SO\left( 12\right) $ & : & $dP_{0}$ & $\cap $ & $dP_{1}$ & $\cap $
& $dP_{2}$ & ,%
\end{tabular}%
\end{equation}%
where $dP_{2}^{\left( 1\right) }$ and $dP_{2}^{\left( 2\right) }$\ stand for
the two del Pezzo surfaces involved in the blown up tetrahedron $\mathcal{T}%
_{2}$.

\begin{figure}[tbph]
\begin{center}
\hspace{0cm} \includegraphics[width=8cm]{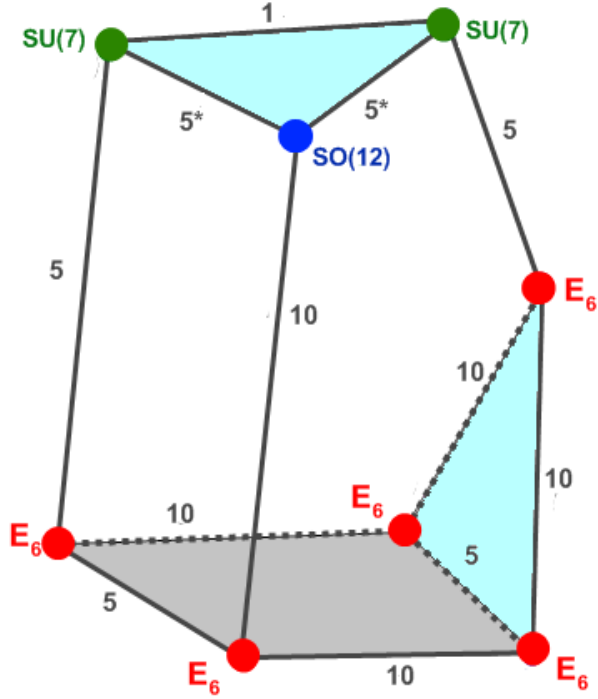}
\end{center}
\par
\vspace{-0.5 cm}
\caption{$SU\left( 5\right) $ GUT- like model based on the $\mathcal{T}_{2}$
geometry with five $E_{6}$ vertices, two $SU\left( 7\right) $ and one $%
SO\left( 12\right) $.}
\label{tx}
\end{figure}

\emph{Model III}\ \newline
In this model, the Yukawa couplings are depicted in the figure (\ref{tx}),

\ \ \newline
This model involves the following tri-fields interactions:
\begin{equation}
\begin{tabular}{ll|lll|l}
{\small Yukawa couplings} &  &  & {\small Singularity} &  & {\small number
of vertices} \\ \hline\hline
\begin{tabular}{l}
$5\otimes 5\otimes 10$ \\
$\bar{5}\otimes \bar{5}\otimes \overline{10}$%
\end{tabular}
&  &  & $E_{6}$ &  &
\begin{tabular}{l}
$5$ \\
$0$%
\end{tabular}
\\ \hline
\begin{tabular}{l}
$5\otimes 5\otimes \overline{10}$ \\
$\bar{5}\otimes \bar{5}\otimes 10$%
\end{tabular}
&  &  & $SO\left( 12\right) $ &  &
\begin{tabular}{l}
$1$ \\
$0$%
\end{tabular}
\\ \hline
\begin{tabular}{l}
$5\otimes \bar{5}\otimes 1$%
\end{tabular}
&  &  & $SU\left( 7\right) $ &  & $%
\begin{tabular}{l}
$2$%
\end{tabular}%
$ \\ \hline
\end{tabular}%
\end{equation}%
The five vertices with an $E_{6}$ singularity is given by two kinds of tri-
intersection of the del Pezzo surfaces. One vertex is given by the tri-
intersection of a projective plane with two del Pezzo surfaces dP$_{2}$
while the five others are given by the intersection of a projective line and
the del Pezzo surfaces $dP_{1}$ and $dP_{2}$:%
\begin{equation}
\begin{tabular}{llllllll}
$1\times E_{6}$ & : & $dP_{0}$ & $\cap $ & $dP_{2}^{\left( 1\right) }$ & $%
\cap $ & $dP_{2}^{\left( 2\right) }$ & , \\
$5\times E_{6}$ & : & $dP_{0}$ & $\cap $ & $dP_{1}$ & $\cap $ & $dP_{2}$ & ,%
\end{tabular}%
\end{equation}%
Regarding the vertex with an $SO\left( 12\right) $ singularity, we have:%
\begin{equation}
\begin{tabular}{llllllll}
$1\times SO\left( 12\right) $ & : & $dP_{0}$ & $\cap $ & $dP_{1}$ & $\cap $
& $dP_{2}$ & ,%
\end{tabular}%
\end{equation}%
and for the two vertices with a $SU\left( 7\right) $ singularity, the tri-
intersections are as follows:%
\begin{equation}
\begin{tabular}{llllllll}
$1\times SU\left( 7\right) $ & : & $dP_{0}$ & $\cap $ & $dP_{1}$ & $\cap $ &
$dP_{2}$ & , \\
$1\times SU\left( 7\right) $ & : & $dP_{0}$ & $\cap $ & $dP_{2}^{\left(
1\right) }$ & $\cap $ & $dP_{2}^{\left( 2\right) }$ & .%
\end{tabular}%
\end{equation}

\section{Conclusion}

In this paper we have studied a class of 4D $\mathcal{N}=1$ supersymmetric
quiver gauge models that describe gauge theory limits of 12D F-theory
compactification on local tetrahedron. In these supersymmetric models; we
have mainly focused on GUT-type gauge symmetries in particular on the $%
SU\left( 5\right) $ symmetry with the SO$\left( 12\right) $, SU$\left(
7\right) $ and E$_{6}$ gauge enhancements. These 4D\ gauge\ models should be
thought of as a first step for building non minimal supersymmetric GUT- type
models along the line of the BHV theory. The other steps are to require the
conditions for realistic supersymmetric GUT models building such as GUT
breaking and doublet/triplet splitting via hypercharge flux, the absence of
bare $\mu $ term and dangerous dimension 4 proton decay operators. \newline
Like in BHV theory, our quiver gauge models are based on using seven branes
wrapping 4-cycles in the framework of twelve dimensional F-theory
compactified on local elliptic K3 fibered Calabi Yau four- folds
\begin{equation}
\begin{tabular}{llll}
Y & $\rightarrow $ & $X_{4}$ &  \\
&  & $\downarrow \pi _{_{S}}$ &  \\
&  & $S$ &
\end{tabular}%
\end{equation}%
where now the base surface $S$ is given by the complex tetrahedral surface $%
\mathcal{T}$ and its blown ups $\mathcal{T}_{n}$. The relation between $%
\mathcal{T}_{n}$ and $\mathcal{T}$\ should be thought of in the same manner
the del Pezzo surfaces $dP_{n}$ are linked to the complex projective plane.
In fact, the complex surfaces $dP_{n}$ are particular sub- geometries of $%
\mathcal{T}_{m}$s; a property which make these CY4- manifolds somehow
extending the local geometry used in the BHV theory.\newline
The engineering of the non abelian gauge symmetry that is visible at the
level of the \emph{4D} $\mathcal{N}=1$ supersymmetric effective GUT model is
achieved through singularities in the K3 fiber of the local CY4- folds $%
X_{4} $. \newline
In the complex base surface $S$, it generally lives a non abelian rank $r$
bulk gauge symmetry $G_{r}$. This gauge invariance gets enhanced on the
matter curves along which seven branes intersect to $G_{r+1}\supset G_{r}$.
It gets further enhanced to $G_{r+2}\supset G_{r+1}$ at isolated points of $%
S $ where matter curves meet and where live tri-fields Yukawa couplings. The
three gauge groups satisfy the embedding property
\begin{equation}
\begin{tabular}{llllll}
$G_{r+2}$ & $\supset $ & $G_{r+1}\times U\left( 1\right) $ & $\supset $ & $%
G_{r}\times U\left( 1\right) \times U\left( 1\right) $ & .%
\end{tabular}%
\end{equation}%
In the 4D $\mathcal{N}=1$ supersymmetric $SU\left( 5\right) $ GUT-like
models, these gauge symmetries should be thought of as follows%
\begin{equation}
\begin{tabular}{ll|lll|ll}
\begin{tabular}{llll}
&  & $G_{r+2}$ &
\end{tabular}
&  &  &
\begin{tabular}{lll}
& $G_{r+1}$ &
\end{tabular}
&  &  & $G_{r}$ \\ \hline
\begin{tabular}{lll}
$E_{6},$ & $SO\left( 12\right) ,$ & $SU\left( 7\right) $%
\end{tabular}
&  &  &
\begin{tabular}{ll}
$SO\left( 10\right) ,$ & $SU\left( 6\right) $%
\end{tabular}
&  &  & $SU\left( 5\right) $ \\ \hline
\end{tabular}%
\end{equation}%
The decomposition of the adjoint representation of $G_{r+2}$ in terms of
representations of the $G_{r}\times U\left( 1\right) \times U\left( 1\right)
$ allows to generate chiral matter in representations other than the adjoint
ones. In our construction, the extra abelian $U\left( 1\right) \times
U\left( 1\right) $ gauge invariance is interpreted in terms of symmetries of
the toric fiber of the complex base surface. Recall that complex surfaces $S$
exhibits a natural toric fibration,
\begin{equation}
\begin{tabular}{lll}
$\mathbb{T}^{2}$ & $\longrightarrow $ & $S$ \\
&  & $\downarrow \pi _{_{S}}$ \\
&  & $B_{S}$%
\end{tabular}%
\end{equation}%
where $\mathbb{T}^{2}$ is the usual fiber of the toric geometry and $B_{S}$
is a real two dimension base which is nicely represented by a toric graph $%
\Delta _{S}$. In the case of the complex tetrahedral surface $\mathcal{T}$,
the corresponding toric graph $\Delta _{_{\mathcal{T}}}$ is given by the
tetrahedron of the figure (\ref{by}). The toric fibration has remarkable
shrinking features on the edges of the tetrahedron and at the vertices.%
\newline
Using the power of toric geometry in the complex base $\mathcal{T}$ and the
degeneracy of its torus fibration, we have engineered a class of $SU\left(
5\right) $ GUT-like models based on local tetrahedron $\mathcal{T}$ and the
two first elements of its blow ups family $\mathcal{T}_{n}$. These $SU\left(
5\right) $ GUT-type models building extend naturally for generic 4D $%
\mathcal{N}=1$ supersymmetric quiver gauge theories that are embedded in
F-theory on local CY4- folds based on blown ups of the tetrahedron; in
particular for the interesting class of GUT- type models using flipped $%
SU\left( 5\right) $ and $SO\left( 10\right) $ gauge symmetries.\newline
In the end of this conclusion, we would like to emphasize that our interest
in GUT-type models buildings based on the complex tetrahedral surface and
its toric blown ups has been motivated by a number of remarkable features;
in particular the two following:\newline
(\textbf{1}) there is an intimate link between the complex tetrahedral
surface and the projective plane $dP_{0}$. The tetrahedron is precisely
given by the four projective planes $dP_{0}^{\left( 1\right) }$, $%
dP_{0}^{\left( 2\right) }$, $dP_{0}^{\left( 3\right) }$\ and $dP_{0}^{\left(
4\right) }$ describing the basic divisors of the complex three dimension
projective space $\mathbb{P}^{3}$ while its blow ups are given by a union of
the del Pezzo surfaces. \newline
For the case of the blow up of $\mathcal{T}$ at a vertex by a projective
plane, we have
\begin{equation}
\begin{tabular}{llll}
$\mathcal{T}_{1}$ & $=$ & $dP_{0}^{\left( 1\right) }\cup dP_{0}^{\left(
2\right) }\cup dP_{0}^{\left( 3\right) }\cup dP_{0}^{\left( 4\right) }\cup
dP_{0}^{\left( 5\right) }$ & , \\
$\Sigma _{ij}$ & $=$ & $dP_{0}^{\left( i\right) }\cap dP_{0}^{\left(
j\right) }$ & , \\
$P_{ijk}$ & $=$ & $dP_{0}^{\left( i\right) }\cap dP_{0}^{\left( j\right)
}\cap dP_{0}^{\left( k\right) }$ & ,%
\end{tabular}%
\end{equation}%
where the complex curves $\Sigma _{ij}$ stand for the nine edges of $%
\mathcal{T}_{1}$ and the six isolated points $P_{ijk}$ for its vertices.
These intersections can be read from eqs(\ref{C4}-\ref{C5}) or directly from
their toric graphs given by the respective figures (\ref{te}) and (\ref{t1}%
). From this view, the blown ups of the tetrahedron contain several copies
of del Pezzo surfaces $dP_{n}^{\left( i\right) }$ as special components on
which may be engineered the \textrm{BHV} theory. Recall that the del Pezzo
complex surfaces $dP_{n}$ play a central role in the BHV theory for F-theory
GUT models building. These complex $dP_{n}$s are strongly linked to the
projective plane $\mathbb{P}^{2}$ since they are precisely given by its
blown ups at eight isolated points,%
\begin{equation}
\begin{tabular}{llllll}
$dP_{0}=\mathbb{P}^{2}$ & , & $dP_{n}$ & , & $n=1,...,8$ & ,%
\end{tabular}%
\end{equation}%
(\textbf{2}) the special degeneracy properties of the fiber $\mathbb{T}^{2}$
of the toric fibration of the tetrahedron $\mathcal{T}\sim \Delta _{_{%
\mathcal{T}}}$ $\times $ $\mathbb{T}^{2}$. The 1- cycle shrinking loci of
the 2- torus fiber down $\mathbb{S}_{_{\Sigma }}^{1}$ allows to host in a
natural way the engineering of the seven branes intersections along the
edges. Moreover, the shrinking of the $\mathbb{T}^{2}$ fiber down to zero
allows to engineer Yukawa couplings at the vertices.\newline
In a future occasion, we give further refinements of this construction and
seek for non minimal quasi-realistic F-theory-GUT models building based on
local tetrahedral geometry.

\begin{acknowledgement}
\qquad {\small \ \ \ }\newline
{\small This research work is supported by the program Protars III D12/25.}
\end{acknowledgement}

\section{Appendix}

To engineer chiral matter transforming in representations of the gauge
invariance G$_{S}$ other than the \emph{adjG}$_{S}$, we have two ways:
either by switching on a gauge bundle E with structure group $H_{S}$ that
breaks the gauge symmetry like $G_{S}\rightarrow H_{S}\times G$, or modify
the base geometry $S$ of the local Calabi-Yau 4- folds into a larger surface
containing at least two intersecting 4- cycles $S_{a}$ and $S_{b}$ like,%
\begin{equation}
S=S_{a}\cup S_{b}\qquad ,\qquad S_{a}\cap S_{b}=\Sigma _{ab}\neq \emptyset ,
\label{int}
\end{equation}%
with $\Sigma _{ab}$\ standing for a intersecting complex curve where
localize bi-fundamental matter. In this way the bulk gauge symmetry $G_{S}$
gets broken down like $G_{S}\rightarrow G_{S_{a}}\times G_{S_{b}}$. To make
an idea on how these deformations work, we review below the key idea behind
these methods.\newline
In the case of deformation by switching on fluxes, the adjoint
representation $ad\left( G_{S}\right) $ decomposes as%
\begin{equation}
adG_{S}=\left( adH_{S},1\right) \dbigoplus \left( 1,adG\right) \dbigoplus
\left[ \dbigoplus_{i}\left( \rho _{i},U_{i}\right) \right] ,
\end{equation}%
where $\rho _{i}$ and $U_{i}$ stand respectively for representations of $%
H_{S}$ and $G$ respectively and where
\begin{equation}
\dim \left[ adG_{S}\right] =\dim \left[ adH_{S}\right] +\dim \left[ adG%
\right] +\sum_{i}\left( \dim \rho _{i}\right) \times \left( \dim
U_{i}\right) .
\end{equation}%
The switching of the bundle E induces then a deformation in the complex
surface $S$ and may be interpreted as splitting the winding of the bulk
seven branes wrapping $S$ into two intersecting stacks; one stack, to which
we refer as matter brane, with gauge symmetry $H_{S}$ and the second stack
with gauge invariance $G$. Along the intersection of the two stacks of seven
branes (matter and bulk), which corresponds geometrically to a curve $\Sigma
$ in $S$, the gauge symmetry is obviously given by $G_{S}$; but outside $%
\Sigma $, the symmetry is $H_{S}\times G$. \newline
The number $N_{i}$ of chiral fields $\phi _{i}$ (resp. $N_{i}^{\ast }$ of
anti-chiral fields $\phi _{i}^{\ast }$ ) transforming in the representation $%
U_{i}$ (resp. $U_{i}^{\ast }$) of the subgroup $G$ is determined by the
bundle-valued Euler characteristics,%
\begin{equation}
N_{i}=\mathcal{\chi }_{S}\left( \mathcal{R}_{i}\right) \qquad ,\qquad
N_{i}^{\ast }=\mathcal{\chi }_{S}\left( \mathcal{R}_{i}^{\ast }\right)
\end{equation}%
where $\mathcal{R}_{i}$ and $\mathcal{R}_{i}^{\ast }$ denote the bundles
transforming in $U_{i}$ and $U_{i}^{\ast }$ respectively. On the del Pezzo
surface dP$_{8}$, the numbers $N_{i}$ and $N_{i}^{\ast }$ are easily
computed by help of the relation
\begin{equation}
\mathcal{\chi }_{S}\left( \mathcal{R}\right) =1-\frac{1}{2}\Omega
_{S}.c_{1}\left( \mathcal{R}\right) +\frac{1}{2}c_{1}\left( \mathcal{R}%
\right) .c_{1}\left( \mathcal{R}\right)
\end{equation}%
where $\Omega _{S}$ denotes the canonical class of $S$. \newline
Notice that this analysis is particularly interesting when the gauge
subgroup $H_{S}$ is abelian; that is an $U^{r_{0}}\left( 1\right) $ abelian
subgroup of the Cartan subalgebra of $G_{S}$. In this case, the deformation
by fluxes has a nice geometric description in terms of deformation of of the
ADE singularity. For instance, by taking as a bulk gauge symmetry $%
G_{S}=SU\left( N+1\right) $ which is described by the local geometry of the
fiber
\begin{equation}
u^{2}+v^{2}+z^{N+1}=0
\end{equation}%
and which represents a bulk brane wrapping the surface N times, the
switching of a $U\left( 1\right) $ gauge bundle yields the deformation%
\begin{equation}
u^{2}+v^{2}+z^{N}\left( z-t\right) =0.
\end{equation}%
Here $t$ is a non zero complex number behaving as $z$ and represents a non
zero vev of a scalar Higgs field in the adjoint. Under this deformation, the
original bulk stacks of wrapped seven branes at $z=0$ gets split to two
stacks: one at $z=0$, with gauge symmetry $SU\left( N\right) $ and the other
with gauge invariance $U\left( 1\right) $ at%
\begin{equation}
z=t,\qquad t\in \mathbb{C}.
\end{equation}%
These two stacks intersect along the curve $\left\{ z=0\right\} \cap \left\{
z=t\right\} $ where gauge symmetry gets enhanced to $SU\left( N+1\right) $.
\newline
On this particular example, which applies as well to D7 seven branes of type
IIB superstring, one can also read the bi-fundamental matter by decomposing
the adjoint of $U\left( N+1\right) $ $=$ $U_{S}\left( 1\right) $ $\times $ $%
SU\left( N+1\right) $, describing the gauge symmetry of $\left( N+1\right) $
parallel D7 branes, with respect to $U\left( N\right) \times U_{\Sigma
}\left( 1\right) $,%
\begin{equation}
\begin{tabular}{llllllllll}
$\underline{\left( N+1\right) ^{2}}$ & $=$ & $\underline{N^{2}}_{0}$ & $%
\oplus $ & $1_{0}$ & $\oplus $ & $\underline{N}_{q}$ & $\oplus $ & $%
\overline{N}_{-q}$ & ,%
\end{tabular}
\label{nq}
\end{equation}%
where the charge $q=N+1$. The charged components in this decomposition
namely $\underline{N}_{q}$ and $\overline{N}_{-q}$ describe precisely
charged matter in bi-fundamentals. Notice that after the rotation of one D7-
brane; say the a-th brane with with gauge group $U_{a}\left( 1\right) $, the
bi-fundamentals $\underline{N}_{q}$ and $\overline{N}_{-q}$ carry the charge
$\left( N,-1\right) $ and $\left( -N,+1\right) $ under the abelian group $%
U_{S}\left( 1\right) \times U_{a}\left( 1\right) $. Comparing with eq(\ref%
{nq}), we find that $U_{\Sigma }\left( 1\right) $ should be identified with
the specific linear combination%
\begin{equation}
q_{\Sigma }=q_{S}-q_{a}
\end{equation}%
In the second case, we consider F-theory on a local CY four-folds with base
surface (\ref{int}) consisting of at least two components surfaces $S_{a}$
and $S_{b}$ with non trivial intersection along a complex curve $S_{a}\cap
S_{b}=\Sigma _{ab}$. So the seven branes wrapping the respective surfaces
\emph{S}$_{a}$ and \emph{S}$_{b}$ intersect in a six-dimensional space $%
\mathbb{R}^{1,3}\times \Sigma _{ab}$. Along $\Sigma _{ab}$, the singularity
in the fiber gets enhanced to $G_{\Sigma _{ab}}$ with new bi-fundamental
matter localized on the curve $\Sigma _{ab}$ determined by decomposing $%
adG_{\Sigma _{ab}}$ with respect to the representation of the bulk gauge
symmetries $G_{_{\emph{S}_{a}}}\times G_{_{\emph{S}_{b}}}$, that is%
\begin{equation}
adG_{\Sigma }=\left( adG_{_{\emph{S}_{a}}},1\right) \text{ \ }\dbigoplus
\text{ \ }\left( 1,adG_{_{\emph{S}_{b}}}\right) \text{ \ }\dbigoplus \text{
\ }\left[ \dbigoplus_{i}\left( \mathcal{U}_{i}^{a},\mathcal{U}%
_{i}^{b}\right) \right]  \label{gi}
\end{equation}%
where $\left( \mathcal{U}_{i}^{a},\mathcal{U}_{i}^{b}\right) $ determine the
\emph{bi-fundamentals} under which matter on $\Sigma _{ab}$ transform.
Notice the two following features:\newline
(a) there is a strong link between the flux deformation of the base geometry
of the Calabi-Yau four-fold and the use of intersecting 4-cycles. \ Indeed,
by setting $G_{\emph{S}_{a}}=G_{\emph{S}}$ and $G_{\emph{S}_{b}}=H_{S}$, the
description using intersecting 4-cycles \emph{S}$_{a}$ and \emph{S}$_{b}$
may be viewed as having a complex surface \emph{S}$_{a}=S$ together with a
gauge bundle E with structure group $H_{S}$.\newline
(b) the intersecting 4-cycles construction and the deformation by fluxes may
be combined altogether. By switching on $U\left( 1\right) $- gauge bundles $%
\mathcal{L}_{a}$ and $\mathcal{L}_{b}$ on the surfaces \emph{S}$_{a}$ and
\emph{S}$_{b}$, the respective gauge symmetries $G_{\emph{S}_{a}}$ and $G_{%
\emph{S}_{b}}$ get broken down to subgroups as shown below%
\begin{equation}
\begin{tabular}{llll}
$G_{\emph{S}_{a}}$ & $\longrightarrow $ & $G_{_{a}}\times U_{a}\left(
1\right) $ & , \\
$G_{\emph{S}_{b}}$ & $\longrightarrow $ & $G_{_{b}}\times U_{b}\left(
1\right) $ & .%
\end{tabular}%
\end{equation}%
This breaking leads to a further decomposition of the bi-fundamental
representations $\left( \mathcal{U}_{i}^{a},\mathcal{U}_{i}^{b}\right) $.
For a given representation $\left( \mathcal{U}^{a},\mathcal{U}^{b}\right) $
in eq(\ref{gi}), we have the typical decomposition%
\begin{equation}
\left( \mathcal{U}^{a},\mathcal{U}^{b}\right) =\dbigoplus_{j}\left(
r_{j}^{a},r_{j}^{b}\right) _{{\small q}_{{\small j}}^{a}{\small ,q}_{{\small %
j}}^{b}}\equiv \dbigoplus_{j}\left( r_{j},\tilde{r}_{j}\right) _{{\small q}_{%
{\small j}}{\small ,p}_{{\small j}}}
\end{equation}%
where $\left( q_{j}^{a},q_{j}^{b}\right) $ are $U_{a}\left( 1\right) \times
U_{b}\left( 1\right) $ charges while $r_{j}^{a}$ and $r_{j}^{b}$ are
representations of $G_{_{a}}$ and $G_{_{b}}$ respectively. Moreover,
following \textrm{\cite{H1,H2,H4}}, the number $N_{\left( r_{j},\tilde{r}%
_{j}\right) }$ of zero modes transforming in the representation $\left(
r_{j},\tilde{r}_{j}\right) _{{\small q}_{{\small j}}{\small ,p}_{{\small j}%
}} $ is given by the bundle cohomology%
\begin{equation}
N_{\left( r_{j},\tilde{r}_{j}\right) _{{\small q}_{{\small j}}{\small ,p}_{%
{\small j}}}}=h^{0}\left( \Sigma ,K_{\Sigma }^{1/2}\otimes \mathcal{L}%
_{a}^{q_{j}}|_{\Sigma }\otimes \mathcal{L}_{b}^{p_{j}}|_{\Sigma }\right) ,
\end{equation}%
where $\mathcal{L}_{a}|_{\Sigma }$ and $\mathcal{L}_{b}|_{\Sigma }$ are the
restriction of the of the bundles $\mathcal{L}_{a}$ and $\mathcal{L}_{b}$ to
the curve $\Sigma $.

\end{document}